\documentclass [a4paper, 11pt] {article}

\usepackage{mathrsfs}
\usepackage{graphics}
\usepackage{graphicx}
\usepackage{subfigure}
\usepackage{latexsym}
\usepackage{euscript}
\usepackage{epsfig,amsmath,amssymb}
\usepackage{slashed}

\long\def\symbolfootnote[#1]#2{\begingroup%
\def\thefootnote{\fnsymbol{footnote}}\footnote[#1]{#2}\endgroup} 

\begin{document}
\begin{center}

{\huge \bf D-term inflation after spontaneous symmetry breaking}

\vspace*{7mm} {\ Bart Clauwens$^{a}$
\symbolfootnote[1]{{E-mail:bartclauwens@gmail.com}}
and Rachel Jeannerot$^{b}$}
\symbolfootnote[2]{E-mail:jeannerot@lorentz.leidenuniv.nl}
\vspace*{.25cm}

${}^{a)}${\it Instituut-Lorentz for Theoretical Physics \& Sterrewacht Leiden,
Niels Bohrweg 2, 2333 CA Leiden, The Netherlands}\\
\vspace*{.1cm} 
${}^{b)}${\it Instituut-Lorentz for Theoretical Physics,
Niels Bohrweg 2, 2333 CA Leiden, The Netherlands}

\end{center}

\begin{abstract}
We show that 1-loop quantum corrections to the potential energy density in supersymmetric hybrid inflation, outside the inflationary valley, can not be neglected. A method is presented to calculate these 1-loop corrections and they are applied to the case of D-term hybrid inflation, where a significant amount of inflation is shown to occur after spontaneous symmetry breaking. Taking this into account improves the agreement with WMAP measurements. A gauge coupling of up to 0.3 is still consistent with the CMB density perturbation. The spectral index is predicted in between 0.98 and 1.00 and the cosmic string contribution to the CMB anisotropy is sufficiently reduced.
\end{abstract}

\tableofcontents

\section{Introduction}

It is widely believed that the early universe goes through a period of accelerated expansion called `inflation' \cite{cil}. This believe has been successfully confirmed by WMAP measurements of the Cosmic Microwave Background radiation, which validate inflation as a mechanism to produce the initial irregularities of our universe \cite{WMA}. Furthermore inflation has been shown to solve a number of cosmological problems, such as the `horizon problem', the `flatness problem' and the `monopole problem'. There are many different inflationary models \cite{lyt}. Experimentally these can be distinguished by their predictions for the main observable quantities: the density perturbation at decoupling, the spectral index of the CMB anisotropy and the cosmic string contribution to the CMB anisotropy \cite{lai}.

These observables are derived from the precise shape of the effective scalar potential of a particular inflation model. Only potentials that contain a nearly flat direction give rise to inflation. Supersymmetric models contain these nearly flat directions naturally (without fine-tuning parameters), because of large cancellations of bosonic and fermionic 1-loop quantum contributions to the effective potential, which give a slight tilt to classically flat directions. In supersymmetric hybrid inflation models, inflation occurs while the scalar fields are on the so called `inflationary valley' and it ends abruptly with a period of spontaneous symmetry breaking. On the inflationary valley the `Coleman-Weinberg' formula \cite{wei} can be applied to get the 1-loop quantum correction to the effective potential \cite{dva}. This gives the shape of the potential from which the aforementioned observables can be derived. For this derivation to be valid, it is crucial that inflation ends abruptly at the end of the inflationary valley, when spontaneous symmetry breaking occurs. This is usually assumed.

To make precise statements about this assumption, one needs to know the 1-loop corrections outside the inflationary valley, during the spontaneous symmetry breaking phase. We will show that one cannot simply generalise the `Coleman-Weinberg' formula to this case, because this introduces a non-renormalizable cut-off dependence and moreover it gives a 1-loop correction on the inflationary valley at odds with earlier calculations. The cause of these problems is that the `Goldstone' boson which enters the Higgs mechanism is not massless anymore for field space points outside the inflationary valley. If we include this `Goldstone' mass in the calculation of the 1-loop corrections, the aforementioned problems disappear.

Using these off-valley 1-loop corrections to the D-term hybrid inflation potential, we show that it is in general not allowed to neglect inflation after spontaneous symmetry breaking. Bearing this in mind, a new analysis is made of all relevant observables, which are in good agreement with the WMAP data for a large portion of the D-term parameter space.

This paper is organised as follows. Section \ref{D-terminflation} explains the basic ideas behind D-term hybrid inflation. The body of the paper is in section \ref{sec1lc}. It contains a way to calculate the 1-loop quantum corrections to the full D-term potential, which is essential to a better understanding of its implications for WMAP observables. Section \ref{colweinfor} introduces the `Coleman-Weinberg formula' for the 1-loop corrections. Then section \ref{fdmasses} goes on with the calculation of the field dependent masses, needed for the Coleman-Weinberg formula. Section \ref{cal1lc} explains a variety of subtleties in applying this formula and stumbles upon fundamental problems in calculating the 1-loop corrections. Section \ref{problemssolution} hopes to resolve these problems and finally in section \ref{results} the result for the 1-loop corrections is given. Section \ref{seciassb} explores the usefulness of these 1-loop corrections. We will see that there is a possibility of inflation even after the spontaneous symmetry breaking has started. In chapter \ref{newbounds} we apply the 1-loop corrections to get new cosmological constraints on the D-term hybrid inflation model. If you are only interested in new parameter bounds for D-term inflation, consult section \ref{newbounds}. This article will end with a small conclusion in section \ref{conclusion}.

\section{D-term inflation}
\label{D-terminflation}

The simplest possible superpotential that can generate D-term inflation is \cite{dte,hal}:
\begin{equation}
W =\lambda S \Phi_{+} \Phi_{-}
\label{eqsp}
\end{equation}
The superfields $\Phi_{+}$ and $\Phi_{-}$ have charge 1 and -1 under a $U(1)_{FI}$ gauge symmetry. $S$ is uncharged and $\lambda$ is a coupling parameter.
The corresponding tree-level effective scalar potential
\footnote{Quantum-mechanically the effective potential can be interpreted as the energy-density of the quantum-state which minimises
this energy-density, subject to the condition that the field expectation values (in this case $<\Phi_{\pm}>, <S>$) are as given. So for every point in field-space you can define
an effective potential value. There are however subtleties in the interpretation of the effective potential in case of a spontaneous symmetry breaking.
These will be discussed in section (\ref{concave}).}
is:
\begin{equation}
V_{tree} = \lambda^{2}|S|^{2}\left(|\Phi_{+}|^{2} + |\Phi_{-}|^{2}\right)
+ \lambda^{2} |\Phi_{+}|^{2} |\Phi_{-}|^{2} + \frac{g^{2}}{2} \left(|\Phi_{+}|^{2}
- |\Phi_{-}|^{2} + \xi\right)^{2}
\label{eqep}
\end{equation}
$S$ is called the inflaton-field, $\xi$ is the Fayet-Illiopoulos gauge term (taken to be positive) and $g$ is the $U(1)_{FI}$ gauge coupling. $\Phi_{+}=\Phi_{-}=0$ represents a flat valley of local minima, because when both
 $\Phi_{+}$ and $\Phi_{-}$ vanish, the tree-level potential does not depend on $S$. In this case the potential energy density equals:
\begin{equation}
V =\frac{g^{2}\xi^{2}}{2}
\label{eqVVac}
\end{equation}
This is the false vacuum energy density, which drives D-term inflation. This inflationary solution ($\Phi_{+}=\Phi_{-}=0$)
can be either stable or unstable, depending on the value of the $S$-field.

For $|S| > S_{c}$ the $\Phi_{-}$-direction corresponds to a (stable) local minimum of the potential, 
but for $|S| < S_{c}$ the $\Phi_{-}$-direction corresponds
to an (unstable) local maximum, which will cause spontaneous symmetry breaking. The fields will settle down in the true global
minimum ($V=0$, $S=0$, $\Phi_{+}=0$, $|\Phi_{-}|=\sqrt{\xi}$) and inflation will stop. The $\Phi_{+}$-direction cannot cause a similar
effect, because it
always corresponds to a (stable) local minimum. The critical value of the inflaton-field is given by:
\begin{equation}
S_{c} \equiv \frac{g \sqrt{\xi}}{\lambda}
\label{eqSc}
\end{equation}

At tree-level (classically) there is no term in the potential which drives the fields toward the true global minimum.
Thus assuming that the fields start out with random values after the Planck-era, they will either settle down in the global minimum
($V=0$, $S=0$, $\Phi_{+}=0$, $|\Phi_{-}|=\sqrt{\xi}$) with no appreciable inflation,
or in the local minimum ($V=\frac{g^{2}\xi^{2}}{2}$, $|S|>S_{c}$, $\Phi_{+}=0$, $\Phi_{-}=0$), in which case inflation goes on forever.

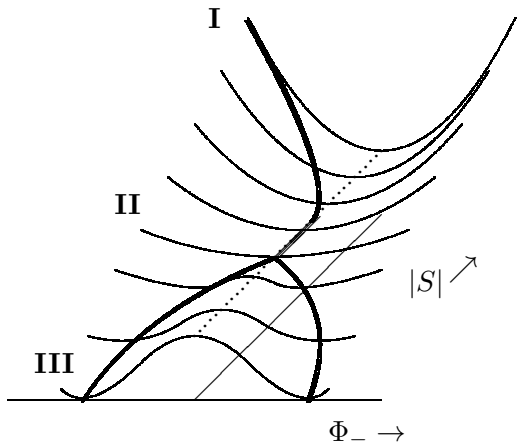
\begin{figure}[htbp]
\begin{center}
\begin{picture}(300,200)
\qbezier(150,180)(200,80)(250,180)
\qbezier(140,160)(190,80)(240,160)
\qbezier(130,140)(180,80)(230,140)
\qbezier(120,120)(170,80)(220,120)

\qbezier(110,100)(160,80)(210,100)

\qbezier(140,80)(150,85)(160,80)
\qbezier(160,80)(170,75)(200,85)
\qbezier(140,80)(130,75)(100,85)

\qbezier(125,65)(140,75)(155,65)
\qbezier(155,65)(170,55)(190,60)
\qbezier(125,65)(110,55)(90,60)

\qbezier(110,50)(130,70)(150,50)
\qbezier(150,50)(170,30)(180,40)
\qbezier(110,50)(90,30)(80,40)

\qbezier(160,90)(185,70)(172,36)
\qbezier(160,90)(110,70)(88,36)
\qbezier(161,89)(186,69)(173,35)
\qbezier(160,89)(110,69)(88,35)
\qbezier(160,89)(185,69)(172,35)

\put(160,90){\line(1,1){16}}
\put(161,89){\line(1,1){16}}
\put(160,89){\line(1,1){16}}
\qbezier(174,104)(186,116)(150,180)
\qbezier(173,103)(185,115)(149,179)
\qbezier(174,103)(186,115)(150,179)

\put(130,60){\circle*{1}}
\put(134,64){\circle*{1}}
\put(138,68){\circle*{1}}
\put(142,72){\circle*{1}}
\put(146,76){\circle*{1}}
\put(150,80){\circle*{1}}
\put(154,84){\circle*{1}}
\put(158,88){\circle*{1}}
\put(162,92){\circle*{1}}
\put(166,96){\circle*{1}}
\put(170,100){\circle*{1}}
\put(174,104){\circle*{1}}
\put(178,108){\circle*{1}}
\put(182,112){\circle*{1}}
\put(186,116){\circle*{1}}
\put(190,120){\circle*{1}}
\put(194,124){\circle*{1}}
\put(198,128){\circle*{1}}

\put(132,62){\circle*{1}}
\put(136,66){\circle*{1}}
\put(140,70){\circle*{1}}
\put(144,74){\circle*{1}}
\put(148,78){\circle*{1}}
\put(152,82){\circle*{1}}
\put(156,86){\circle*{1}}
\put(160,90){\circle*{1}}
\put(164,94){\circle*{1}}
\put(168,98){\circle*{1}}
\put(172,102){\circle*{1}}
\put(176,106){\circle*{1}}
\put(180,110){\circle*{1}}
\put(184,114){\circle*{1}}
\put(188,118){\circle*{1}}
\put(192,122){\circle*{1}}
\put(196,126){\circle*{1}}
\put(200,130){\circle*{1}}

\put(130,36){\line(1,1){70}}
\put(60,36){\line(1,0){140}}

\put(135,180){\makebox(40,0)[l]{\textbf{I}}}
\put(100,110){\makebox(40,0)[l]{\textbf{II}}}
\put(70,50){\makebox(40,0)[l]{\textbf{III}}}
\put(210,80){\makebox(40,0)[l]{$|S|$}}
\put(225,85){\makebox(40,0)[l]{$\nearrow$}}
\put(180,23){\makebox(40,0)[l]{$\Phi_{-}\rightarrow$}}
\end{picture}
\caption{\footnotesize{This figure depicts the different stages for a typical time evolution of the fields in D-term inflation. In region I the $\Phi$-fields
quickly roll down to zero. Then slow-roll inflation starts in region II (also called the inflationary valley). Finally in region III spontaneous symmetry breaking
occurs and the fields roll down to the true global minimum, whereby $\Phi_{-}$ attains a vacuum expectation value.}}
\label{regions}
\end{center}
\end{figure}

However, as shown in Ref. \cite{dva}, (1-loop)-quantum-corrections to the tree-level potential will add a slight slope to the D-flat direction\footnote{Loop corrections only contribute if supersymmetry is broken. If supersymmetry is not broken, the bosonic and fermionic contributions to the loop correction will cancel, so that the total loop-correction vanishes. For this model supersymmetry is broken during inflation and then it is restored at the end of inflation in the global minimum.}. Now the typical evolution of the fields after the Planck-era will consist of three stages: I) First the $\Phi_{+}$ and $\Phi_{-}$ fields 
will quickly roll down to the D-flat valley ($|S|>S_{c}$, $\Phi_{+}=0$, $\Phi_{-}=0$). II) Then a period of 
slow-roll-inflation starts. The S-field rolls down until $S_{c}$ is reached. III) 
Now the spontaneous symmetry breaking puts a natural halt to the inflationary phase and the 
fields settle in the true global minimum, see figure (\ref{regions}).

For inflation-model builders, the most important part of the potential is the valley ($|S|>S_{c}$, $\Phi_{+}=0$, $\Phi_{-}=0$),
which causes inflation (region II). Nevertheless the other parts of the potential can also be of interest.
Region I is of interest if you want to study the initial conditions after the Planck-era \cite{pan,pat}. In fact not all homogeneous initial
conditions give inflation, but most will, because the tree-level potential is very steep in the direction
toward the inflationary valley. Region III is of interest if you want to study reheating. The initial value problem will probably not
lead to observable consequences, because everything happening more than about 60 e-folds before the end of inflation will not have entered our observable universe today (it will have left its imprint on scales larger than our observable universe).
On the other hand all processes occurring in region III will be of observational interest, because they involve length-scales that are within our past light-cone.

The study of quantum-corrections to the effective potential in these two regimes however is still far from finished.
Actually the only part for which we know the 1-loop-corrections for certain is the inflationary valley
($|S|>S_{c}$, $\Phi_{+}=0$, $\Phi_{-}=0$) (See Ref. \cite{dva}). Work has been done however in the preheating region, Ref. \cite{jea,mac}. In the following we will try to generalise their approach and to point out possible errors.

\section{1-loop corrections}
\label{sec1lc}

\subsection{Coleman Weinberg formula for calculating 1-loop corrections to the effective potential}
\label{colweinfor}

To compute the 1-loop correction to the effective potential people normally use the Coleman-Weinberg formula. 
It is the total contribution to the effective potential of all Feynman diagrams, which have truncated external scalar propagators with zero momentum and one
internal loop with any possible particle going around (in this case one of the scalar particles, a fermionic super-partner, 
a gaugefield or a gaugino).
For the derivation in a general (non-SUSY) context see Ref. \cite{wei}. It is given by:
\begin{equation}
\Delta V = \frac{1}{64 \pi^{2}}\sum_{i}\left(-1\right)^{F}m_{i}^{4} ln\left(\frac{m_{i}^{2}}{\Lambda_{m}^{2}} \right)\; ,
\label{eqcw}
\end{equation}
with $m_{i}$ being the mass of a given particle, the sum goes over all particles,
$F$ can be taken 0 for bosons and 1 for fermions and $\Lambda_{m}$ is the renormalization mass.
But this formula is really a simplification of the more general \cite{fer}:
\begin{displaymath}
\Delta V = \frac{1}{64 \pi^{2}}Str\left(M^{0}\right)\Lambda_{c}^{4}ln\left(\frac{\Lambda_{c}^{2}}{\mu^{2}}\right) +
\end{displaymath}
\begin{equation}
\frac{1}{32\pi^{2}}Str\left(M^{2}\right)\Lambda_{c}^{2}+\frac{1}{64 \pi^{2}}Str\left(M^{4}ln\left(\frac{M^{2}}{\Lambda_{c}^{2}}\right)\right)+... \; ,
\label{eqf}
\end{equation}
with $\Lambda_{c}$ being a momentum cut-off and $\mu$ the scale parameter. The dots stand for $\Lambda_{c}$-independent contributions. $Str$ is called the supertrace and
is defined as follows:
\begin{equation}
Str\left(M^{n}\right)=\sum_{i}\left(-1\right)^{2J_{i}}\left(2J_{i}+1\right)m_{i}^{n} \; ,
\end{equation}
with $m_{i}$ being the field dependent mass-eigenvalues, $J_{i}$ being the spin of the corresponding particles. 
The supertrace weights the number of degrees of freedom and
gives fermions the opposite sign of bosons.

It is important to emphasise that the two $\Lambda$'s are not the same. They are sometimes confused because they appear in the same form (that's why there is an added subscript).
The renormalization mass $\Lambda_{m}$ is an arbitrary mass-scale the theory should not dependent on, 
which is guaranteed via the renormalization group equations. The momentum cut-off $\Lambda_{c}$ on the other hand
 will go to infinity after renormalization,
 so the potential had better not depend on this after renormalization. Both $\Lambda$'s appear in a similar way in the formulae.
 In fact equation (\ref{eqcw}) comes from the last term in equation (\ref{eqf}) after renormalizing (which makes $\Lambda_{c}$-dependence
 of this term disappear, but introduces a `nicer' $\Lambda_{m}$-dependence, which happens to be of the same form).
 
 Under what circumstances can we use equation (\ref{eqcw}) rather than equation (\ref{eqf})? The first term in (\ref{eqf}) will drop out in any supersymmetric theory, because
 $Str (M^{0})$ equals the number of bosonic degrees of freedom minus the number of fermionic degrees of freedom, which will be zero
 in any supersymmetric theory. What about the second term? In Ref. \cite{fgp} it is derived that under suitable requirements (meaning no anomalies) for
 supersymmetric theories this term also vanishes. If however we look at the derivation of this cancellation we see that explicit use has
 been made of the fact that all first order partial derivatives of the tree-level effective potential to the fields vanish. This shrinks the
 region of applicability of the simple Coleman-Weinberg formula. Only at extremum (or saddle) points are we justified to throw 
 away the second term of equation (\ref{eqf}). For the case
 at hand this means that we can only use equation (\ref{eqcw}) in the global minimum and along the inflationary valley,
 because here all derivatives in the tree-level potential vanish.

 We stress this point, because these formulae are not always used with the appropriate care. In Ref. \cite{mac} the 1-loop corrections to the
 reheating-part of the potential are calculated, using the simplified formula, without checking whether the other part really drops out.
  Let us now look at this case in full detail.
  
\subsection{Calculation of the field dependent masses in D-term inflation}
\label{fdmasses}

\subsubsection{Introduction}

To calculate the 1-loop correction we first have to find the masses of the fundamental particles in the theory.
These will be dependent on the field-values and so in the different regimes we may have different masses. Mass-terms
 are all terms in Lagrangian that are of second order in the fields, if we expand around any
 given point in field-space. The most appropriate expansion is as follows:
 \begin{equation}
 S=\frac{1}{\sqrt{2}}(S_{1}+s_1+i(S_{2}+s_{2}))
\label{eq1}
\end{equation}
\begin{equation}
 \Phi_{+}=\frac{1}{\sqrt{2}}(\Phi_{+1}+\phi_{+1}+i(\Phi_{+2}+\phi_{+2}))
\label{eq2}
 \end{equation}
 \begin{equation}
 \Phi_{-}=\frac{1}{\sqrt{2}}(\Phi+\phi)e^{i(\Theta+\theta)} \; .
 \label{eqfm}
 \end{equation}
All capital letters denote the fixed values of the fields around which we make an expansion.
The small letters can be interpreted as the field-excitations with a certain mass. Of course other choices
can be made, for example $\Phi_{+}$ and $\Phi_{-}$ are treated differently.

The well known expansion around
the inflationary valley consists of setting $\Phi_{+1}$, $\Phi_{+2}$, $\Phi$ and $\Theta$ to zero and taking $|S|>S_{c}$.
If we want to know the loop-corrections off-valley in principle all fixed values are nonzero. I will comment later on this
possibility, but for the moment it suffices to say that this will give very messy calculations, 
so we should be more restrictive in our choice of fixed field values. In fact we can set $\Phi_{+1}$ and $\Phi_{+2}$ to
zero, if we restrict our attention to regions II and III, because the spontaneous symmetry 
breaking part for the D-term potential will be in the $\Phi_{-}$-sector. As discussed
before the $\Phi_{+}$ will always be in a stable local maximum $\Phi_{+}=0$.
One could also argue that nothing should depend on the phases
of the $S$- and $\Phi_{-}$-fields, since the tree-level potential is also not dependent on these phases\footnote{unlike in the F-term inflation case, where 
the tree-level potential does depend on the phases of the $\Phi_{+}$- and $\Phi_{-}$-fields.}.
So one could take $S$- and $\Phi_{-}$-fields that are real.
Here we will take them complex from the start, but we will indeed see in the end that the 1-loop correction is independent
on these phases.

\subsubsection{The scalar masses}
\label{scalar}

Plugging the expansion from equations (\ref{eq1},\ref{eq2},\ref{eqfm}) into the effective potential, equation (\ref{eqep}), we can immediately read of the mass of the $\phi_{+1}$ and $\phi_{+2}$ fields. 
For the $s_{1}$, $s_{2}$ and $\phi$ fields we first have to diagonalize the second order part of the Lagrangian in order to
find the physical particles (which should have simple propagators, so no cross-terms):
\begin{equation}
\frac{1}{2}
\left( \begin{array}{ccc}
s_{1} & s_{2} & \phi
\end{array} \right)
\left( \begin{array}{ccc}
\frac{1}{2}\lambda^{2}\Phi^{2} & 0 & \lambda^{2}\Phi S_{1} \\
0 & \frac{1}{2}\lambda^{2}\Phi^{2} & \lambda^{2}\Phi S_{2} \\
\lambda^{2}\Phi S_{1} & \lambda^{2}\Phi S_{2} & \frac{1}{2}\lambda^{2}(S_{1}^2+S_{2}^{2})-g^{2}\xi+\frac{3}{2}g^{2}\Phi^{2}
\end{array} \right)
\left( \begin{array}{ccc}
s_{1} \\
s_{2} \\
\phi
\end{array} \right) \; .
\end{equation}
The corresponding $(mass)^{2}$-eigenvalues and (non-normalised) eigenstates are given in table (\ref{tabdiagsec}),
\begin{table}[htbp]
\begin{center}
\begin{tabular}{|c|c|}
\hline
$(mass)^{2}$-eigenvalue & (non-normalised) eigenstate \\ \hline
$\frac{1}{2}\lambda^{2}\Phi^{2}$ &
$\frac{1}{\sqrt{S_{1}^{2}+S_{2}^{2}}}
\left( \begin{array}{ccc}
-S_{2}\\
S_{1}\\
0
\end{array} \right)$
 \\ \hline
$\frac{1}{2}\left(B-\sqrt{B^{2}+4C}\right)$ &
$\left( \begin{array}{ccc}
2\lambda^{2}\Phi S_{1}\\
2\lambda^{2}\Phi S_{2}\\
D-\sqrt{B^{2}+4C}
\end{array} \right)
$ \\ \hline
$\frac{1}{2}\left(B+\sqrt{B^{2}+4C}\right)$ &
$\left( \begin{array}{ccc}
2\lambda^{2}\Phi S_{1}\\
2\lambda^{2}\Phi S_{2}\\
D+\sqrt{B^{2}+4C}
\end{array} \right)$ \\ \hline
\end{tabular}
\caption{\footnotesize{Diagonalization of the ($s_{1},s_{2},\phi$)-sector}}
\label{tabdiagsec}
\end{center}
\end{table}
where we used the following definitions:
\begin{equation}
B=\frac{1}{2}\lambda^{2}(S_{1}^{2}+S_{2}^{2}+\Phi^{2})+\frac{3}{2}g^{2}\Phi^{2}-g^{2}\xi
\label{eqB}
\end{equation}
\begin{equation}
D=\frac{1}{2}\lambda^{2}(S_{1}^{2}+S_{2}^{2}-\Phi^{2})+\frac{3}{2}g^{2}\Phi^{2}-g^{2}\xi
\end{equation}
\begin{equation}
C=\frac{3}{4}\lambda^{4}\Phi^{2}(S_{1}^{2}+S_{2}^{2})+\frac{1}{2}\lambda^{2}g^{2}\xi\Phi^{2}-\frac{3}{4}\lambda^{2}g^{2}\Phi^{4} \; .
\label{eqC}
\end{equation}
Now we can give a complete list of all scalar-masses in the theory, see table (\ref{tabelsca}).
\begin{table}[htbp]
\begin{center}
\begin{tabular}{|c|c|c|}
\hline
field & d.o.f. & $(mass)^{2}$ \\ \hline
$\phi_{+1}$ & 1 & $\frac{1}{2}\lambda^{2}(S_{1}^{2}+S_{2}^{2}+\Phi^{2})-\frac{1}{2}g^{2}\Phi^{2}+g^{2}\xi$\\ \hline
$\phi_{+2}$ & 1 &...same...\\ \hline
lin. comb. of $s_{1},s_{2}$ & 1 & $\frac{1}{2}\lambda^{2}\Phi^{2}$ \\ \hline
lin. comb. of $s_{1},s_{2},\phi$ & 1 & $\frac{1}{2}\left(B-\sqrt{B^{2}+4C}\right)$ \\ \hline
lin. comb. of $s_{1},s_{2},\phi$ & 1 & $\frac{1}{2}\left(B+\sqrt{B^{2}+4C}\right)$ \\ \hline

\end{tabular}
\caption{\footnotesize{Mass and degrees of freedom of the scalar fields}}
\label{tabelsca}
\end{center}
\end{table}

\subsubsection{The gauge boson mass}

In the previous section we did not get a mass for the $\theta$-field. Actually this is because we have chosen to write the complex
field $\Phi_{-}$ in (\ref{eqfm}) in terms of a modulus $\phi$ and an angular part $\theta$. This $\theta$-field has a nontrivial
kinetic term, which does not allow us to interpret the second-order-part as a mass, like we did for the other fields. However the reason for writing things out
like this is that the interaction of the $\theta$-field with the massless gauge boson $A^{\mu}$ will cancel all $\theta$-dependence
of the Lagrangian in favour of a gauge boson $B^{\mu}$, which acquires mass. This is the so called Higgs-mechanism.

$\Phi_{-}$ has a charge $-1$ under the local gauge transformation\footnote{under the gauge transformation $\Phi \to e^{i q \xi}\Phi$
and $A_{\mu} \to A_{\mu} + \frac{1}{g} \partial_{\mu}\xi$, with q the charge of the fields, so $q = \pm 1,0$ for $\Phi_{\pm},S$.}. Hence the gauge-invariant kinetic term for $\Phi_{-}$ is:
\begin{displaymath}
\mathscr{L}_{kin. \Phi_{-}}=(D_{\mu}\Phi_{-})^{\dagger}D^{\mu}\Phi_{-}
=(\partial_{\mu}-i g A_{\mu})\Phi_{-}^{*}(\partial^{\mu}+i g A^{\mu})\Phi_{-}
\end{displaymath}
\begin{equation}
=\frac{1}{2}\partial_{\mu}\phi\partial^{\mu}\phi+\frac{1}{2}(\Phi+\phi)^{2}\left(\partial_{\mu}\theta\partial^{\mu}\theta+g^{2}
A_{\mu}A^{\mu}+2 g \partial_{\mu}\theta A^{\mu}\right) \; .
\label{eqkin}
\end{equation}
There is a `nasty' kinetic term for $\theta$ and an unwanted coupling of $\partial_{\mu}\theta$ to $A^{\mu}$. This quadratic coupling means we have not identified the physical
particles correctly (it represents a mixed propagator). What we should do is write everything in terms of the gauge-invariant
combination $B_{\mu}=A_{\mu}+\frac{1}{g}\partial_{\mu}\theta$. In doing so, equation (\ref{eqkin}) gets a simple form:
\begin{equation}
\mathscr{L}_{kin. \Phi_{-}}=\frac{1}{2}\partial_{\mu}\phi\partial^{\mu}\phi+\frac{1}{2}g^{2}(\Phi+\phi)^{2}B_{\mu}B^{\mu} \; .
\label{eqkin2}
\end{equation}
All $\theta$-dependence has cancelled and the gauge field has acquired a mass (a mass-term for a gauge boson looks like
$\frac{1}{2}m^{2}A_{\mu}A^{\mu}$). The massive $B_{\mu}$-field now has three degrees of freedom instead of two for the massless $A_{\mu}$-field.
The $\theta$-field behaves as the longitudinal component of $B_{\mu}$. The kinetic term for the gauge field remains unchanged under the
redefinition, because the partial derivatives in $F_{\mu\nu}=(\partial_{\mu}A_{\nu}-\partial_{\nu}A_{\mu})$ arising from the shift commute. The mass of the gauge field is included in table (\ref{tabelgau}). 
\begin{table}[htbp]
\begin{center}
\begin{tabular}{|c|c|c|}
\hline
field & d.o.f. & $(mass)^{2}$ \\ \hline
$B_{\mu}$ & 3 & $g^{2}\Phi^{2}$ \\ \hline
\end{tabular}
\caption{\footnotesize{Mass and degrees of freedom of the gauge field}}
\label{tabelgau}
\end{center}
\end{table}

In principle the $\Phi_{+}$-field should be treated in the same way, but in this case it gives just simple kinetic- and mass-terms,
because we expanded around the point $\Phi_{+}=0$. If you expand around another point, you will get a more complicated interaction, but it will still be possible to define a Higgs-mechanism which works for all field-space points.

Note that this derivation is done at any field value $\Phi$, so not only in the global minimum, as is done usually. The striking
thing is that the $\theta$-dependence really drops out everywhere. A possible exception being the origin $\Phi = 0$, because here
there is a coordinate singularity, which makes $B_{\mu}$ ill defined.

\subsubsection{The fermion masses}

In any supersymmetric theory all scalar bosons have fermionic super-partners and all gauge-fields have corresponding fermionic
gaugino-fields, see for example Ref. \cite{mar}.
The terms quadratic in the fermionic fields are summarised in the following matrix, which must again be diagonalized
in order to find the field-dependent fermionic masses of the theory:
\begin{equation}
\left( \begin{array}{cccc}
\psi_{A} & \psi_{-} & \psi_{+} & \psi_{S} \\
\end{array} \right)
\left( \begin{array}{cccc}
0 & g\Phi e^{i\Theta} & 0 & 0\\
g\Phi e^{-i\Theta} & 0 & \frac{\lambda}{\sqrt{2}}(S_{1}+i S_{2}) & 0\\
0 & \frac{\lambda}{\sqrt{2}}(S_{1}-i S_{2}) & 0 & \frac{\lambda}{\sqrt{2}}\Phi e^{i\Theta} \\
0 & 0 & \frac{\lambda}{\sqrt{2}}\Phi e^{-i\Theta} & 0 \\
\end{array} \right)
\left( \begin{array}{cccc}
\psi_{A} \\
\psi_{-} \\
\psi_{+} \\
\psi_{S} \\
\end{array} \right)
\end{equation}
$\psi_{-}, \psi_{+}$ and $\psi_{S}$ are the Weyl-fermions belonging to the chiral multiplet and $\psi_{A}$ is the gaugino. The corresponding mass-eigenvalues are:
\begin{equation}
\pm\frac{1}{2}\sqrt{E\pm\sqrt{E^{2}-8\lambda^{2}g^{2}\Phi^{4}}} \; ,
\end{equation}
with
\begin{equation}
E=\lambda^{2}(S_{1}^{2}+S_{2}^{2}+\Phi^{2})+2g^{2}\Phi^{2} \; .
\end{equation}
This completes the calculation of the fermionic masses. The results are summarised in table (\ref{tabelfer}).
\begin{table}[htbp]
\begin{center}
\begin{tabular}{|c|c|c|}
\hline
field & d.o.f. & $(mass)^{2}$ \\ \hline
lin. comb. of $\psi_{A},\psi_{-},\psi_{+},\psi_{S}$ & 4 & $\frac{1}{4}\left(E+\sqrt{E^{2}-8\lambda^{2}g^{2}\Phi^{4}}\right)$ \\ \hline
lin. comb. of $\psi_{A},\psi_{-},\psi_{+},\psi_{S}$ & 4 & $\frac{1}{4}\left(E-\sqrt{E^{2}-8\lambda^{2}g^{2}\Phi^{4}}\right)$ \\ \hline
\end{tabular}
\caption{\footnotesize{Mass and degrees of freedom of the fermionic fields}}
\label{tabelfer}
\end{center}
\end{table}

\subsection{Calculating the 1-loop corrections to the D-term potential}
\label{cal1lc}

Insertion of all masses in equation (\ref{eqcw}) in order to get a formula for the total 1-loop correction is in principle straightforward. Some details can be found in the corresponding appendices. first of all massless particles do not contribute (Appendix \ref{secmassless}). For tachyonic masses we should only take the real part of the corresponding 1-loop correction (Appendix \ref{concave}) and imaginary mass-squared values do not occur (Appendix \ref{imaginary}).

\subsubsection{Cut-off dependence}

Since we have obtained well defined masses
for all particles as a function of $S_{1}, S_{2}$ and $\Phi$ we can explicitly calculate the second term in equation (\ref{eqf}):
\begin{equation}
\frac{1}{32\pi^{2}}Str\left(M^{2}\right)\Lambda_{c}^{2} \; .
\end{equation}
The supertrace should be equal to zero in order to make sense of the theory. Otherwise, when $\Lambda_{c}$ goes to infinity (or to a very large value
like the Planck-mass), the whole term blows up.
Indeed in calculating the supertrace most contributions cancel out, but not all. We're left with:
\begin{equation}
Str (M^{2})=- \frac{1}{2}\lambda^{2}(S_{1}^{2}+S_{2}^{2})-\frac{1}{2}g^{2}\Phi^{2}+g^{2}\xi
\label{equv}
\end{equation}
There is no renormalization-counterterm to cancel this field-dependent divergence. So we conclude that the effective potential
is really ultraviolet divergent, except possibly for field-values where equation (\ref{equv}) vanishes. This happens exactly along
the "spontaneous symmetry breaking valley", by which I mean the valley determined by minimising $\Phi$ for a given $S_{1}$ and $S_{2}$:
\begin{equation}
\frac{\partial V_{tree}}{\partial \Phi}= 0 \Longrightarrow
\frac{1}{2}\lambda^{2}(S_{1}^{2}+S_{2}^{2})-\frac{1}{2}g^{2}\Phi^{2}+g^{2}\xi = 0 \vee
\Phi = 0 \; .
\label{eqssbv}
\end{equation}

\subsubsection{Limit to the inflationary valley}

A good check as to whether the method in the enlarged parameter-space makes sense is to check if it gives the same results as before on the inflationary
valley ($\Phi = 0$). Mathematically you could argue that there is a problem in taking the limit $\Phi\rightarrow 0$, because $B_{\mu}$
is not well-defined for $\Phi = 0$. If you insist, take a very small value for $\Phi$, then $B_{\mu}$ will be well defined and
the Lagrangian will have no dependence on $\theta$.
The particle-masses we get along the valley are summarised in table (\ref{tabelinfval}).
\begin{table}[htbp]
\begin{center}
\begin{tabular}{|c|c|c|}
\hline
field & d.o.f. & $(mass)^{2}$ \\ \hline
$\phi_{+1}$ & 1 & $\frac{1}{2}\lambda^{2}(S_{1}^{2}+S_{2}^{2})+g^{2}\xi$\\ \hline
$\phi_{+2}$ & 1 &...same...\\ \hline
$\frac{1}{\sqrt{S_{1}^{2}+S_{2}^{2}}}(-S_{2} s_{1}+S_{1} s_{2})$ & 1 & $0$ \\ \hline
$\frac{1}{\sqrt{S_{1}^{2}+S_{2}^{2}}}(S_{1} s_{1}+S_{2} s_{2})$ & 1 & $0$ \\ \hline
$\phi$ & 1 & $\frac{1}{2}\lambda^{2}(S_{1}^{2}+S_{2}^{2})-g^{2}\xi$ \\ \hline
$B_{\mu}$ & 2 or 3 & 0 \\ \hline
lin. comb. of $\psi_{A},\psi_{-},\psi_{+},\psi_{S}$ & 4 & $\frac{1}{2}\lambda^{2}(S_{1}^{2}+S_{2}^{2})$ \\ \hline
lin. comb. of $\psi_{A},\psi_{-},\psi_{+},\psi_{S}$ & 4 & $0$ \\ \hline
\end{tabular}
\caption{\footnotesize{Mass and degrees of freedom of all fields along the inflationary valley in the limit $\Phi\rightarrow 0$ of the enlarged parameter-space method. The gauge field has 2 degrees of freedom
if $\Phi$ is really zero and three degrees of freedom if $\Phi$ is very small. For the computation of the 1-loop corrections it makes no
difference ($2 \cdot 0 = 3\cdot 0= 0$).}}
\label{tabelinfval}
\end{center}
\end{table}

The 1-loop corrections on the inflationary valley are well known and easy to calculate if we just set $\Phi_{-}$ equal to zero from the very beginning. 
The particle-masses that contribute are given in table (\ref{tabelinfval2}).
\begin{table}[htbp]
\begin{center}
\begin{tabular}{|c|c|c|}
\hline
field & d.o.f. & $(mass)^{2}$ \\ \hline
$\phi_{+1}$ & 1 & $\frac{1}{2}\lambda^{2}(S_{1}^{2}+S_{2}^{2})+g^{2}\xi$\\ \hline
$\phi_{+2}$ & 1 &...same...\\ \hline
$\phi_{-1}$ & 1 & $\frac{1}{2}\lambda^{2}(S_{1}^{2}+S_{2}^{2})-g^{2}\xi$\\ \hline
$\phi_{-2}$ & 1 &...same...\\ \hline
lin. comb. of $\psi_{-},\psi_{+}$ & 4 & $\frac{1}{2}\lambda^{2}(S_{1}^{2}+S_{2}^{2})$ \\ \hline
\end{tabular}
\caption{\footnotesize{Mass and degrees of freedom of all massive fields along the inflationary valley as it is commonly used.}}
\label{tabelinfval2}
\end{center}
\end{table}

Comparison of these two tables points out that the two methods really give different results. In the usual case there is an extra scalar particle
of $(mass)^{2}$ equal to $\frac{1}{2}\lambda^{2}(S_{1}^{2}+S_{2}^{2})-g^{2}\xi$. As a result the 1-loop correction derived from both methods will
also differ. This makes the enlarged parameter-space method not trustable.

\subsubsection{The cut-off dependence and the limit problem}

Obviously the described method similar to the one in Ref. \cite{mac} cannot be correct. Results are unphysical
 because of the cut-off dependence and just wrong because of the contradiction with the previous method along the inflationary valley.
 In the following we will describe a possible method which doesn't suffer from these two problems.
 
 The origin of the limit-problem seems to be the Higgs-mechanism. The Higgs-mechanism dictates that we write the $\Phi_{-}$-field
  in polar coordinates. In doing so the angular $\theta$-field becomes massless.
 This could well be the mass we're missing with respect to the usual case. Furthermore the cut-off dependent supertrace contribution in
 equation (\ref{equv}) also amounts to the same $(mass)^{2}$-value of $\frac{1}{2}\lambda^{2}(S_{1}^{2}+S_{2}^{2})-g^{2}\xi$ 
 in the limit $\Phi$ goes to zero. So maybe we can solve both problems at a time by evaluating the 
 loop-contributions with the masses given by the theory before applying the Higgs-mechanism.
 First I will show that this indeed solves the aforementioned problems. Then we will argue about the physical validity of this method.

 \subsubsection{A possible solution}

 Since we suspect that the problems come from the massless $\theta$-field we will use the following expansion:
  \begin{equation}
 S=\frac{1}{\sqrt{2}}(S_{1}+s_1+i(S_{2}+s_{2}))
\end{equation}
\begin{equation}
 \Phi_{+}=\frac{1}{\sqrt{2}}(\phi_{+1}+i\phi_{+2})
 \end{equation}
 \begin{equation}
 \Phi_{-}=\frac{1}{\sqrt{2}}(\Phi_{-1}+\phi_{-1}+i (\Phi_{-2}+\phi_{-2}))
 \end{equation}
 Also we will set $S_{2}$ and $\Phi_{-2}$ to zero, because otherwise the calculations become unnecessary
 bothersome and we have seen before that these phases really
 make no difference in D-term inflation. Going through the same steps as before in calculating the scalar masses (see section \ref{scalar}) we
 get the following mixing between the $(s_{1}, s_{2}, \phi_{-1}, \phi_{-2})$-fields:
\begin{equation}
\frac{1}{2}
\left( \begin{array}{cccc}
\phi_{-1} & \phi_{-2} & s_{1} & s_{2}
\end{array} \right)
\end{equation}
\begin{displaymath}
\left( \begin{array}{cccc}
\frac{1}{2}\lambda^{2} S_{1}^{2} + \frac{3}{2}g^{2}\Phi_{-1}^{2}-g^{2}\xi & 0 & \lambda^{2} S_{1} \Phi_{-1} & 0 \\
0 & \frac{1}{2}\lambda^{2} S_{1}^{2} + \frac{1}{2}g^{2}\Phi_{-1}^{2}-g^{2}\xi & 0 & 0 \\
\lambda^{2} S_{1} \Phi_{-1} & 0 & \frac{1}{2}\lambda^{2} \Phi_{-1}^{2} & 0\\
0 & 0 & 0 &  \frac{1}{2}\lambda^{2} \Phi_{-1}^{2}
\end{array} \right)
\left( \begin{array}{cccc}
\phi_{-1} \\
\phi_{-2} \\
s_{1} \\
s_{2} \\
\end{array} \right) \; .
\end{displaymath}
Diagonalizing this $(mass^{2})$-matrix and going back to the original notation (replacing $S_{1}$ by $(S_{1}+S_{2})$ and $\Phi_{-1}$ by $\Phi$)
gives the following field dependent particle masses, see table (\ref{scamass}).
\begin{table}[htbp]
\begin{center}
\begin{tabular}{|c|c|c|}
\hline
field (for $S_{2}=0$  and $\Phi_{-2}=0$) & d.o.f. & $(mass)^{2}$ \\ \hline
$\phi_{+1}$ & 1 & $\frac{1}{2}\lambda^{2}(S_{1}^{2}+S_{2}^{2}+\Phi^{2})-\frac{1}{2}g^{2}\Phi^{2}+g^{2}\xi$\\ \hline
$\phi_{+2}$ & 1 &...same...\\ \hline
$s_{2}$ & 1 & $\frac{1}{2}\lambda^{2}\Phi^{2}$ \\ \hline
$\phi_{-2}$ & 1 & $\frac{1}{2}\lambda^{2}(S_{1}^{2}+S_{2}^{2}) +\frac{1}{2}g^{2} \Phi^{2} - g^{2}\xi$ \\ \hline
lin. comb. of $s_{1},\phi_{-1}$ & 1 & $\frac{1}{2}\left(B-\sqrt{B^{2}+4C}\right)$ \\ \hline
lin. comb. of $s_{1},\phi_{-1}$ & 1 & $\frac{1}{2}\left(B+\sqrt{B^{2}+4C}\right)$ \\ \hline
\end{tabular}
\caption{\footnotesize{Mass and degrees of freedom of the scalar fields, without applying the Higgs-mechanism (meaning there are still quadratic mixing-terms with $A_{\mu}$ left).
The fields are given for
$S_{2}$ and $\Phi_{-2}$ set to zero. For other fixed values the $(mass)^{2}$-eigenfields are different, but the mass-spectrum is the same. The definition of $B$ and $C$ is as before, see equations (\ref{eqB}) and (\ref{eqC}).
 }}
\label{scamass}
\end{center}
\end{table}
 
 The mass-spectrum is the same as in table (\ref{tabelsca}) with an extra mass for the $\phi_{-2}$-field. Comparing this extra mass with equation (\ref{equv}) reveals that using this method the $\Lambda_{c}^{2}$-dependence really drops out. Also In the limit $\Phi \rightarrow 0$ the scalar masses in tables (\ref{scamass}) and (\ref{tabelinfval2}) become identical, resulting in a correct  smooth limit to the known 1-loop corrections along the inflationary valley. To be precise, in table (\ref{scamass}) $\phi_{-2}$ and one of the linear combinations of $(s_{1},\phi_{-1})$ get the same mass, the other linear combination becomes massless.
 
 This new method works because the masses of the gauge field and the fermionic fields remain unaffected by this different choice of coordinates, so all other masses remain the same as before. For the fermionic part this is trivial, because the fermionic masses only depend on the fixed values of the scalar fields. For the gauge field this is less trivial, but comparison of equations (\ref{eqkin}) and (\ref{eqkin2}) reveals that the mass of the gauge field is unaffected by the Higgs mechanism. In both cases the $(mass)^{2}$-value equals $g^{2} \Phi^{2}$. In principle the difference between both methods lies in the inclusion of Feynman-graphs with the mixed 'Goldstone'-gauge propagator. In the next section we will present another way of viewing this situation.

 \subsubsection{Validity of the method}

 Of course we can't just change the calculational method, because there is a physical background behind this. The reason for trying this out anyway
 is that there is no clearcut
 answer in the literature which relates the Higgs-mechanism to the Coleman-Weinberg formula and which states explicitly
 what are the masses to be used in applying
 the Coleman-Weinberg formula in this setting.
 This is because normally if you only look at the global minimum it really makes no difference, since here the Goldstone mode is massless anyway and, as is shown in appendix \ref{secmassless}, massless fields do not contribute to the 1-loop correction.

The Coleman-Weinberg formula, as used before, includes all 1-loop Feynman graph contributions to
the effective potential, where the external truncated scalar legs have zero momentum (since the vacuum has no momentum).
The particle in the internal loop can be any particle in the theory that couples to the scalar external legs. In the cut-off dependent $Str(M^{2})$-part
of the Coleman-Weinberg formula it really makes no difference whether we take the $(mass)^{2}$-values of the physical particles, after diagonalizing the
$(mass)^{2}$-matrix, or we take the 'unphysical' particle masses (without diagonalizing this matrix). This is because the trace of a matrix
is invariant under a change of basis
\footnote{Use $Tr(A B) = Tr(B A)$. If matrix M is diagonalized as follows $M = C D C^{-1}$, then
$Tr(M)=Tr(C D C^{-1}) = Tr(C C^{-1} D) = Tr(D)$. Also $Tr(M^{n}) = Tr(D^{n})$.}. Somehow in mixing the scalar 
field with the gauge field via the Higgs mechanism, it seems that it does make a difference which basis we use. 

In appendix \ref{apD} we describe a Higgs mechanism for a simple U(1) invariant potential, using both polar and Cartesian field definitions. Surprisingly for the global U(1) case the `Goldstone' boson has a mass, if it is outside the global minimum. For the local case one usually uses polar fields, because of the Higgs mechanism. However it is also possible to define the equivalent Higgs mechanism with Cartesian fields. Then the mixed propagator term vanishes, as does the kinetic term for the `Goldstone' field, but the `Goldstone' mass-term does not vanish\footnote{Maybe `mass-term' is not the right word to use here, because for the local U(1) case it no longer corresponds to a pole in the propagator. It is the term quadratic in the corresponding `Goldstone' field}.

\begin{figure}[!h]
\center
\includegraphics{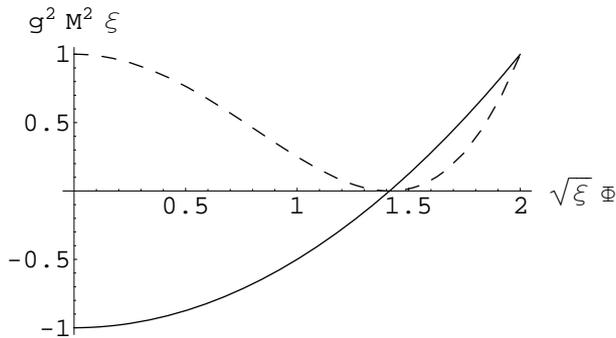}
\caption{\footnotesize{The solid line gives the $(Mass)^{2}$-value of the global `Goldstone' mode as a smooth function of the $\Phi$-field for $S = 0$. The dashed line
depicts the shape of the tree-level effective potential (the vertical axis doesn't refer to this). Note that the Goldstone mode
is indeed massless at the global minimum of the potential.}}
\label{plaatje_1}
\end{figure}

Of course the discussion in appendix \ref{apD} holds equally well for the Mexican hat type U(1) symmetry we're interested in. In this case the $(mass)^{2}$-value of the global
Goldstone mode will be a smooth function of $\Phi$, which goes from a negative value at the top of the Mexican hat smoothly to zero
at the global minimum circle to positive values outside this circle, see figure (\ref{plaatje_1}). Surprisingly the $(mass)^{2}$ of this Goldstone mode
$\frac{1}{2}\lambda^{2}(S_{1}^{2}+S_{2}^{2})+\frac{1}{2}g^{2}\Phi^{2}-g^{2}\xi$ is exactly what we need to solve the cut-off dependence and the limit problem. So using second derivatives of the potential for the classical masses in the Coleman Weinberg formula gives a well-defined 1-loop correction even outside the global minimum and using the tree-level propagator masses doesn't.

There are now two ways of viewing the situation.
\begin{enumerate}
\item We should not include the Goldstone masses in a calculation of the
1-loop corrections to the effective D-term potential outside the global minimum. As a result these 1-loop corrections are ill-defined for almost all fieldvalues.
\item We should include the Goldstone masses
in the calculation of the 1-loop corrections to the effective D-term potential outside the global minimum. The resulting 1-loop corrections agree with previous results along the inflationary valley
and have no $\Lambda_{c}^{2}$-dependence. This is the viewpoint we will take.
\end{enumerate}
There are some field-space points for which both methods agree, because the Goldstone boson is really massless there.
These points lie exactly on the "spontaneous symmetry breaking valley" defined by equation (\ref{eqssbv}). Note that both methods disagree on the inflationary valley.

\subsection{Graphs of the one-loop correction}
\label{results}

This section is merely meant to give the reader a feeling of what the 1-loop corrections look like in comparison with the tree-level potential and what are the possible implications. A quantitative analysis of the different aspects will be given later on. The results are shown with the aid of graphs, because the actual algebraic expression is too complicated\footnote{In fact the slightly broken supersymmetric nature of the 1-loop correction makes it of the form: near-infinity minus near-infinity. Huge bosonic and fermionic contributions nearly cancel and the difference is the correction we are interested in. Many of these cancellations can be done analytically, by tricks of the form: $A^{2}ln(A)-B^{2}ln(B)=A^{2}ln(A/B)+(A^{2}-B^{2})ln(B)$, 
which can make things much simpler if $A$ and $B$ are, nearly equal, huge numbers or formulae. Unfortunately we didn't manage to get rid of all (near-infinity minus near-infinity)-terms in this way, so the final formula for the total 1-loop correction is still too complicated to analyse by hand. The near-infinity minus near-infinity nature can also lead to numerical problems, but these we have solved.\label{nimni}}.

There are four adjustable parameters: $\lambda$, $g$, $\xi$ and $\Lambda_{m}$, although the dependence on the last will be small. The output, being a potential energy density, will have mass dimension four and will always be given in units of $(M_{Pl})^{4}$. Don't get confused by very small numbers, this is just because of the fourth power. First we will look at the case where all these parameters are one, although this will not be the typical case, because normally $\xi$ is much less than one square Planck mass.

\begin{figure}[!h]
\subfigure[tree-level]{\includegraphics[width=0.30\textwidth]{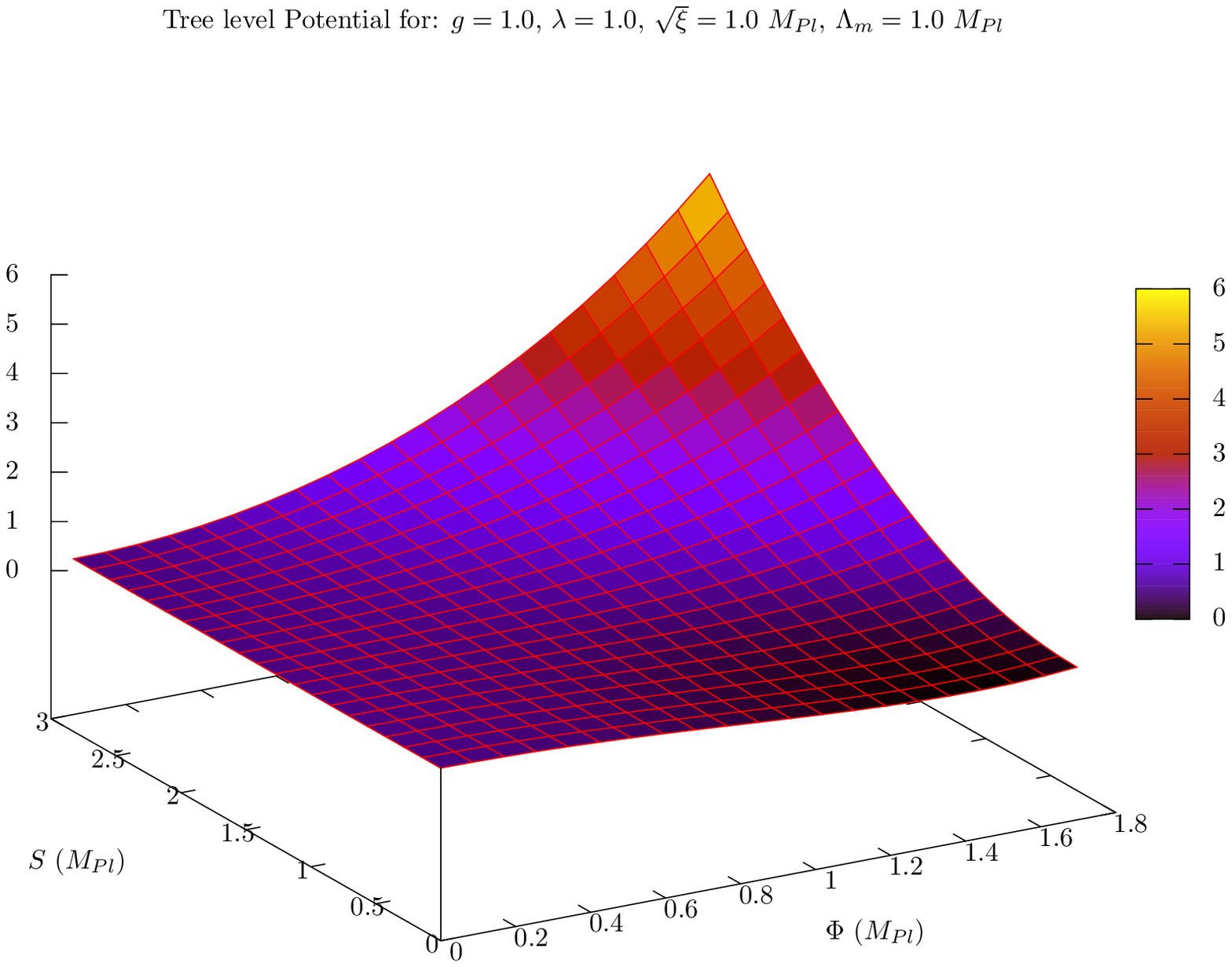}} \quad
\subfigure[1-loop]{\includegraphics[width=0.30\textwidth]{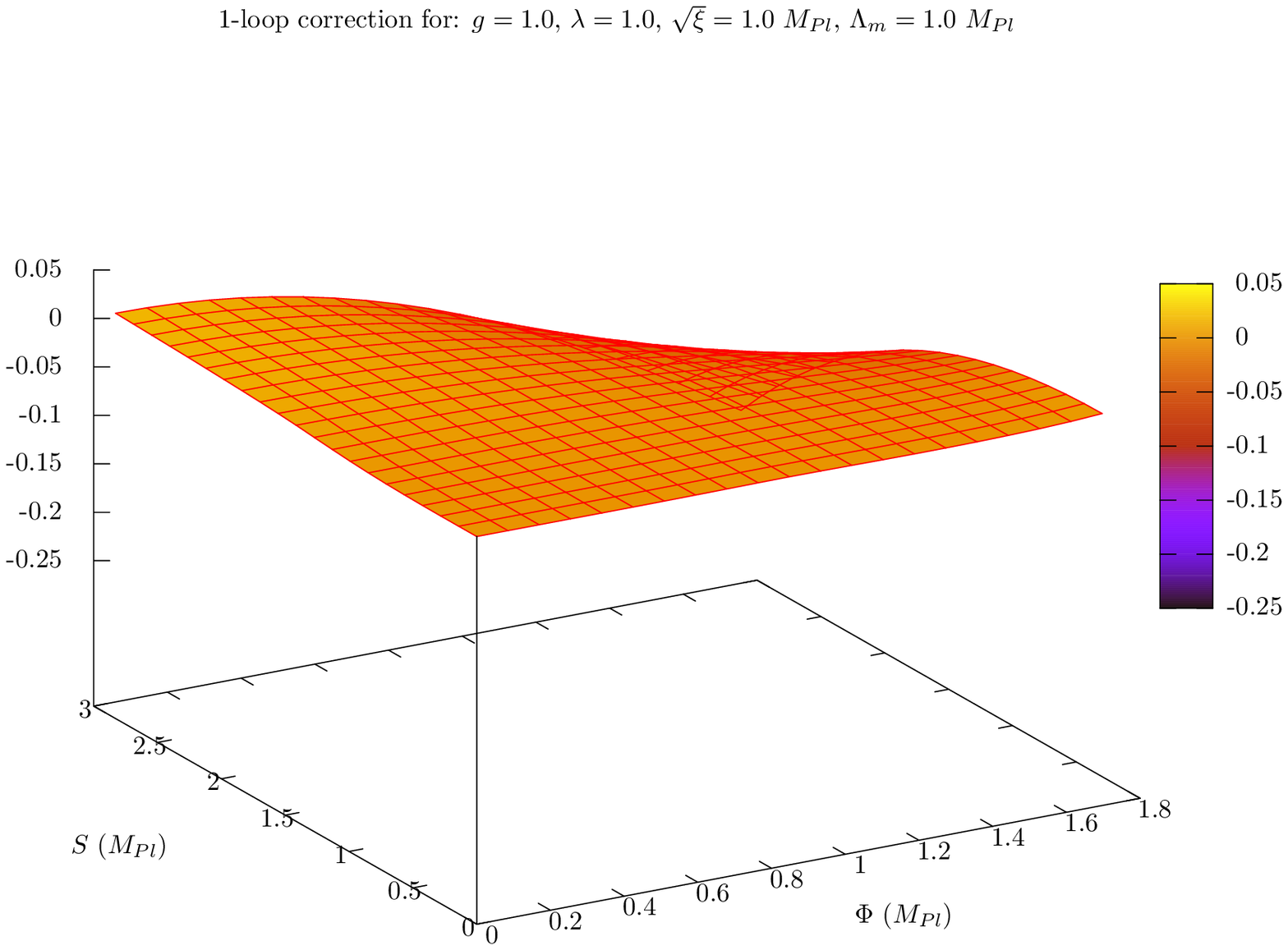}} \quad
\subfigure[total]{\includegraphics[width=0.30\textwidth]{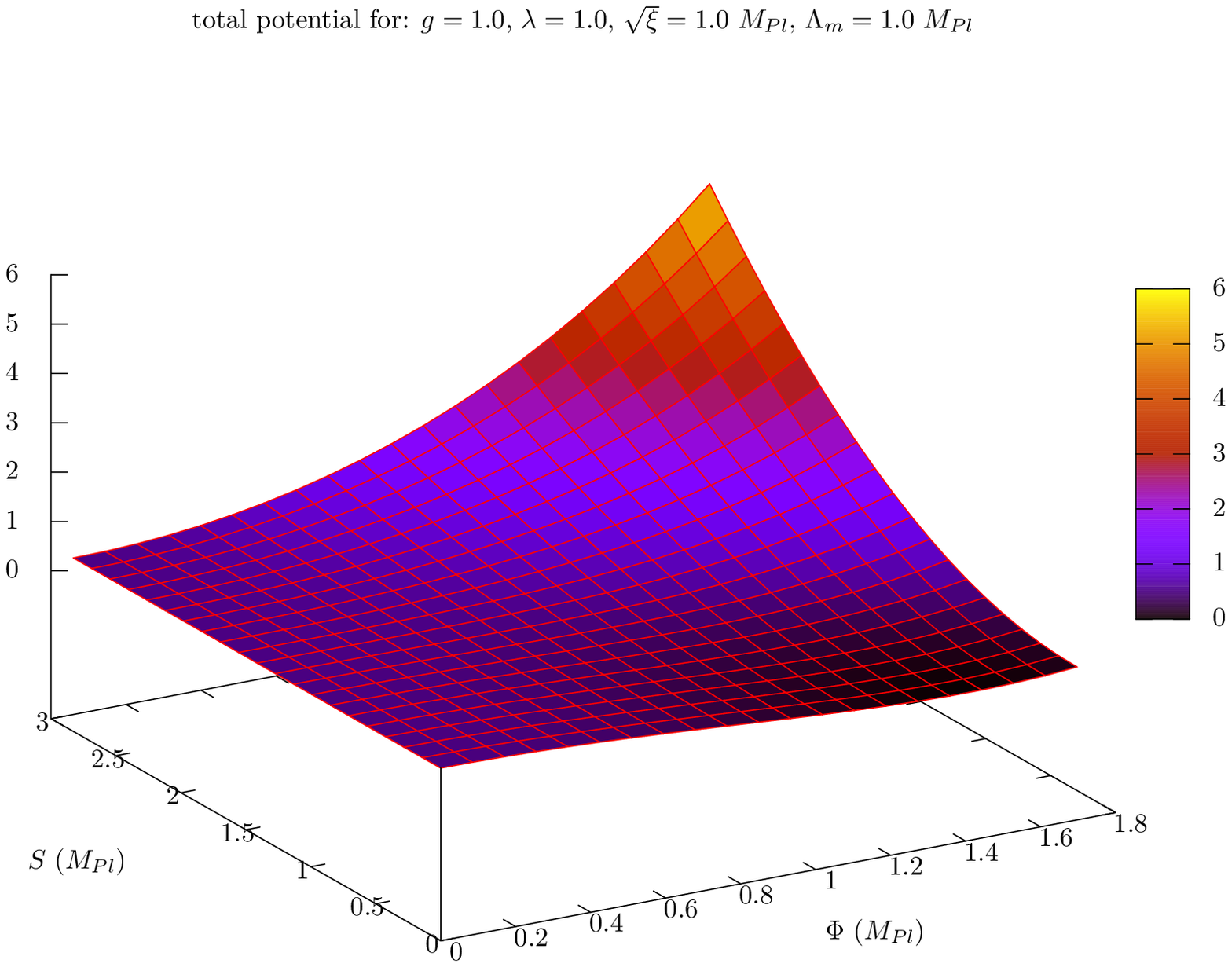}}
\caption{\footnotesize{The tree level potential, 1-loop correction and the sum of both, for $g=1.0$, $\lambda=1.0$, $\sqrt{\xi}=1.0$ $M_{Pl}$, $\Lambda_{m}=1.0$ $M_{Pl}$ on the domain $0<S<2 \cdot S_{c}$, $0<\Phi<1.2 \cdot \Phi_{g. min.}$ The tree-level potential looks similar to figure (\ref{regions}). The 1-loop correction is around a factor 20 smaller than the tree-level potential and it has an opposite shape. The general shape is not changed by the 1-loop corrections, but there is a slight slope on the inflationary valley, which can not really be seen on this scale. It is very small, which results in many efolds: around 2900 in the upper half of the picture.}}
\label{3Dplot-tree1-script.eps}
\end{figure}

Figure (\ref{3Dplot-tree1-script.eps})(a) gives the classical tree-level potential in the region $0<S<2\cdot S_{c}$. The $\Phi$-axis reaches to 1.2 times the global minimum value, so we can see the entire classical path during spontaneous symmetry breaking. Part (b) depicts the 1-loop correction on the same domain. It is around a factor twenty smaller than the tree-level potential and has an opposite shape. Of course we rely on the 1-loop corrections to be (very) small in comparison to the tree-level potential, so that we are hopefully justified to ignore all two and higher order corrections. Part (c) depicts the total (tree + 1-loop) potential. As mentioned in section (\ref{D-terminflation}) the quantum-correction indeed adds a little slope to the inflationary valley ($\Phi=0$). In slowly rolling down from $2 \cdot S_{c}$ to $S_{c}$, this universe will undergo around 2900 efolds.

\begin{figure}[!h]
\subfigure[tree-level]{\includegraphics[width=0.30\textwidth]{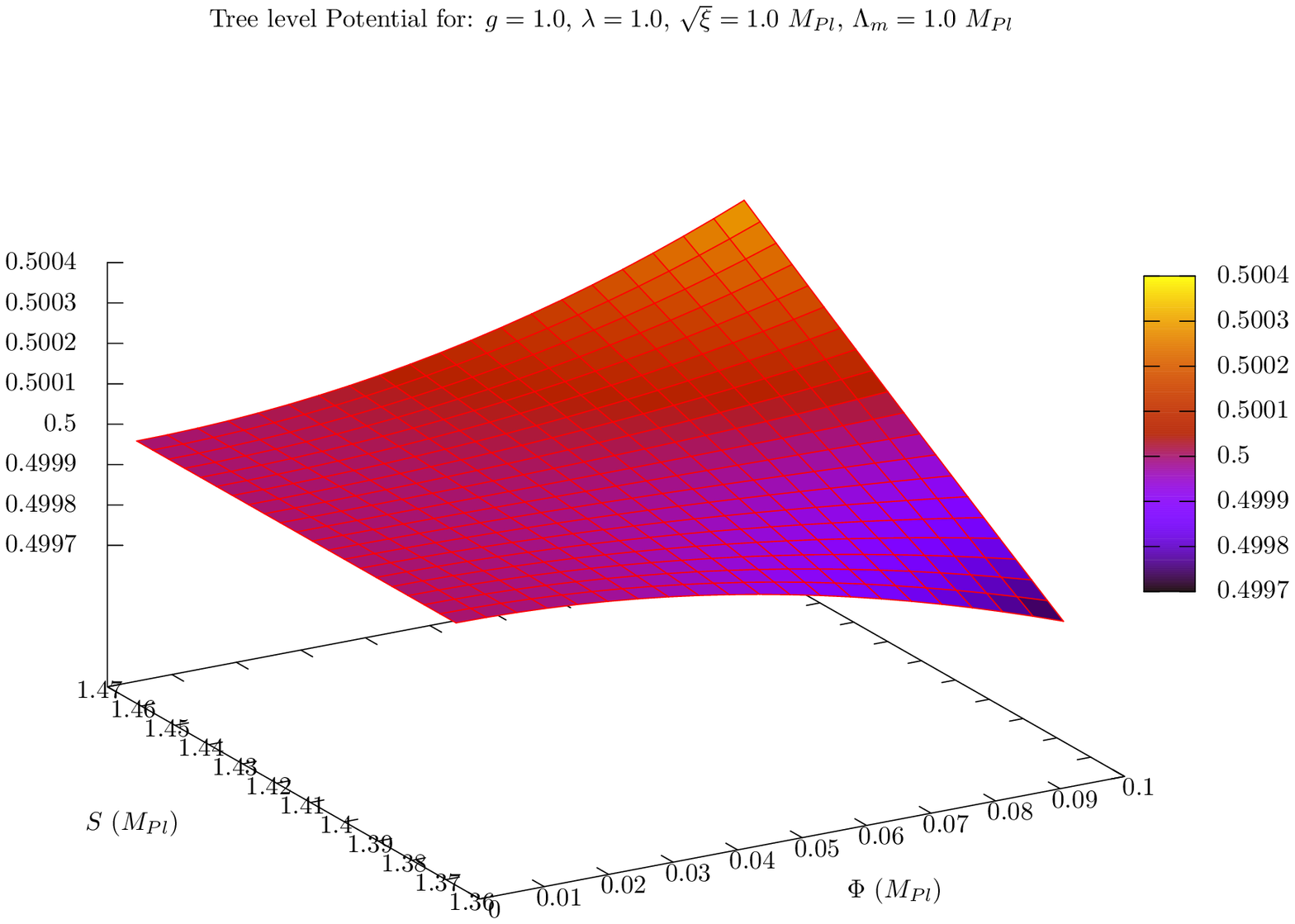}} \quad
\subfigure[1-loop]{\includegraphics[width=0.30\textwidth]{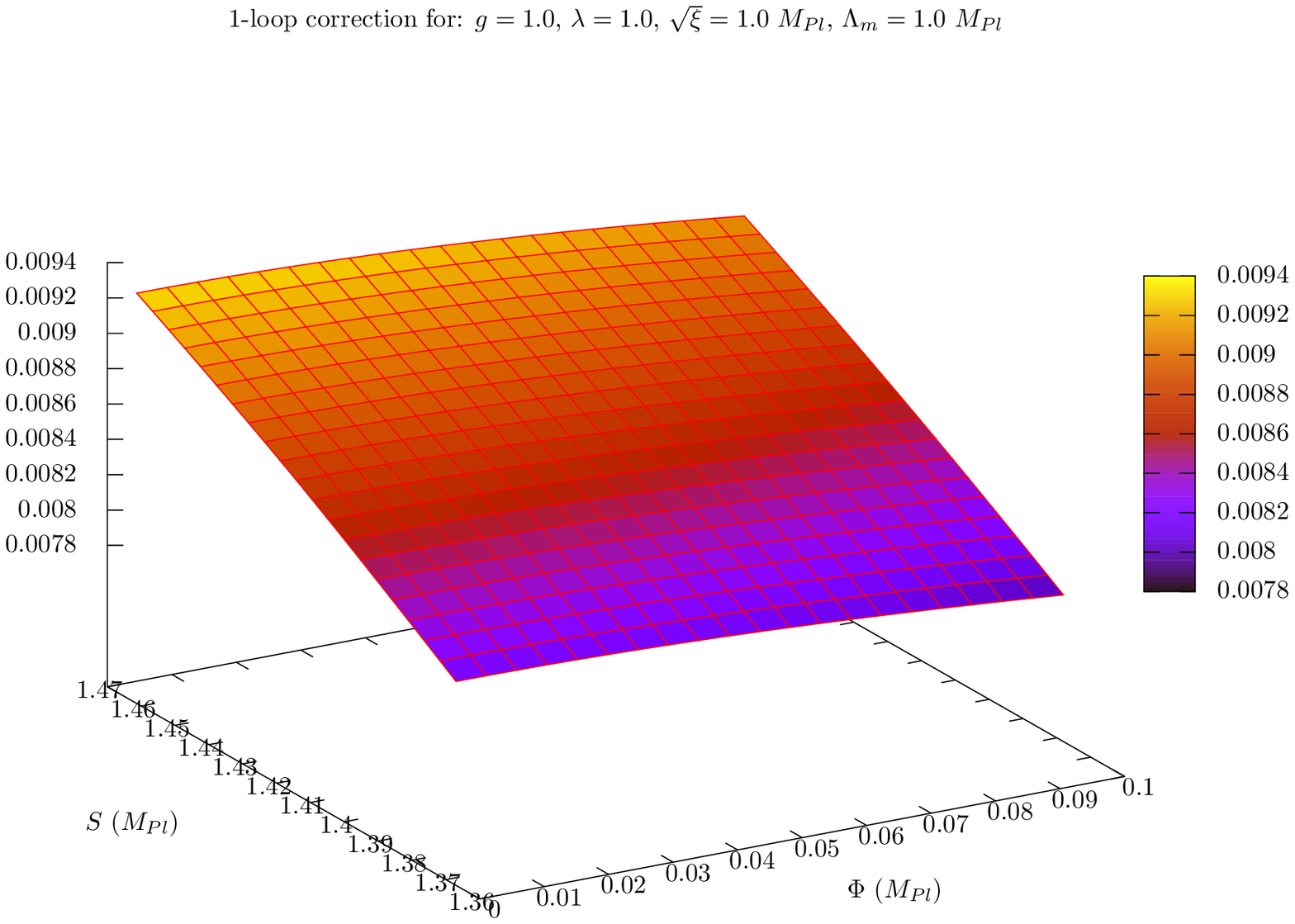}} \quad
\subfigure[total]{\includegraphics[width=0.30\textwidth]{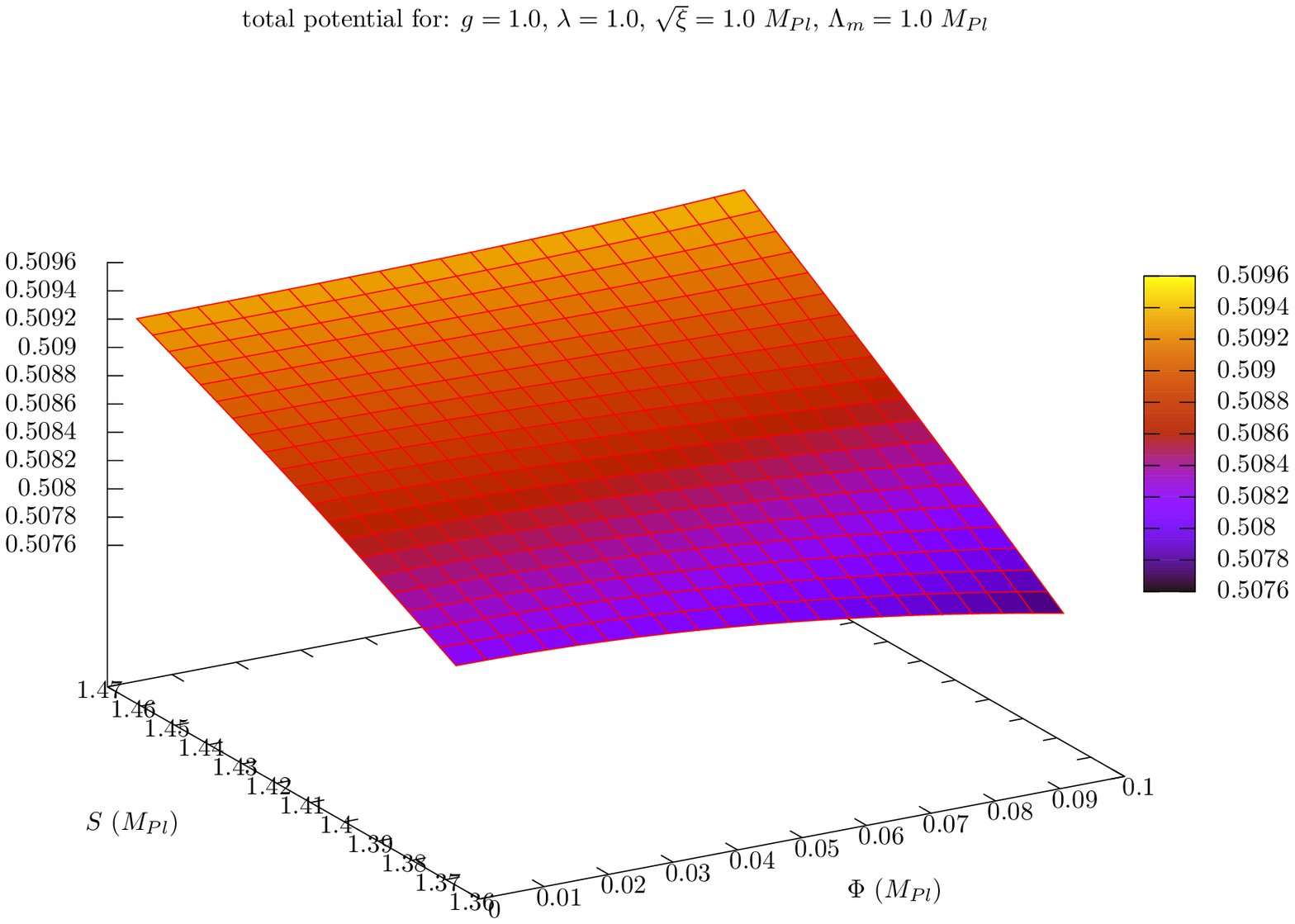}}
\caption{\footnotesize{The tree level potential, 1-loop correction and the sum of both, for the same parameters as in figure (\ref{3Dplot-tree1-script.eps}) on a domain centred around the critical point ($S_{c}$). Both field-axes have the same scale. The slope along the inflationary valley, arising from the 1-loop corrections is clearly visible now. It will give rise to 52 efolds of inflation on the upper half of the picture, which therefore represents the cosmologically interesting region. It is apparent that the variations in the 1-loop corrections dominate the tree-level variations in this region.}}
\label{3Dplot-tree1_zoom-script.eps}
\end{figure}

The 1-loop corrections in this case are small as compared to the tree level potential. In most cases they will be even much smaller, but this doesn't mean that they are unimportant, because sometimes it is the variation that matters. In figure (\ref{3Dplot-tree1_zoom-script.eps}), we have zoomed in on the tree-, 1-loop- and total values of the potential in a region centred around the critical point ($S_{c}$). This part of the inflationary valley corresponds to around 52 efolds. Although the tree-level potential is a factor 60 larger, the 1-loop corrections still dominate the behaviour around the critical point. You can nicely see the slope that the 1-loop corrections generate. The assumption that inflation ends when the fields leave the inflationary valley, which is usually made, is probably not valid here: since the upper half of the picture corresponds to 52 efolds, we can already see that the total potential doesn't change fast enough on the relevant time-scales to end inflation abruptly (note that the scales on both axes are the same). The dependence on the renormalization mass scale ($\Lambda_{m}$) is very small. If we change the value of $\Lambda_{m}$ from 1 $M_{Pl}$ to 0.01 $M_{Pl}$, the slope changes only to give around 58 efolds instead of 52 on the same domain.

\begin{figure}[!h]
\subfigure[tree-level]{\includegraphics[width=0.30\textwidth]{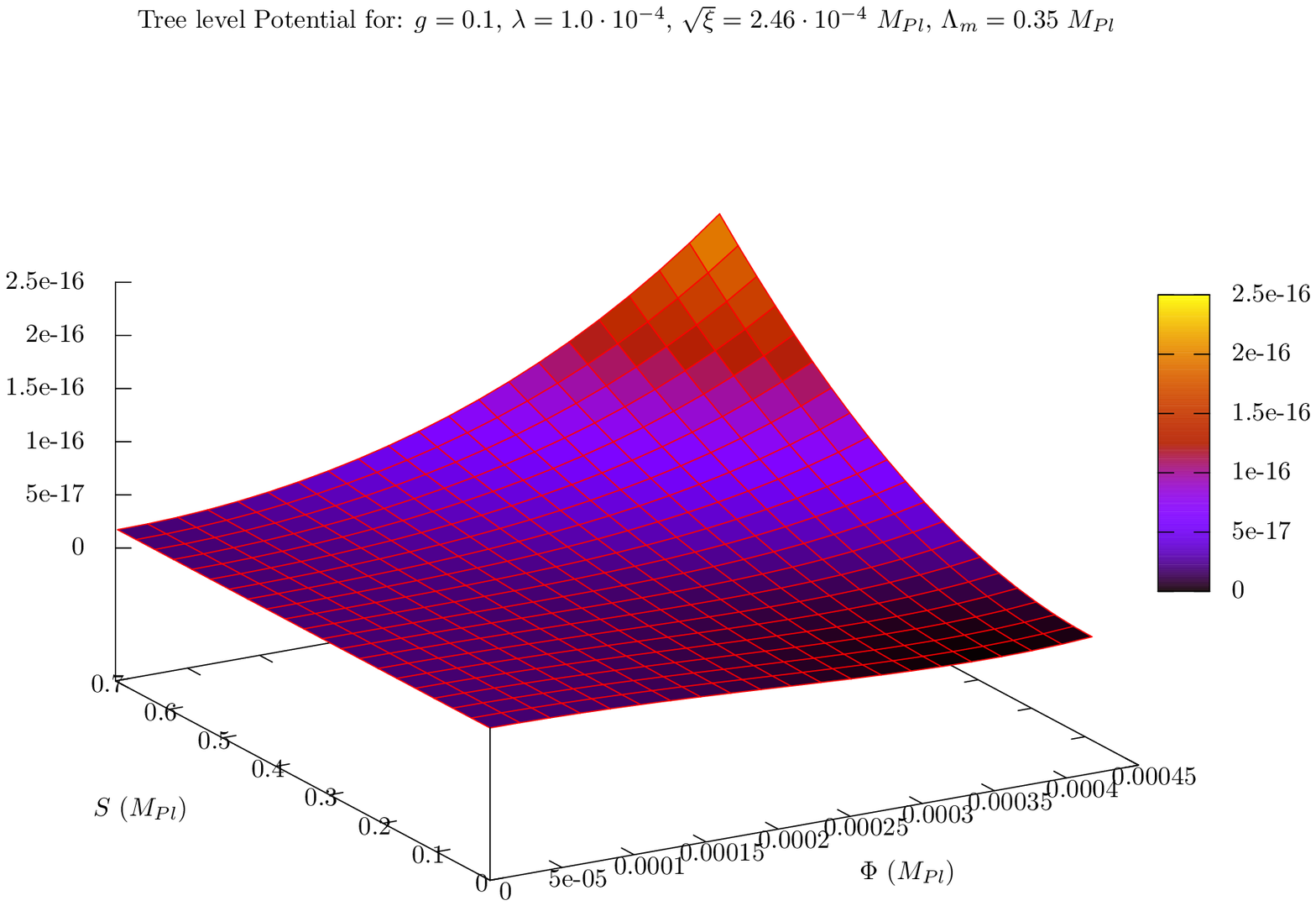}} \quad
\subfigure[1-loop]{\includegraphics[width=0.30\textwidth]{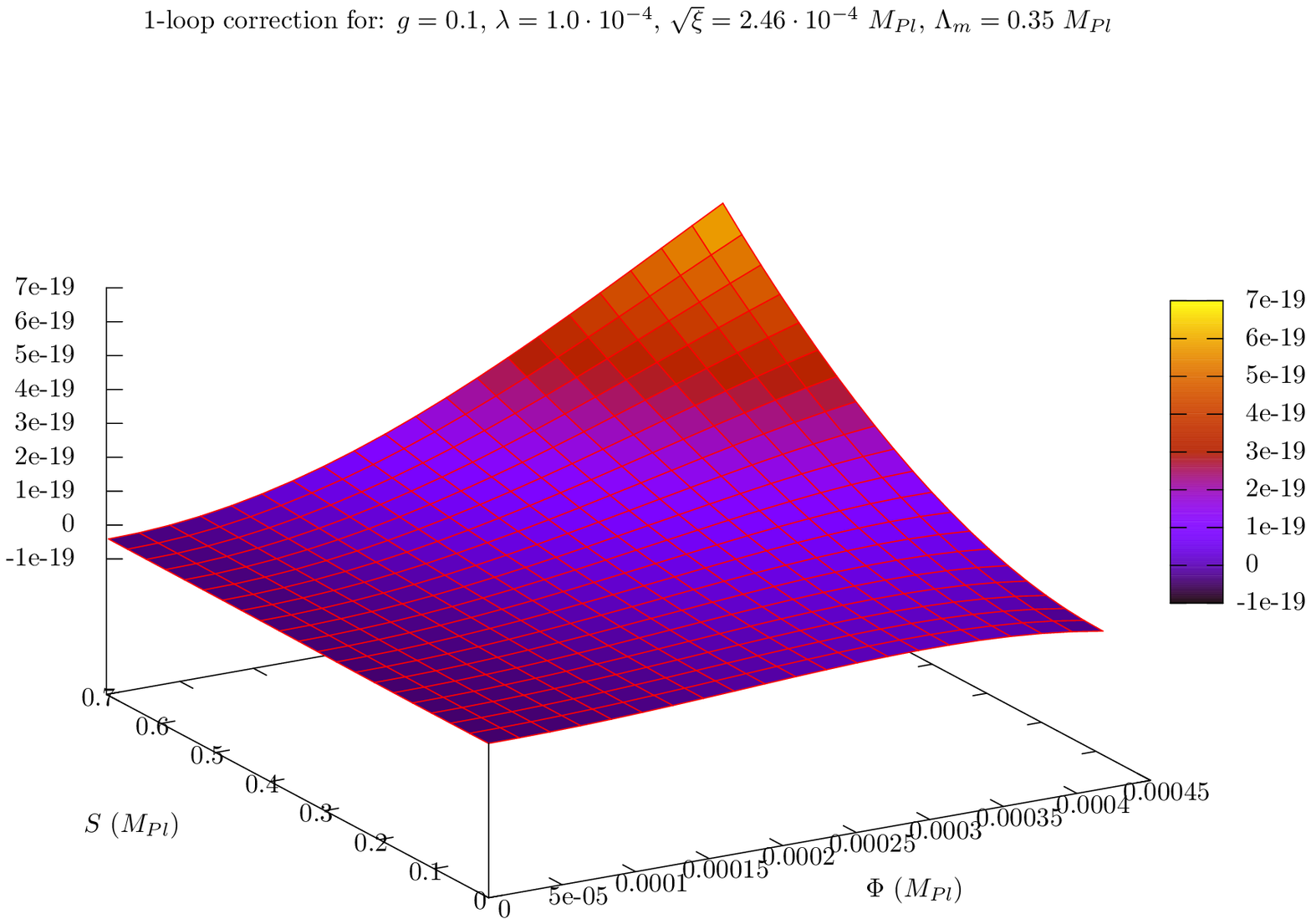}} \quad
\subfigure[total]{\includegraphics[width=0.30\textwidth]{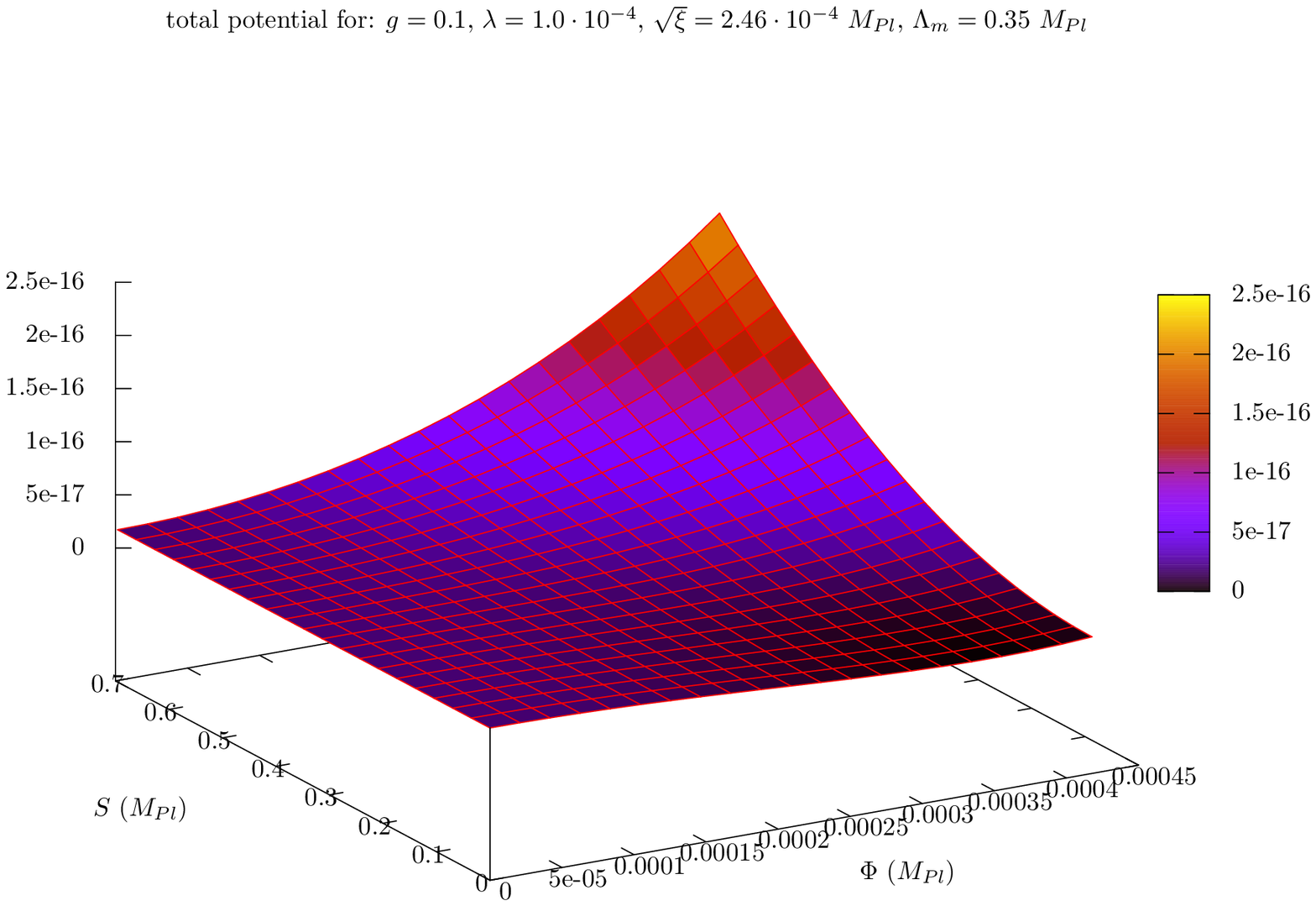}}
\caption{\footnotesize{The tree level potential, 1-loop correction and the sum of both, for $g=0.1$, $\lambda=1.0 \cdot 10^{-4}$, $\sqrt{\xi}=2.46 \cdot 10^{-4}$ $M_{Pl}$ and $\Lambda_{m}=0.35$ $M_{Pl}$ on the domain $0<S<2 \cdot S_{c}$, $0<\Phi<1.2 \cdot \Phi_{g. min.}$. The 1-loop correction is around a factor of 500 smaller than the tree-level potential. The shape is opposite near the inflationary valley. The tilt produced by the 1-loop corrections corresponds to around 18000 efolds in the upper half of the picture.}}
\label{3Dplot-tree6-script.eps}
\end{figure}

\begin{figure}[!h]
\subfigure[tree-level]{\includegraphics[width=0.30\textwidth]{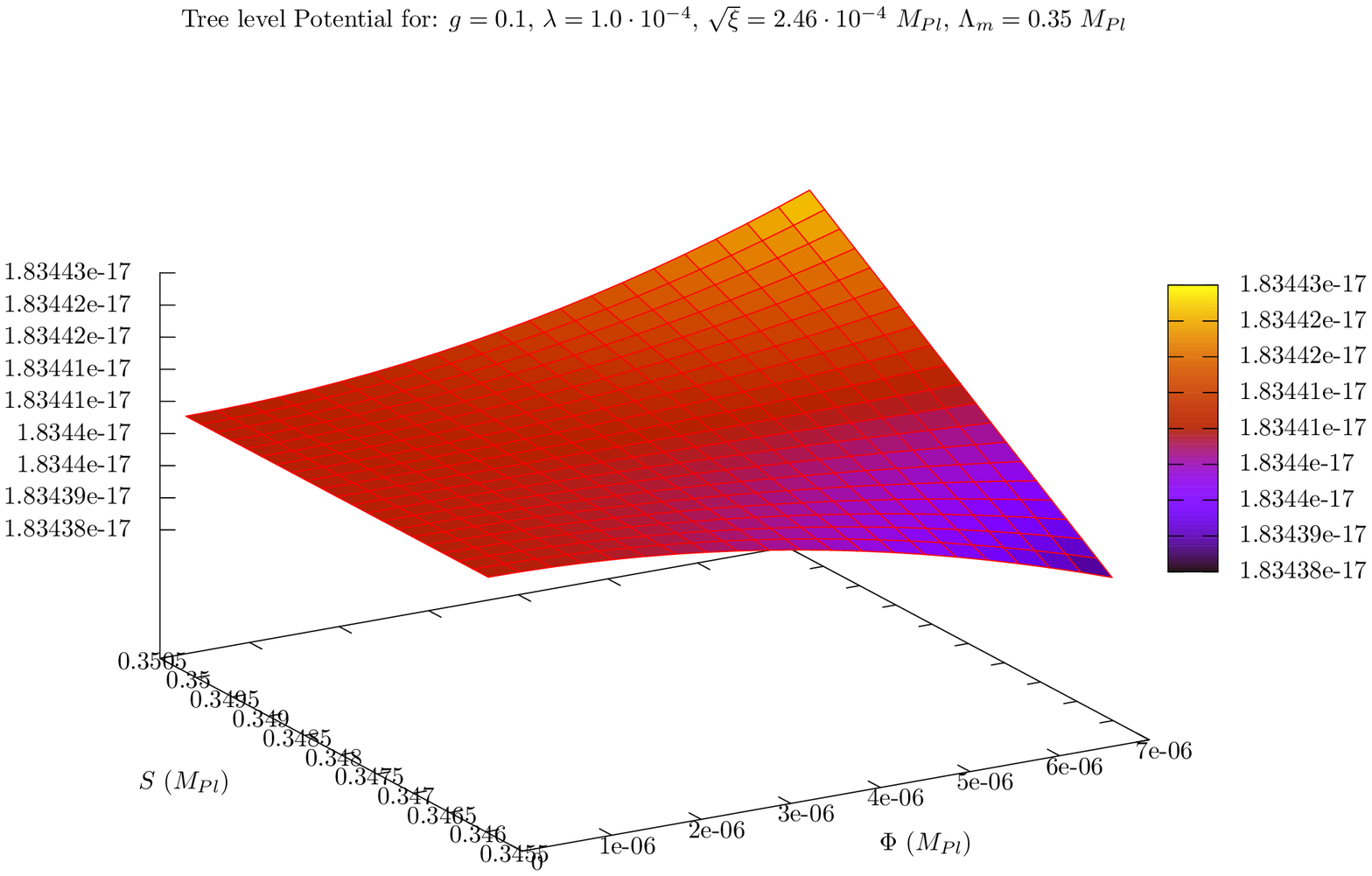}} \quad
\subfigure[1-loop]{\includegraphics[width=0.30\textwidth]{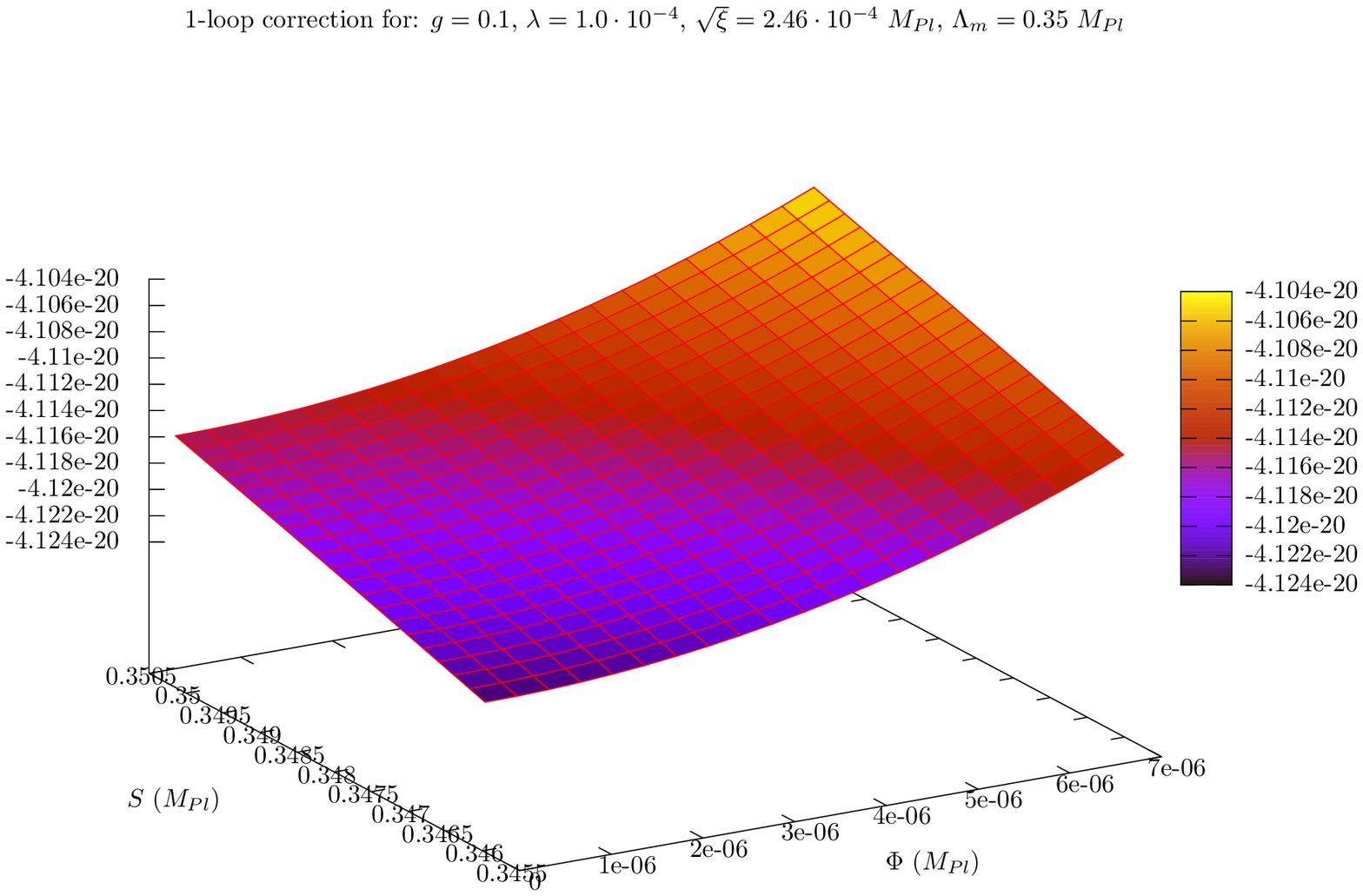}} \quad
\subfigure[total]{\includegraphics[width=0.30\textwidth]{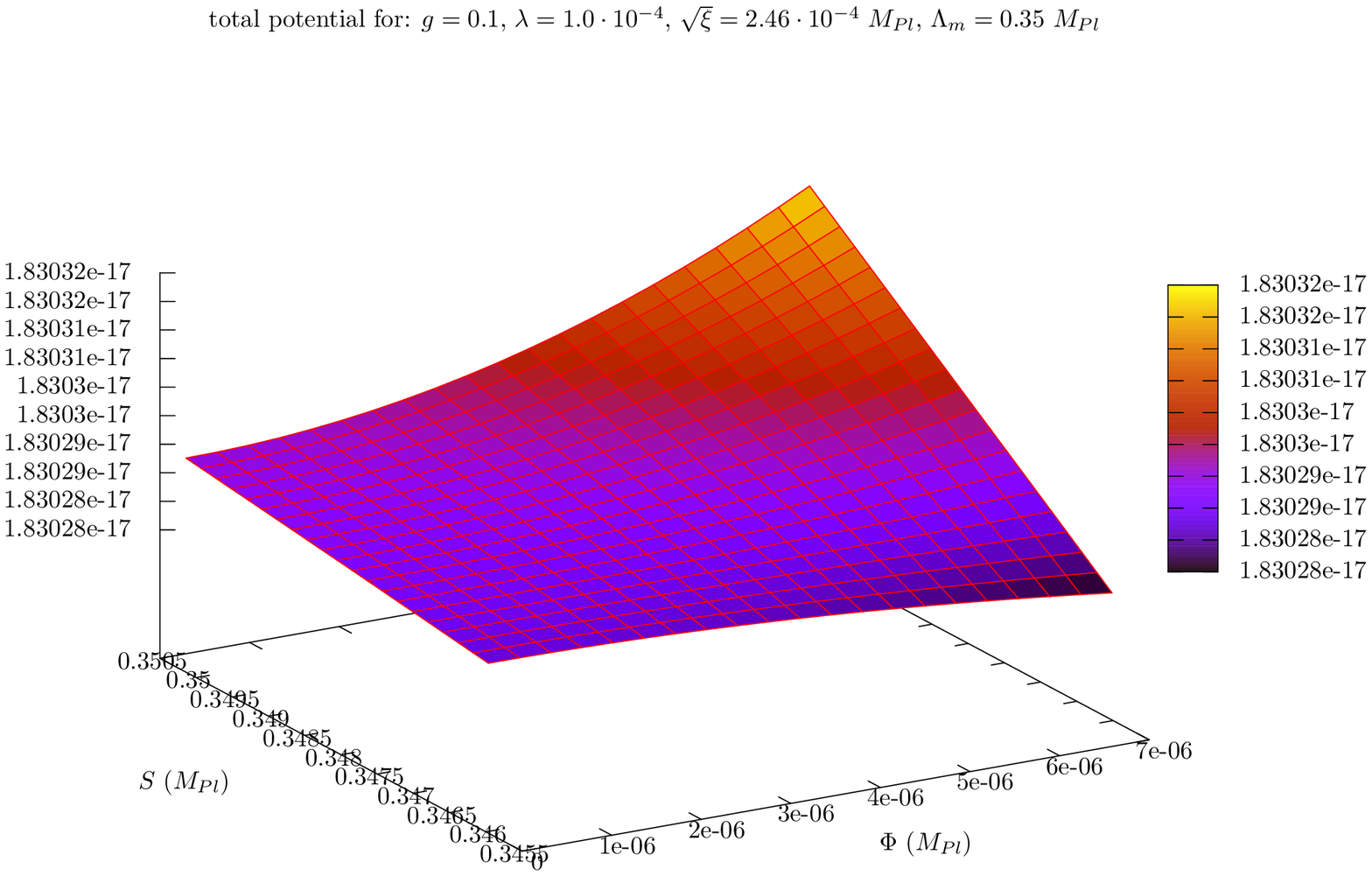}}
\caption{\footnotesize{The tree level potential, 1-loop correction and the sum of both, for the same parameters as in figure (\ref{3Dplot-tree6-script.eps}) zoomed in around $S_{c}$. Both axes have a different scaling. The upper half of the picture corresponds to 58 efolds, so the 1-loop corrections are definitely important on the relevant scales.}}
\label{3Dplot-tree6_zoom-script.eps}
\end{figure}

Figures (\ref{3Dplot-tree6-script.eps}) and (\ref{3Dplot-tree6_zoom-script.eps}) represent the general shape of the 1-loop corrections, which you get for almost all physically relevant parameter combinations. For these figures $g=0.1$, $\lambda=0.0001$, $\sqrt{\xi}=0.00025 $ $M_{Pl}$ and $\Lambda_{m}=0.35$ $M_{Pl}$ (which is equal to the value of $S_{c}$). These values seem a bit out of the blue, but in fact I took a combination which is compatible with WMAP observations, which will be explained in section (\ref{newbounds}). Anyway, this selection doesn't change the overall picture.

The main difference with the previous set of figures is that the 1-loop corrections are even smaller as compared to the tree-level potential (around a factor of 500) and that the 1-loop corrections are only of opposite shape as compared to the tree level potential for small $\Phi$. For large $\Phi$ the 1-loop corrections mimic the tree-level potential. So only in the neighbourhood of the inflationary valley can we expect the corrections to be of interest.

Another thing, which is shared by all possible parameter choices, is that in the global minimum both the tree-level potential and the 1-loop corrections vanish, giving a true, totally supersymmetric, non inflating, ground state.

In the ($S=0$)-plane the 1-loop corrections always counteract the tree-level potential, although normally the effect is too small to have important consequences.

Close to the inflationary valley the 1-loop corrections behave more like, say, a parabola and the tree-level potential more like a fourth order function. So the closer you get to the inflationary valley the more important relatively the 1-loop corrections become.

Again, if we zoom in on the relevant region around the critical point we see that the 1-loop corrections here will definitely have important consequences. This opens an interesting window. It seems that the, up to now neglected, off-valley quantum-corrections will play a major role on all scales which are of cosmological interest. Most surely they will have consequences to the explanation of WMAP\footnote{The Wilkinson Microwave Anisotropy Probe is a satellite which looks in great detail at the cosmic microwave background radiation, see http://map.gsfc.nasa.gov. It's predecessor was the COBE satellite.} observations.

\begin{figure}[!h]
\subfigure[tree-level]{\includegraphics[width=0.30\textwidth]{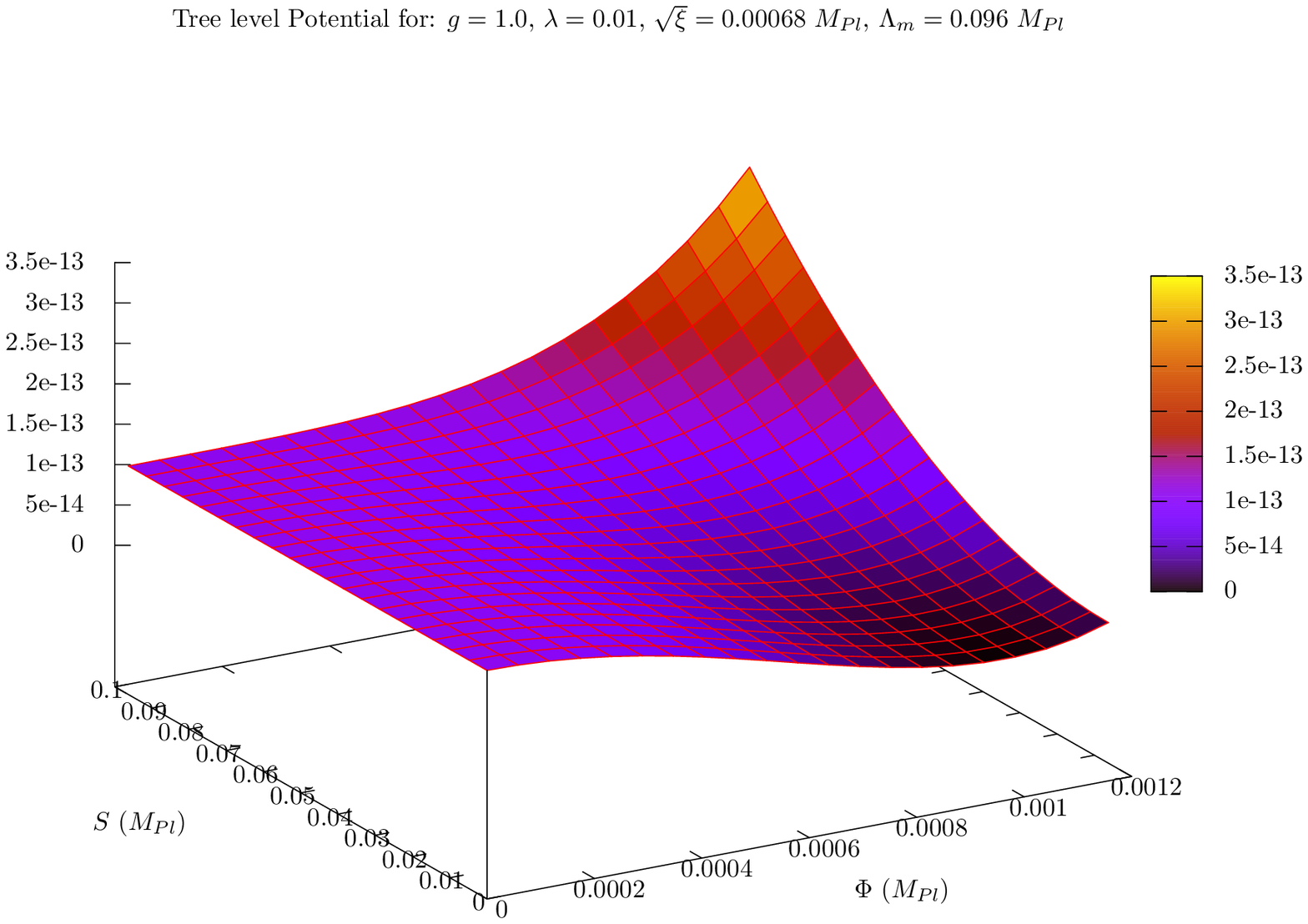}} \quad
\subfigure[1-loop]{\includegraphics[width=0.30\textwidth]{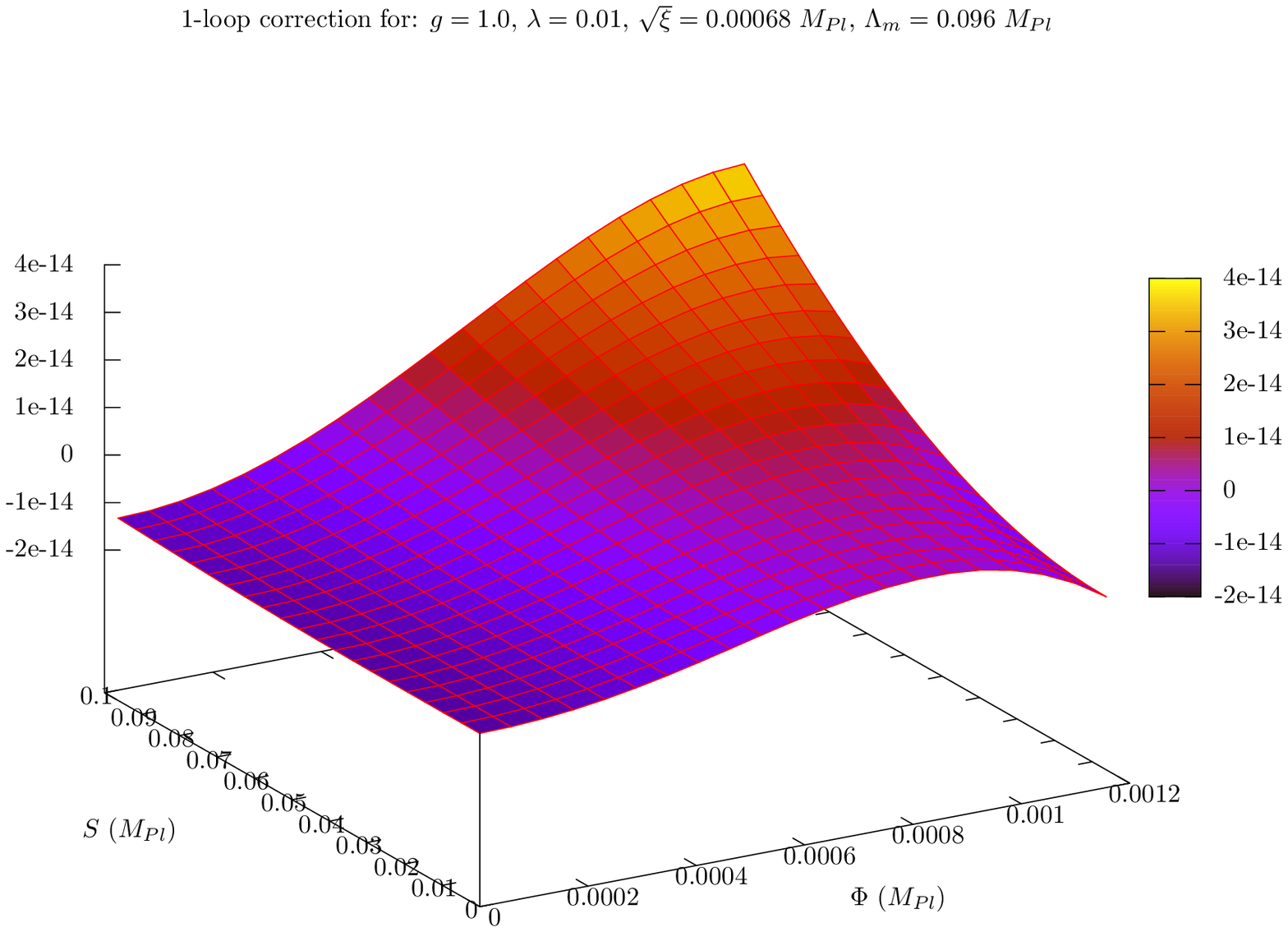}} \quad
\subfigure[total]{\includegraphics[width=0.30\textwidth]{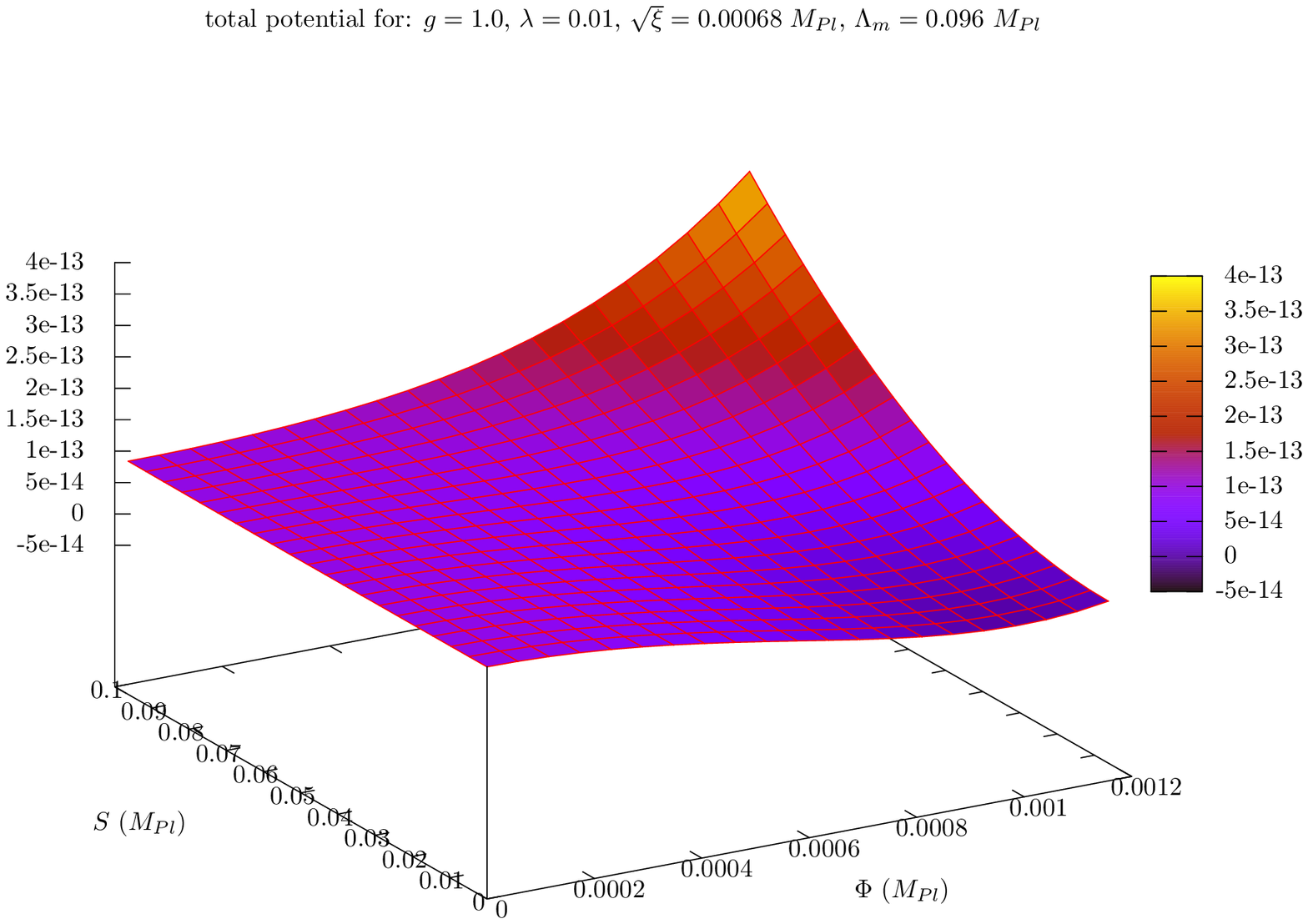}}
\caption{\footnotesize{The tree level potential, 1-loop correction and the sum of both, for $g=1.0$, $\lambda=0.01$, $\sqrt{\xi}=6.8 \cdot 10^{-4}$ $M_{Pl}$ and $\Lambda_{m}=0.096$ $M_{Pl}$ ($=S_{c}$) on the domain $0<S<S_{c}$, $0<\Phi<1.2 \cdot \Phi_{g. min.}$. The 1-loop corrections have lowered the top of the Mexican hat (origin in this picture) considerably and will thus slightly lower the resulting cosmic string tension.}}
\label{3Dplot-tree2_zoom-script.eps}
\end{figure}

\begin{figure}[!h]
\subfigure[tree-level]{\includegraphics[width=0.30\textwidth]{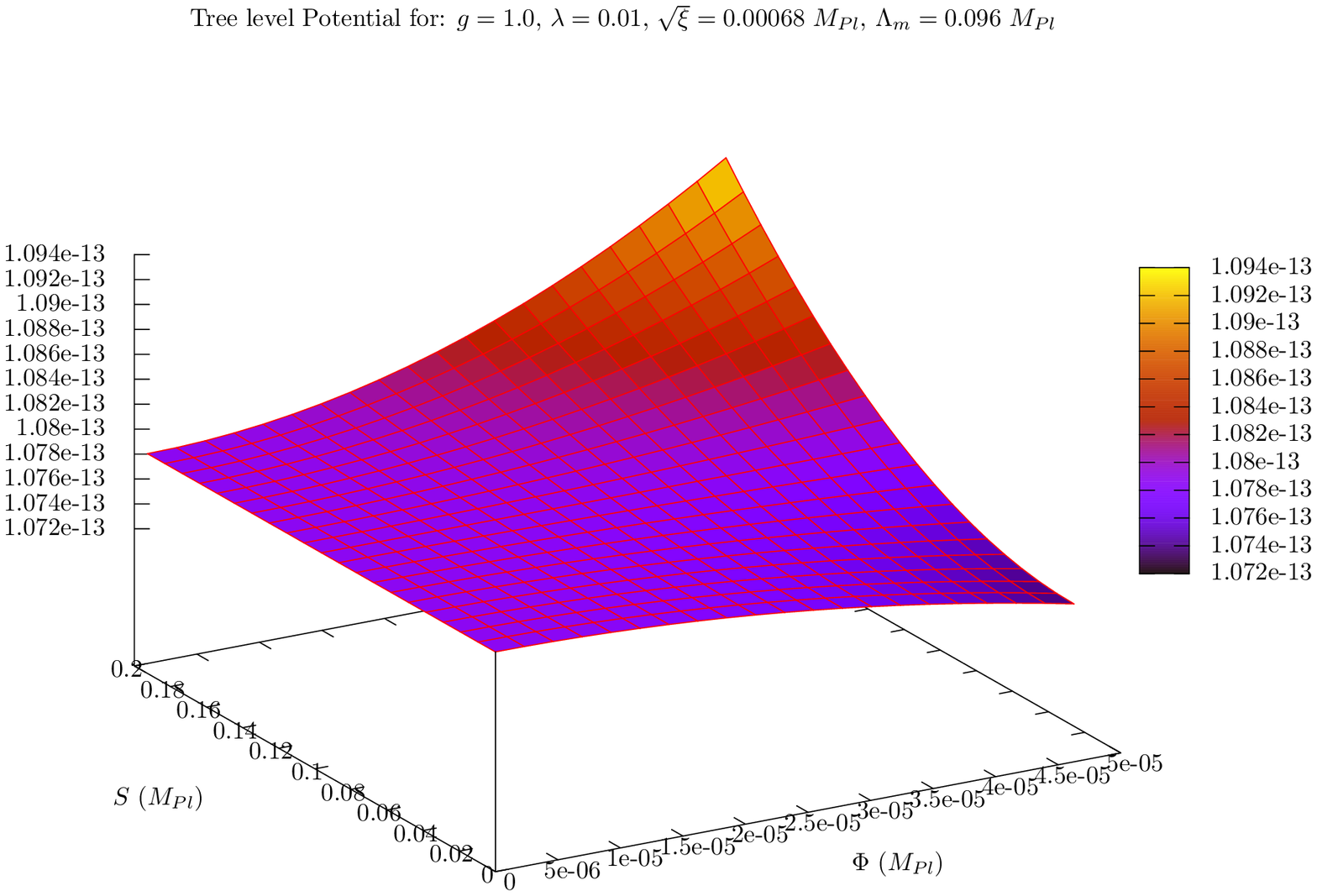}} \quad
\subfigure[1-loop]{\includegraphics[width=0.30\textwidth]{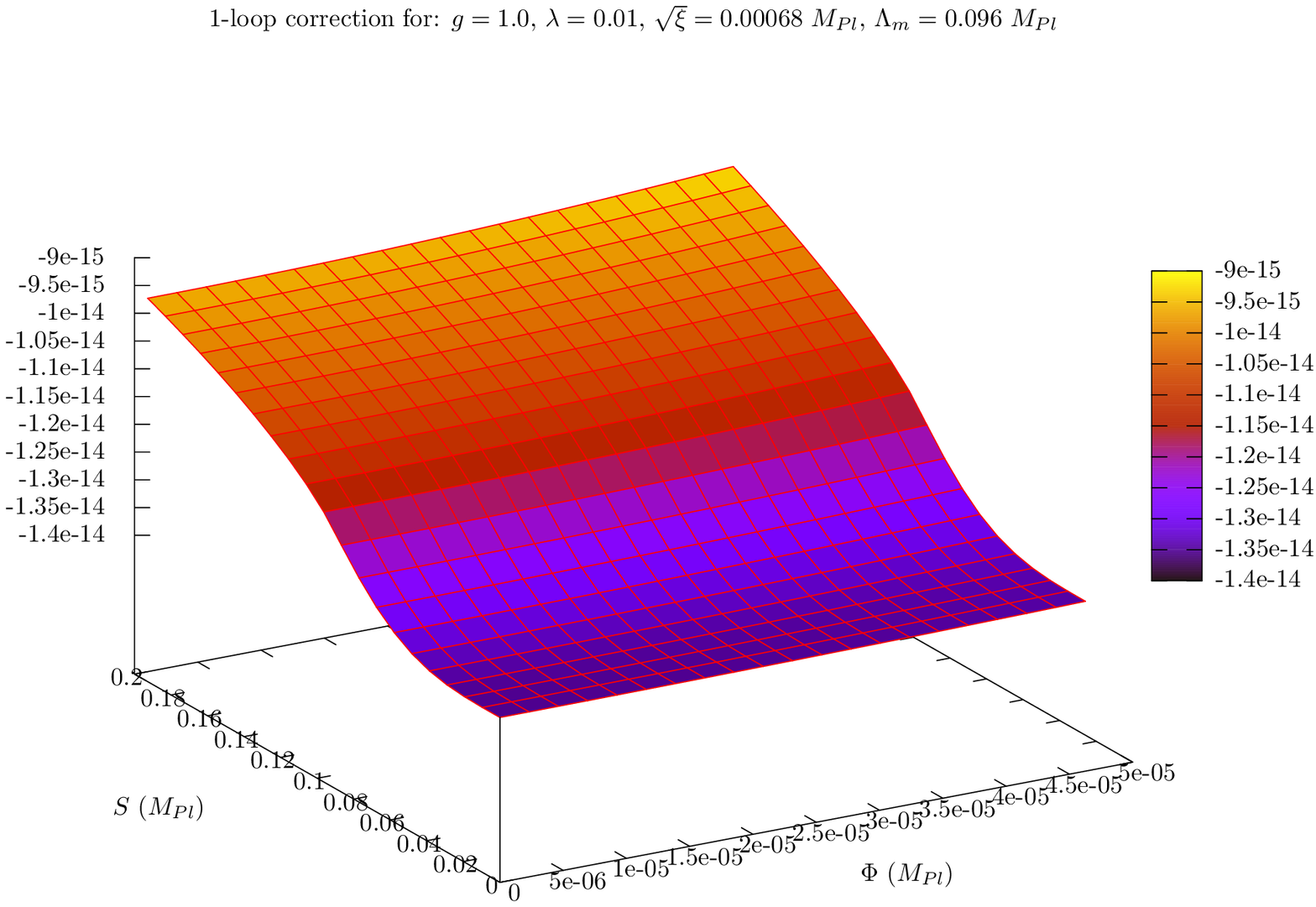}} \quad
\subfigure[total]{\includegraphics[width=0.30\textwidth]{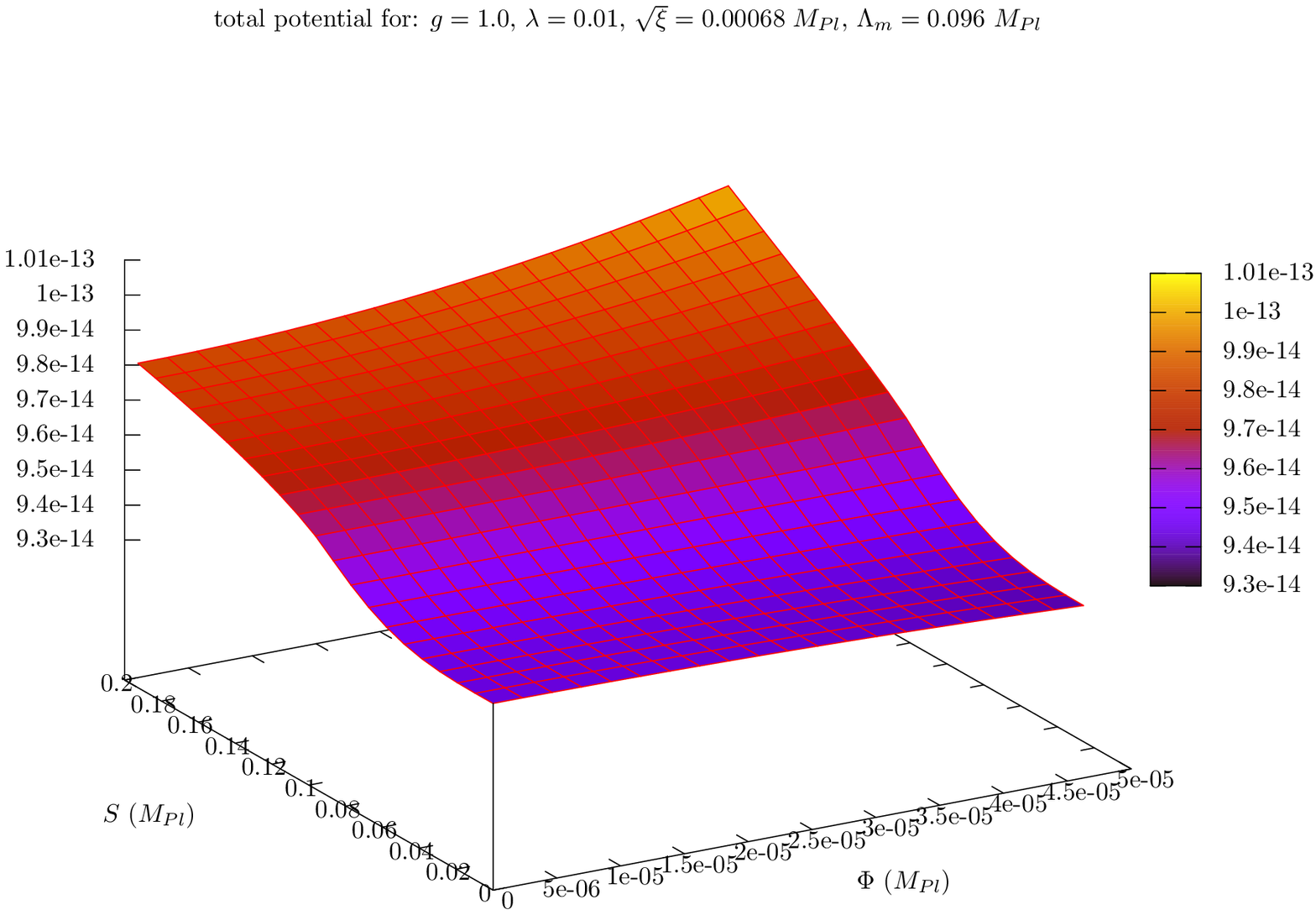}}
\caption{\footnotesize{The tree level potential, 1-loop correction and the sum of both, for the same parameters as in figure (\ref{3Dplot-tree2_zoom-script.eps}) on the domain $0<S<2 \cdot S_{c}$, small $\Phi$. This shows how the height of the Mexican hat goes down as the inflaton field S goes to zero. The 1-loop corrections will slightly decrease the tension of the resulting cosmic strings.}}
\label{3Dplot-tree2_zoomzoom-script.eps}
\end{figure}

For a select region in parameter space it is possible that the negative 1-loop corrections at the ($S,\Phi$) = (0,0) point become relatively large, say order 10\% of the tree-level potential. This less common situation is depicted in figures (\ref{3Dplot-tree2_zoom-script.eps}) and (\ref{3Dplot-tree2_zoomzoom-script.eps}). The lowering of the top of the Mexican hat would probably lead to less energetic cosmic strings. Of course if the 1-loop corrections become too big, we can no longer neglect the second and higher order loop contributions. Again we show a parameter combination which will be compatible with the WMAP density perturbation.

\section{Inflation below the critical inflaton value}
\label{seciassb}

Usually it is assumed that no significant inflation occurs after the inflaton field $S$ reaches the critical value $S_{c}$. In this section we will show that even classically (without 1-loop corrections) this assumption may be false. Then we will calculate explicitly the number of e-folds below $S_{c}$ for the D-term parameter combinations which best fit WMAP measurements, according to Ref. \cite{rem}. We will see that indeed inflation below $S_{c}$ can not be neglected. A whole new analysis is needed in order to compare the predictions of the D-term model with WMAP measurements. This is done in section (\ref{newbounds}).

Substantial inflation after $S_{c}$ is interesting for two reasons. First of all this offers the possibility to dilute the cosmic strings which are produced in the symmetry breaking process. This is a good thing, since WMAP has put stringent bounds on the string contribution to the cosmic background anisotropy. It can not be more than about 10\% \footnote{The cosmic microwave background anisotropies have a fluctuating nature if they arise from quantum fluctuations that are enlarged by inflation. This is because of acoustic oscillations in the period between inflation and reionization. Anisotropies coming from cosmic strings have no specific scale to it. WMAP measurement look very much like the "quantum fluctuation" anisotropies, leaving at most a 10\% contribution to cosmic strings.} \cite{pog,hin}.

Secondly, the cosmologically interesting region, which is around 50 to 60 efolds before the end of inflation, will shift when substantial inflation occurs after $S_{c}$. This region, for example, determines the density perturbation at reionization (which should be around $2 \cdot 10^{-5}$ in order to be compatible with WMAP measurements). So the density perturbation at reionization is dependent on the possible inflation below $S_{c}$.

Inflation takes place whenever the potential energy-density of the scalar fields dominates the kinetic energy density \cite{lyt}. This is possible only if the slope of the potential is very small. In this case the equation of motion becomes dominated by the friction term ($H \partial_{t} \phi$), the fields are in slow-roll and the number of efolds before the end of inflation is effectively determined by how long the fields manage to stay in slow-roll. A good approximation to the number of efolds is \cite{lyt}:
\begin{equation}
N=-8\pi\int \frac{V}{V'}d\phi
\label{eqNef}
\end{equation}
The integral is over the path in field-space, units are in Planck masses and the approximation is valid as long as the slow-roll conditions are satisfied.

One normally assumes that inflation ends either before reaching $S_{c}$, whenever the slow-roll parameters become too big (order one), or at $S_{c}$, because then spontaneous symmetry breaking starts and will speed up the fields very quickly. In the previous chapter we have seen that this is probably not always true, because the potential does not always become very steep in a region around $S_{c}$ corresponding to about 50 to 60 efolds. Partly this is because close to the inflationary valley the 1-loop corrections are just opposite to the classical potential. But also even classically this assumption may be false.

\begin{figure}[!h]
\begin{center}
\begin{picture}(300,200)
\qbezier(140,150)(204,150)(240,190)

\qbezier(120,130)(185,130)(220,150)

\qbezier(120,105)(140,95)(200,110)

\qbezier(100,110)(110,110)(120,105)

\qbezier(110,80)(140,60)(180,70)

\qbezier(80,90)(95,90)(110,80)

\qbezier(100,50)(140,10)(160,30)

\qbezier(60,70)(80,70)(100,50)

\put(60,70){\circle*{1}}
\put(64,74){\circle*{1}}
\put(68,78){\circle*{1}}
\put(72,82){\circle*{1}}
\put(76,86){\circle*{1}}
\put(80,90){\circle*{1}}
\put(84,94){\circle*{1}}
\put(88,98){\circle*{1}}
\put(92,102){\circle*{1}}
\put(96,106){\circle*{1}}
\put(100,110){\circle*{1}}
\put(104,114){\circle*{1}}
\put(108,118){\circle*{1}}
\put(112,122){\circle*{1}}
\put(116,126){\circle*{1}}
\put(120,130){\circle*{3}}

\put(115,145){\makebox(40,0)[l]{$S_{c}$}}

\put(124,134){\circle*{1}}
\put(128,138){\circle*{1}}
\put(132,142){\circle*{1}}
\put(136,146){\circle*{1}}
\put(140,150){\circle*{1}}
\put(60,22){\line(1,1){78}}
\put(60,22){\line(1,0){140}}
\put(60,22){\line(0,1){100}}

\put(127,16){\makebox(40,0)[l]{$\longrightarrow$}}
\put(60,16){\makebox(40,0)[l]{$\longleftarrow$}}
\put(101,16){\makebox(40,0)[l]{$a$}}

\put(50,27){\makebox(40,0)[l]{$\downarrow$}}
\put(50,45){\makebox(40,0)[l]{$b$}}
\put(50,64){\makebox(40,0)[l]{$\uparrow$}}

\put(61,86){\makebox(40,0)[l]{$\swarrow$}}
\put(84,108){\makebox(40,0)[l]{$c$}}
\put(102,127){\makebox(40,0)[l]{$\nearrow$}}

\put(141,110){\makebox(40,0)[l]{$S$}}
\put(151,120){\makebox(40,0)[l]{$\nearrow$}}
\put(205,52){\makebox(40,0)[l]{$\Phi\rightarrow$}}
\put(55,132){\makebox(40,0)[l]{$V$}}
\put(57,144){\makebox(40,0)[l]{$\uparrow$}}
\end{picture}
\caption{\footnotesize{This picture depicts the typical shape of the spontaneous symmetry breaking part of the classical D-term potential. Inflation during spontaneous symmetry breaking is favourable if the potential is flat, meaning large a, small b and large c.}}
\label{abc}
\end{center}
\end{figure}
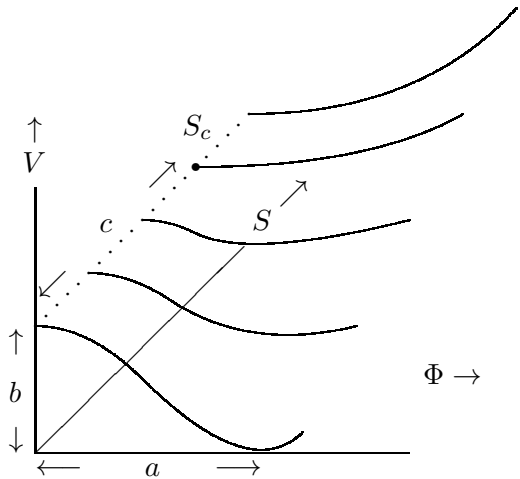

Figure (\ref{abc}) gives the general shape of the tree-level potential in terms of three parameters $a,b,c$ which correspond respectively to the width and height of the Mexican hat and to the value of $S_{c}$. The relation with the D-term parameters is:
\begin{equation}
\begin{array}{lll}
a=\sqrt{2\xi} \\
b=\frac{g^{2}\xi^{2}}{2} \\
c=\frac{g\sqrt{2\xi}}{\lambda} 
\end{array} \qquad \textrm{or} \qquad \begin{array}{lll}
\xi=\frac{1}{2}a^{2} \\
g=\frac{2\sqrt{2b}}{a^{2}} \\
\lambda=\frac{2\sqrt{2b}}{ac}
\end{array} \; .
\label{eqvanalles}
\end{equation}
The conditions that favour inflation during spontaneous symmetry breaking are a large $a$, small $b$ and large $c$, because these make the potential flat. Since the dependence on the physical parameters is fairly simple this translates to a large $\xi$, small $g$ and small $\lambda$. We can even get a simple bound on when inflation happens approximately all the way down to the global minimum. Let's assume that $a$ is much smaller than $c$ \footnote{This is quite natural to assume, because big values of $\xi$ will give cosmic strings with a too large tension \cite{hin}.}  and the field follows a straight line in going from $S_{c}$ to the global minimum, then we have the following average values:
\begin{equation}
<V>=\frac{g^{2}\xi^{2}}{4}
\end{equation}
\begin{equation}
<V'>=\frac{g^{2}\xi^{2}}{2\sqrt{a^{2}+c^{2}}} \approx \frac{g^{2}\xi^{2}}{2c}
\end{equation}
\begin{equation}
<\epsilon>=\frac{1}{16\pi}\left(\frac{<V'>}{<V>}\right)^{2} \approx \frac{1}{4\pi c^{2}} \; ,
\end{equation}
with $\epsilon$ being one of the slow-roll parameters \cite{lyt}. The condition for inflation to persist approximately all the way down becomes very simple:
\begin{equation}
\epsilon<1 \, , \qquad S_{c}>\frac{1}{\sqrt{4\pi}} \approx 0.28 M_{Pl} \; .
\label{eqendinf}
\end{equation}
Inflation will happen during a large part of the spontaneous symmetry breaking if $S_{c}$ is bigger than 0.28 $M_{Pl}$. Otherwise it is still possible that the area around $S_{c}$ is flat enough to give substantial inflation in the beginning of the symmetry breaking, but certainly not during this whole stage. 

We will now look at the parameter space which give the right magnitude of density perturbation according to Ref. \cite{rem}. Figure 2 in Ref. \cite{rem} gives various parameter combinations that lead to the right density perturbation\footnote{Note that $\kappa$ in Ref. \cite{rem} is $\lambda$ in my notation.}. Moreover combinations that lie under the horizontal line give a cosmic string contribution less than 10\%.

\begin{figure}[!h]
\includegraphics{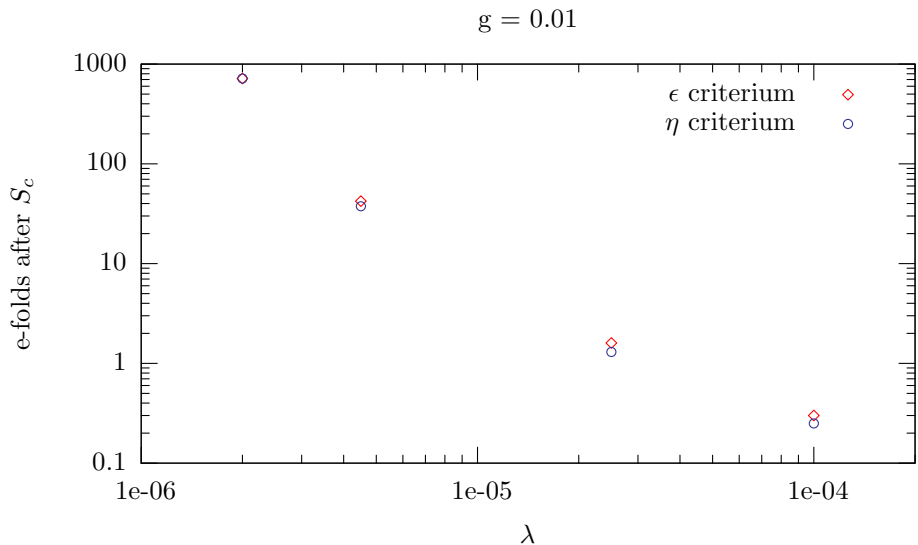}
\caption{\footnotesize{This figure depicts the number of efolds after $S_{c}$ using datapoints from Ref. \cite{rem} for $g=0.01$. In one simulation inflation stops when $\epsilon$ becomes one, in the other case when $\eta$ becomes one. This gives approximately the same results.}}
\label{steepestdescent-2-script.eps}
\end{figure}

Using their parameters as input, we are able to see if there is any inflation after $S_{c}$ (which is not included in their computation). We calculate numerically the steepest path down, using of course the total (tree-level + 1-loop) potential and, with the aid of equation (\ref{eqNef}), we attain the number of efolds after $S_{c}$. Figure (\ref{steepestdescent-2-script.eps}) gives the results for several datapoints with $g=0.01$. Inflation stops when either $\epsilon$ or $\eta$ comes close to one. You can see that it doesn't really matter which criterion we use. In general $\eta$ will become large just before $\epsilon$ does so and even if there is a substantial difference in the field values at which respectively the $\eta$ and $\epsilon$ criterions fail, this will not give a substantial difference in efolds, because when $\eta$ becomes large the slope is already to steep to have a considerable contribution from equation (\ref{eqNef}). However $\epsilon$ is a more reliable numerical quantity than $\eta$, since it is dependent on the first derivative of $V$ instead of the second derivative. It is important to note that in all simulations the difference between both criteria never exceeds about 3 efolds. Since the $\eta$ criterion can sometimes be off due to an enhanced numerical error, we will stick to the $\epsilon$ criterion from now on.

\begin{figure}[!h]
\includegraphics{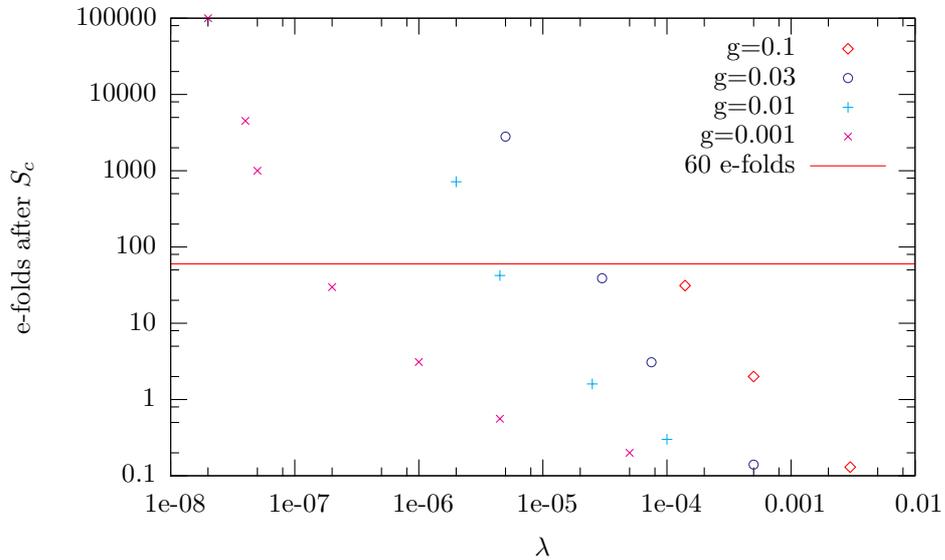}
\caption{\footnotesize{This figure depicts the number of efolds after $S_{c}$ using datapoints from Ref. \cite{rem} spanning the whole range of parameters. As you can see in approximately half of the cases the number of efolds exceeds 10, which will certainly give detectable consequences. In some cases the number of efolds even exceeds 60, meaning that the whole region of inflation relevant to cosmology happens after $S_{c}$.}}
\label{steepestdescent-all-script.eps}
\end{figure}

Figure (\ref{steepestdescent-all-script.eps}) gives the results for all sampled parameter combinations. The remarkable thing is that the number of e-folds after $S_{c}$ can be as high as 100000, so we are definitely not always justified to ignore this. It is hard to extract further physical results from this, because as soon as the number of efolds after $S_{c}$ become substantial, it could well be that this messes up the WMAP agreement in the density perturbation, which is dependent on the region about 60 efolds before the end of inflation. In other words, the relation between $\lambda$, $g$ and $\xi$ used in Ref. \cite{rem} could be wrong. In the next section I will try to find the corrected relation between these parameters, that is compatible with WMAP.

\clearpage
\section{New bounds on SUSY D-term inflation parameters}
\label{newbounds}

We have found a way to compute 1-loop quantum corrections to the D-term effective potential in section (\ref{sec1lc}). In section (\ref{graphs}) we have seen that these 1-loop corrections might be more important than thought previously. Especially in section (\ref{seciassb}) we saw that, using these 1-loop corrections and looking at data from \cite{rem}, there can be a huge amount of inflation after the critical $S$-value (where inflation is assumed to stop). The purpose of this section is to re-examine the agreement with the WMAP data \cite{WMA}, given the new insight of inflation after spontaneous symmetry breaking.

\subsection{Matching with WMAP-data}
\label{intronewb}

There are three free parameters $g$, $\lambda$ and $\xi$. $g$ is the dimensionless gauge coupling. Its most natural value would be somewhat smaller than unity. $\lambda$ is a dimensionless coupling and $\xi$ is the Fayet-Illiopoulos gauge term in units of $(M_{Pl})^{2}$. Its squareroot gives the field-value of the $\Phi_{-}$ field after spontaneous symmetry breaking, see equation (\ref{eqvanalles}), which directly determines the tension of the resulting cosmic strings, see equation (\ref{eqstrten}). The energy scale at which spontaneous symmetry breaking occurs is equal to $\sqrt{\xi}$ up to a factor $g / \lambda$, see equation (\ref{eqSc}).

The formulae for the amplitude of the (scalar) density perturbations $A_{S}$ and gravitational waves (or tensor perturbations) $A_{T}$ are\cite{bun}:
\begin{equation}
A_{S}=\sqrt{\frac{512\pi}{75}}\; \frac{V^{3/2}}{V'}
\label{eqA_S}
\end{equation}
\begin{equation}
A_{T}=\sqrt{\frac{32}{75}}\; V^{1/2}
\label{eqA_T}
\end{equation}
The input is again in Planck masses and both should be evaluated at 60 efolds before the end of inflation (or later if you want to know the perturbations on smaller scales). The value of the potential energy density $V$ and its derivative in the direction of steepest descent $V'$ are coming from the total (tree level  plus 1-loop) potential. As long as $A_{T}$ is small in comparison with $A_{S}$ it can be neglected and $A_{S}$ should be approximately $2 \cdot 10^{-5}$ according to CMB measurements. In all simulations, the amplitude of the gravitational waves is very small and thus can be neglected.

We proceed as follows: we fix a given combination of $\lambda$ and $g$, set $\Lambda_{m}$ equal to $S_{c}$ so that we can compare results with Ref. \cite{rem}\footnote{Normally the 1-loop corrections are very small in comparison to the tree-level potential. That is why people don't include the 1-loop corrections in the value of $V$ plugged in to equations like (\ref{eqA_S}) and (\ref{eqA_T}). The nice thing is that derivatives of $V$ are independent of $\Lambda_{m}$. Neglecting the 1-loop corrections in $V$ is practically the same as setting $\Lambda_{m}$ equal to the energy-scale of interest, since this makes the 1-loop corrections very small in the direct neighbourhood of this energy-scale. As we saw before, the $\Lambda_{m}$-dependence is very small anyway, so different choices of $\Lambda_{m}$ don't give dramatically different results.}, then we guess a value of $\xi$. We do the whole simulation with this value of $\xi$, see whether we get the wanted value of $2\cdot 10^{-5}$ for the density perturbation, equation (\ref{eqA_S}). If not, an algorithm makes a good second guess, and so on, until the density perturbation converges to the wanted limit\footnote{There is really no hope to do such a thing analytically, because the 1-loop corrections are way too complicated, and also the trajectory is not a straight line, as it is if you stick to values of $S$ bigger than $S_{c}$.}.

The main differences between our analysis and the analysis in \cite{rem} are:
\begin{enumerate}
\item{Their simulation stops at $S_{c}$, because they only have the 1-loop corrections on the inflationary valley. This new simulation takes into account the number of efolds after $S_{c}$ and the off-valley 1-loop corrections.}
\item{Because of aforementioned numerical problems of the kind "near-infinity minus near-infinity", see footnote (\ref{nimni}), in Ref. \cite{rem} they use an approximate formula for the 1-loop corrections. We solved these numerical difficulties and use the real formula.}
\item{The simulation in Ref. \cite{rem} includes the cosmic string contribution in the determination of $\xi$. Our simulation doesn't. This is justified since we can not assume right away that cosmic strings contribute, due to inflation after spontaneous symmetry breaking, see section (\ref{diluting}). If the cosmic strings do contribute this will give a difference of at most 10\% for datapoints consistent with WMAP.}
\end{enumerate}

\begin{figure}
\center
\includegraphics[width=1.0\textwidth]{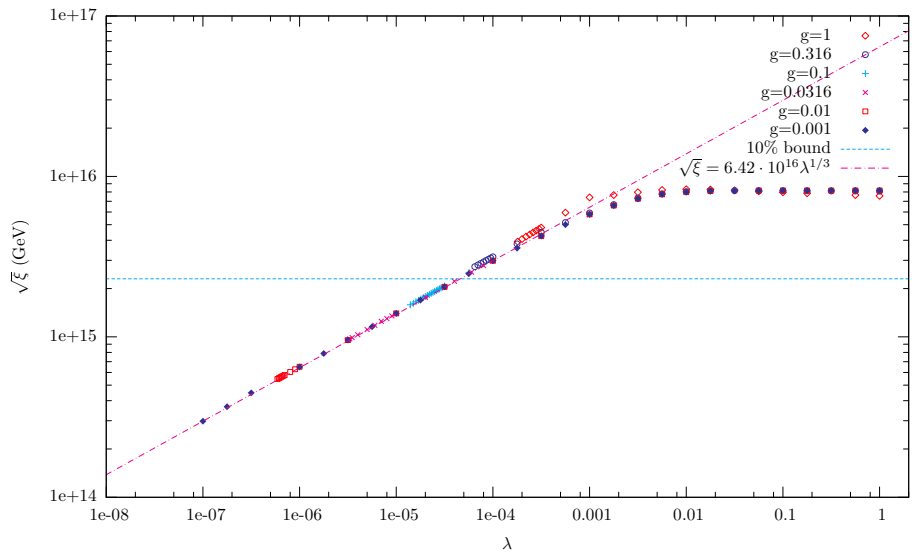}
\caption{\footnotesize{This graph shows the parameter-combinations that are compatible with the WMAP density perturbation. Points below the 10\% cosmic string contribution line are also compatible with the WMAP measurements of the spectrum of the CMB-anisotropies. There seems to be almost no dependence on the gauge parameter $g$. Datapoints for different $g$ overlap perfectly. The formula: $\sqrt{\xi}=6.42 \cdot 10^{16} \lambda^{1/3}$ (GeV) gives a perfect fit for $\lambda<5 \cdot 10^{-4}$}}
\label{xideterminationplotall_fit-script}
\end{figure}

This now happens to be one of the rare occasions in which increasing the complexity of the computation actually makes the results more simple, as can be seen by comparing figure (\ref{xideterminationplotall_fit-script}) with figure 2 in Ref. \cite{rem}. It depicts the whole range of values of the three parameters $\lambda$, $g$ and $\xi$ that give the right density perturbation of $2 \cdot 10^{-5}$. The parameter-range is larger than in Ref. \cite{rem}; There is an extra dataset for $g=0.3$ and $g=1.0$. Although a lot of computerwork is put in to get all these datapoints, the result is astonishingly simple. In Ref. \cite{rem} there is a large dependence on the gauge parameter $g$. This dependence has totally disappeared. Only for $g=1.0$ can we detect a slightly different path. For $\lambda$-values up to about $5 \cdot 10^{-4}$ the remaining ($\lambda$,$\xi$)-dependence is perfectly described by:
\begin{equation}
\sqrt{\xi}=6.42 \cdot 10^{16} \lambda^{1/3} \; \textrm{(GeV)} \; .
\label{fitxi}
\end{equation}
We cannot explain why a very involved simulation puts out such a nice precise number as 1/3.

\begin{figure}[!h]
\center
\subfigure[$g=1.0$]{\includegraphics[width=0.45\textwidth]{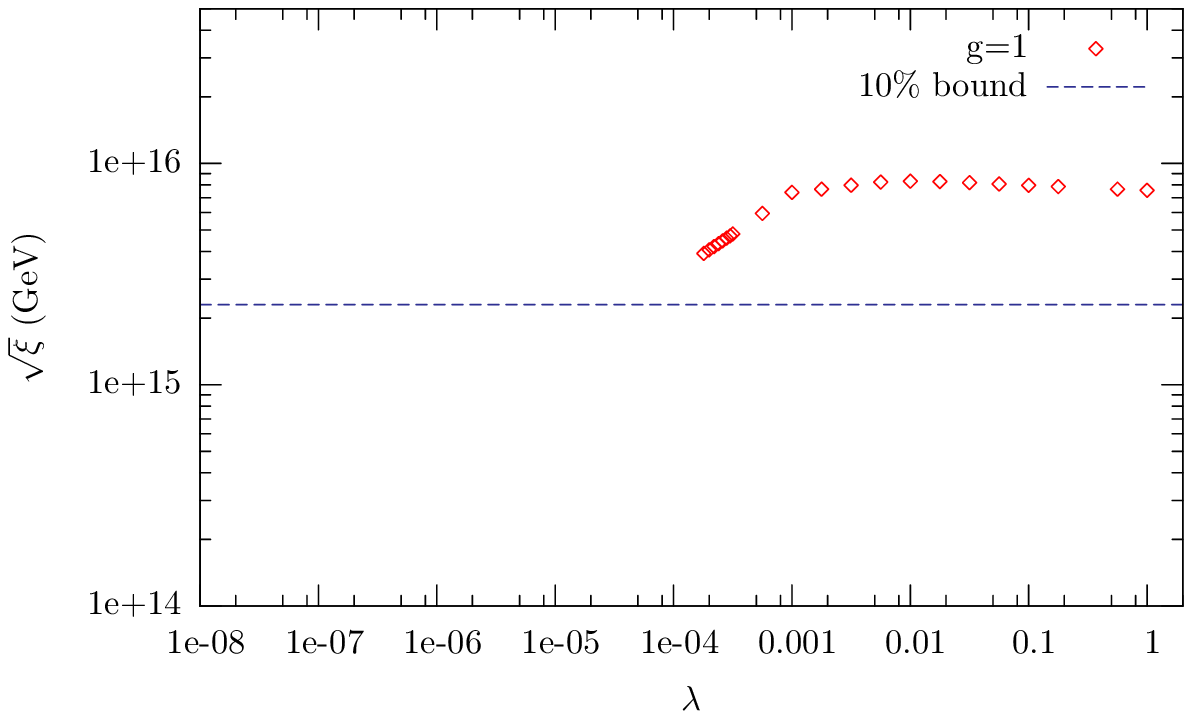}} \quad
\subfigure[$g=0.1\cdot \sqrt{10}$]{ \includegraphics[width=0.45\textwidth]{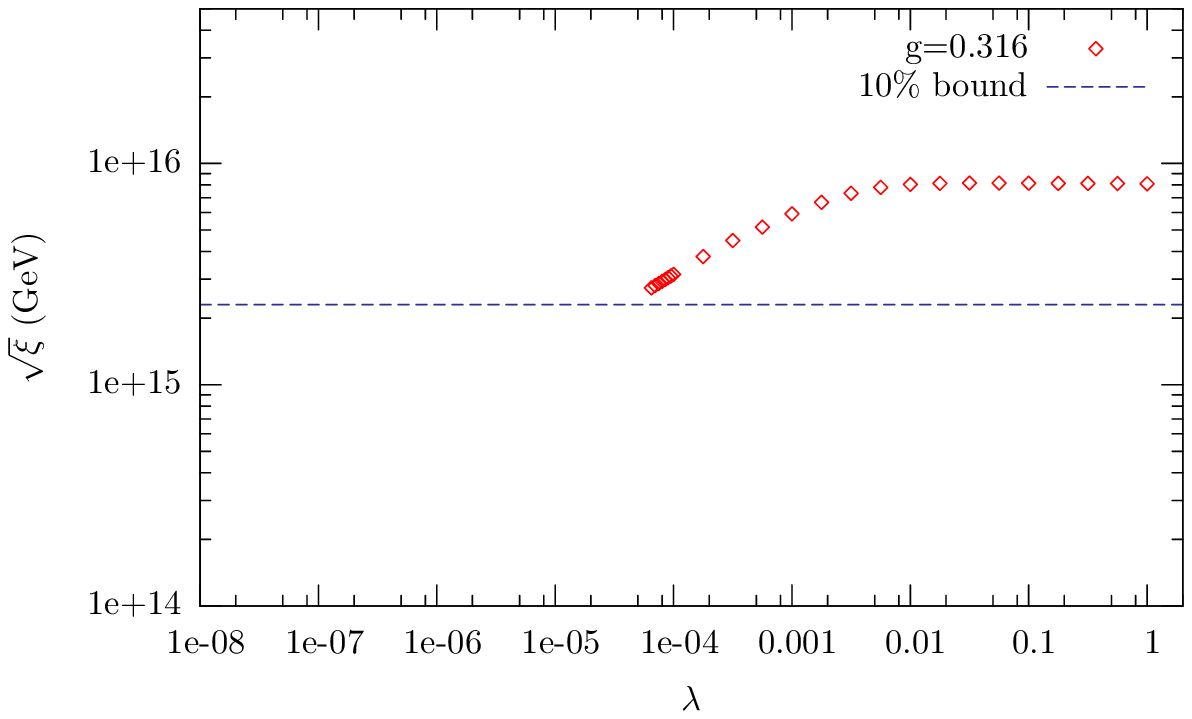}} \quad
\subfigure[$g=0.1$]{\includegraphics[width=0.45\textwidth]{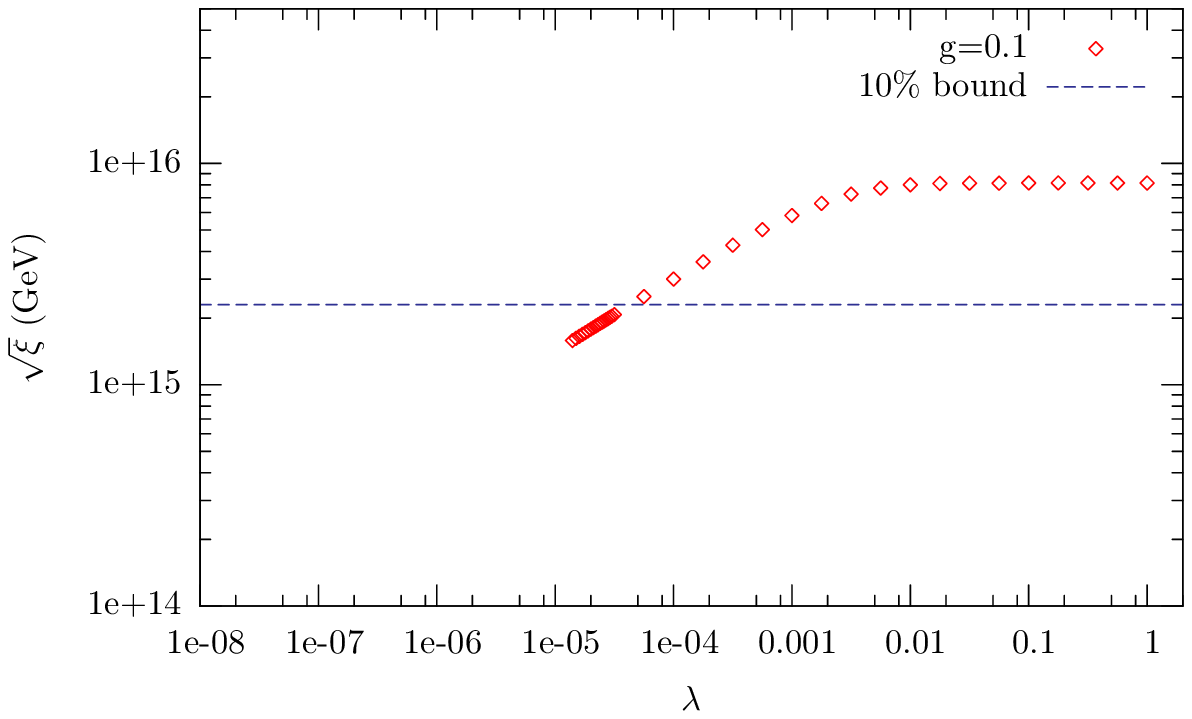}} \quad
\subfigure[$g=0.01\cdot\sqrt{10}$]{\includegraphics[width=0.45\textwidth]{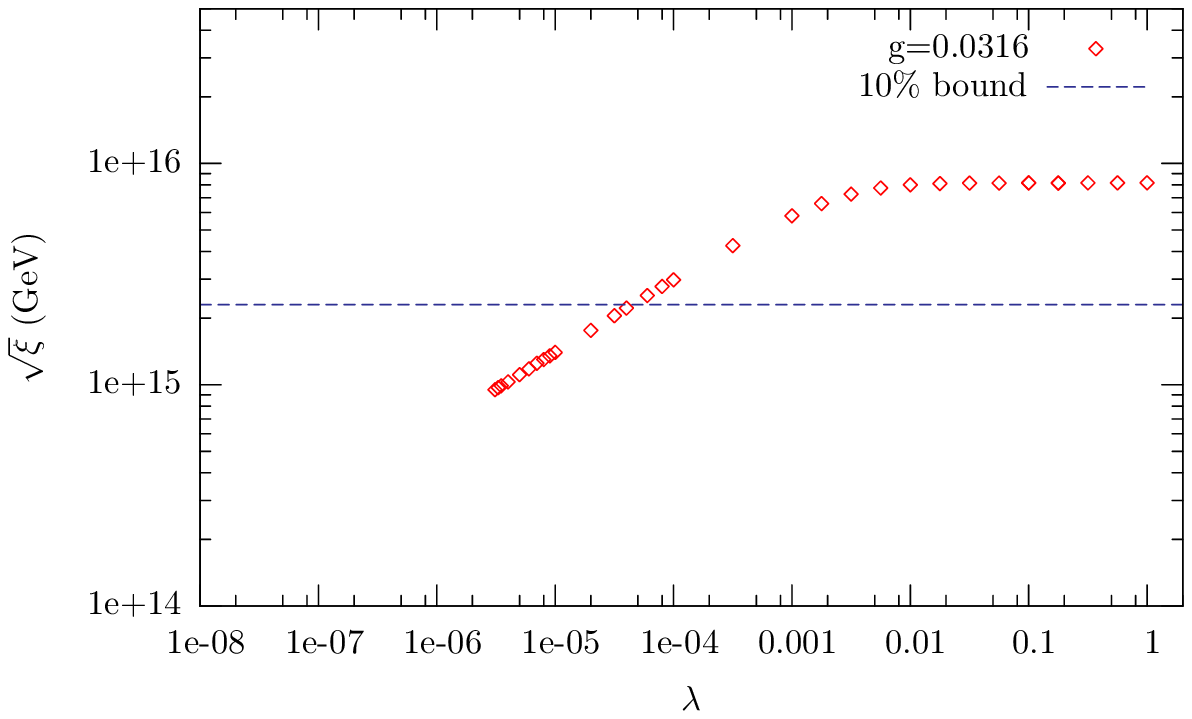}} \quad
\subfigure[$g=0.01$]{\includegraphics[width=0.45\textwidth]{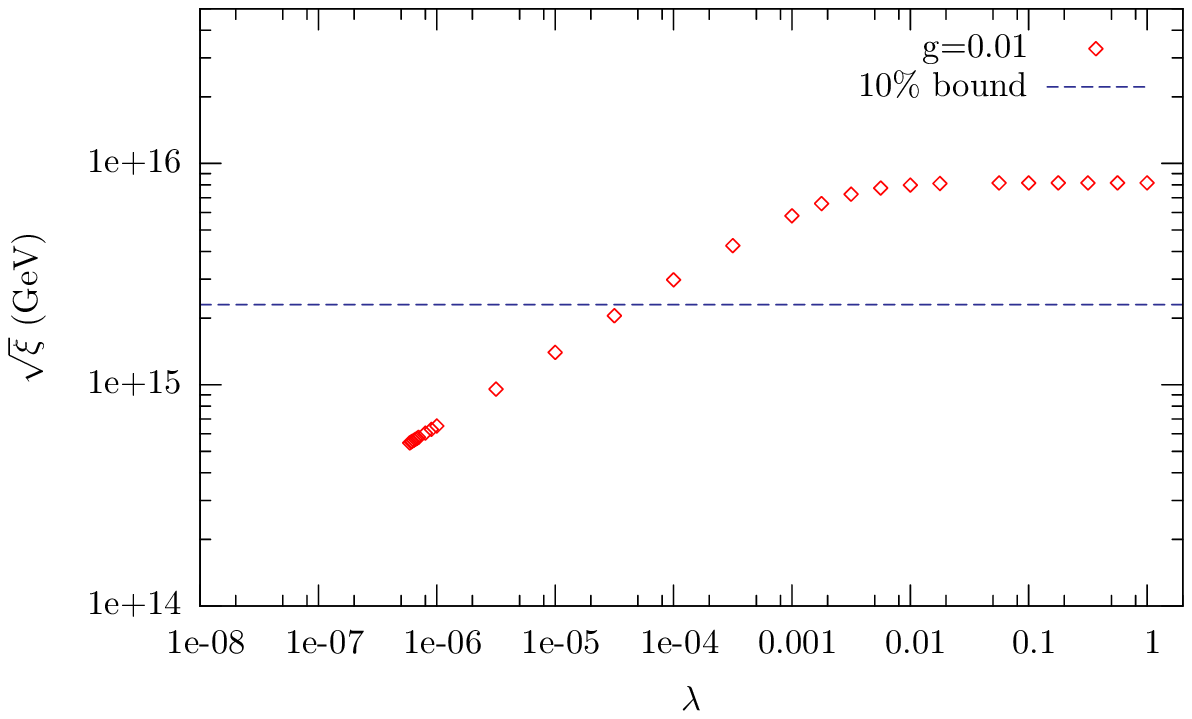}} \quad
\subfigure[$g=0.001$]{\includegraphics[width=0.45\textwidth]{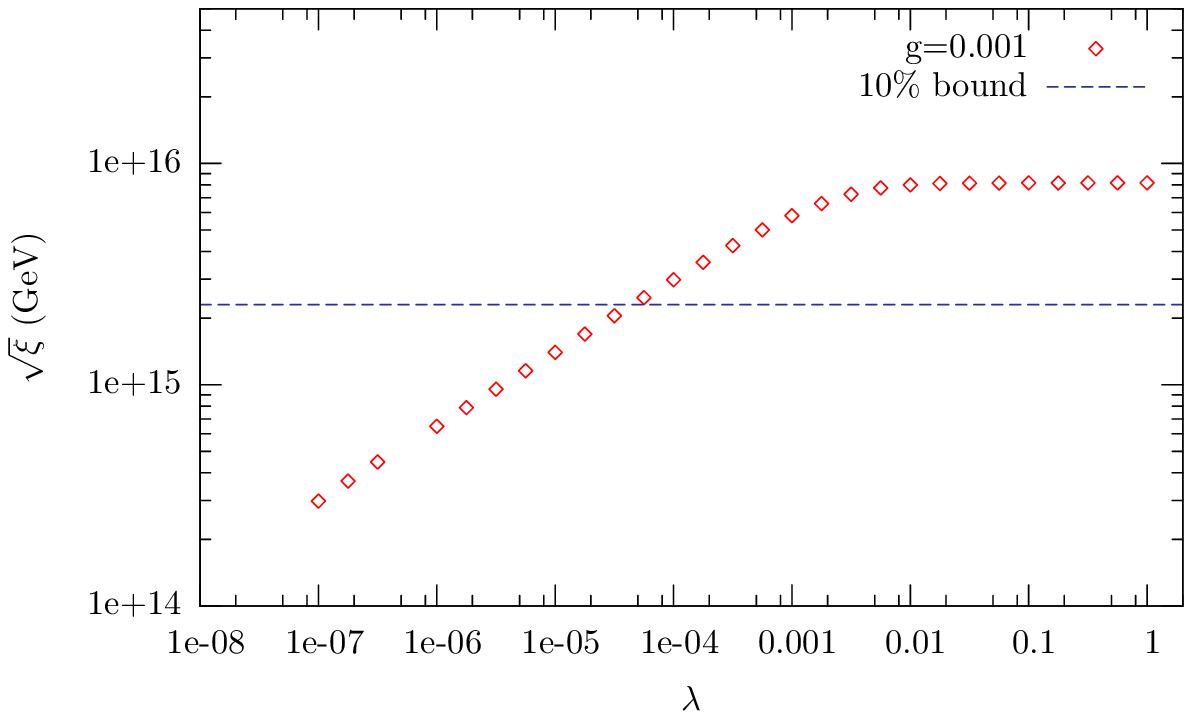}}
\caption{\footnotesize{The same dataset as in figure (\ref{xideterminationplotall_fit-script}) for different values of the gauge coupling $g$ separately. The cut-off at the lower $\lambda$-value corresponds to about 60 off-valley efolds. Taking smaller $g$-values shifts this cut-off to consecutively smaller $\lambda$-values.}}
\label{xideterminationplot1-script}
\end{figure}

 The value of the gauge coupling $g$ has no influence on relation (\ref{fitxi}), but it does determine at which point the number of off-valley efolds become important. This can be seen in figure (\ref{xideterminationplot1-script}). For each value of $g$ the simulation stops when the number of off-valley efolds comes close to 60. The value of $\lambda$ at which this occurs becomes smaller if $g$ becomes smaller. The simulation stops at this point, because when the 60th efold is off-valley the determination of the density perturbation becomes numerically very difficult and more importantly, my algorithm to find the desired $\xi$-value consistent with WMAP does not converge any more. We do not exclude the possibility that there are still parameter-combinations with smaller $\lambda$-values compatible to with WMAP, but they don't lie in the neighbourhood of the other points. In mentioning the number of off-valley efolds I really mean the number of efolds after the $\Phi$-field leaves the valley ($\Phi=0$). In general the 1-loop corrections first extend the steepest descent path a bit along the $S$-axis and then spontaneous symmetry breaking start quite abruptly, see for example figure (\ref{3Dplot-tree6_zoom-script.eps})(c), but note that the scaling is different on both axes. This is the starting point for what I call 'off-valley inflation'.

The reason why the $g=1$ plot does not perfectly overlap with the other $g$-values is probably that for $g=1.0$ already off-valley inflation becomes important around $\lambda=5 \cdot 10^{-4}$, before the curve bends around the corner.

The resulting parameter-regime that is compatible with WMAP is thus much larger than previously estimated. For small $\lambda$ values all datapoints remain at low $\xi$ values, which puts them safely below the 10\% cosmic string contribution line at $\sqrt{\xi}=2.3 \cdot 10^{15}$ GeV. But even more important, beforehand the maximum value of $g$ which still had some datapoints below the 10\% limit was $g=0.03$ and this has gone up to about $g=0.3$. In principle gauge couplings close to unity are more natural and they could arise from string theory. Even for gauge-couplings $g$ bigger than 0.3, there might be combinations compatible with WMAP, because for all g's it is possible to get substantial off-valley inflation which might dilute the cosmic strings enough to suppress its contribution to the density perturbation at reionization (an effect which is not included in the 10\% bound in figure (\ref{xideterminationplotall_fit-script})). If, for example, as a first guess we assume that there is one cosmic string produced on average per Hubble volume at the moment of symmetry breaking, then 60 efolds of off-valley inflation would mean that there is on average about one cosmic string left in the whole of our observable universe. The contribution to the density perturbation would then be way less than 10\%.

\clearpage

\subsection{Number of e-folds}
\label{numberofefolds}

The reason why the results in section (\ref{intronewb}) have simplified so dramatically, is because of the inclusion of a descent calculation of the amount of inflation below the critical inflaton field value. This shifts the cosmologically interesting region, which is 60 efolds before the end of inflation and therefore changes all observables, like the density perturbation and the spectral index, see section (\ref{secspec}). The effect will prove to be the largest for big values of $g$ and small values of $\lambda$. let us now look at the number of e-foldings in more detail.

As mentioned before in section (\ref{seciassb}) the 1-loop corrections will lengthen the path travelled on the inflationary valley. There are now two variables to look at: the number of efolds after $S_{c}$, something which was neglected before, but more interesting even is the number of efolds after the symmetry breaking, because this opens the possibility to dilute the cosmic strings after they are produced, which will be discussed in section (\ref{diluting}).

\begin{figure}
\center
\includegraphics[width=1.0\textwidth]{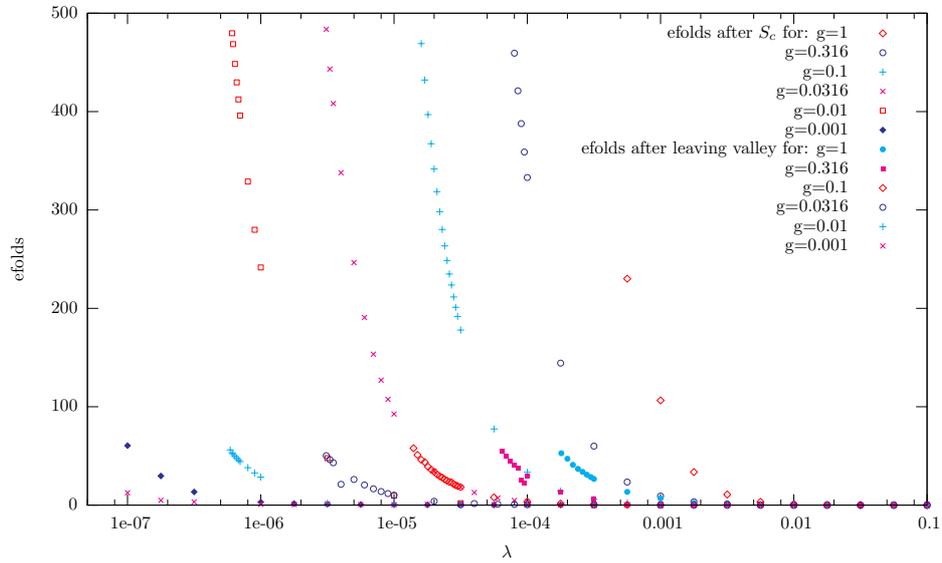}
\caption{\footnotesize{This graph shows the amount of inflation both after $S_{c}$ and after leaving the $(\Phi=0)$-valley for all datapoints.}}
\label{xideterminationplot_efolds_nolog-script}
\end{figure}

\begin{figure}
\center
\includegraphics[width=0.9\textwidth]{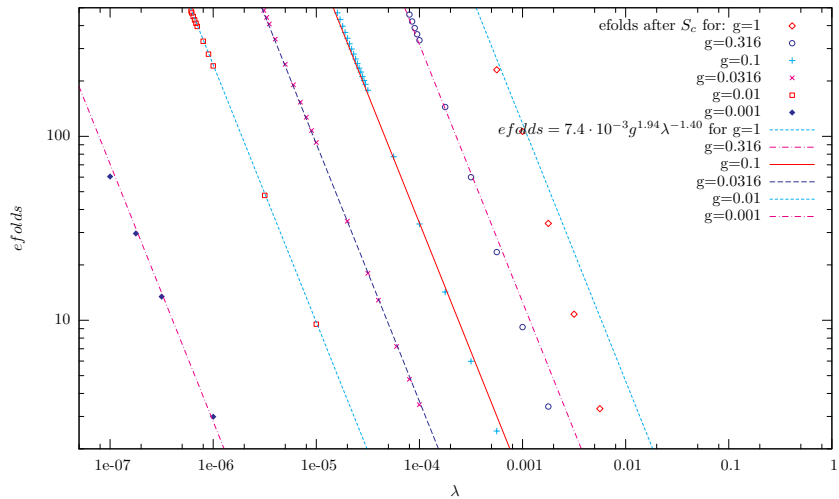}
\caption{\footnotesize{This graph shows the amount of inflation after $S_{c}$ for all datapoints. $N_{Sc}=7.4\cdot 10^{-3} g^{1.94} \lambda^{-1.40}$ is a good fit to all datapoints except $g=1.0$.}}
\label{xideterminationplot_efolds_Sc-script}
\end{figure}

\begin{figure}
\center
\includegraphics[width=0.9\textwidth]{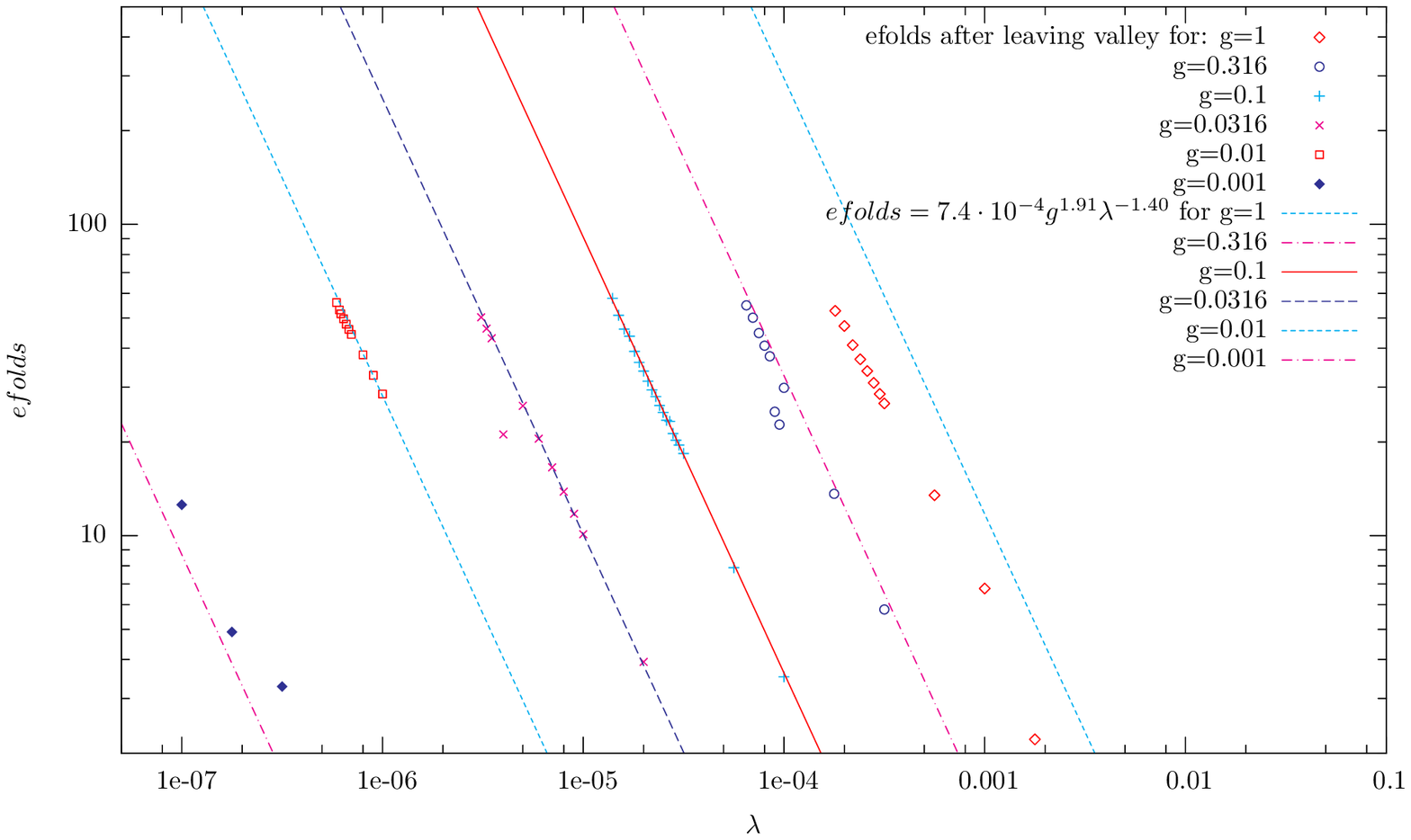}
\caption{\footnotesize{This graph shows the amount of inflation after leaving the $(\Phi=0)$-valley for all datapoints. $N_{Sc}=7.4\cdot 10^{-4} g^{1.91} \lambda^{-1.40}$ is a good fit to all datapoints except $g=1.0$. The occasional point that is off the line is probably due to a slight mistake in the automatic recognition of the point where $\Phi$ really leaves the neighbourhood of $(\Phi=0)$, which can be tricky.}}
\label{xideterminationplot_efolds_valley-script}
\end{figure}

In order to compute the number of efolds we use the integral in equation (\ref{eqNef}). Figures (\ref{xideterminationplot_efolds_nolog-script}) to (\ref{xideterminationplot_efolds_valley-script}) show the number of e-folds as a function of $g$ and $\lambda$ and the fitted $\xi$. Again the results are very simple. The following formulae give a good fit, except for $g=1.0$.
\begin{equation}
N_{Sc}=7.4\cdot 10^{-3} g^{1.94} \lambda^{-1.40}
\end{equation}
\begin{equation}
N_{val}=7.4\cdot 10^{-4} g^{1.91} \lambda^{-1.40}
\end{equation}
$N_{Sc}$ and $N_{val}$ give the number of efolds respectively after $S_{c}$ and after leaving the ($\Phi=0$) valley. It is quite surprising that the fit works well in both the $g$ and $\lambda$ direction and moreover both functions are very similar, with about a factor of 10 difference.

\begin{figure}
\center
\includegraphics[width=1.0\textwidth]{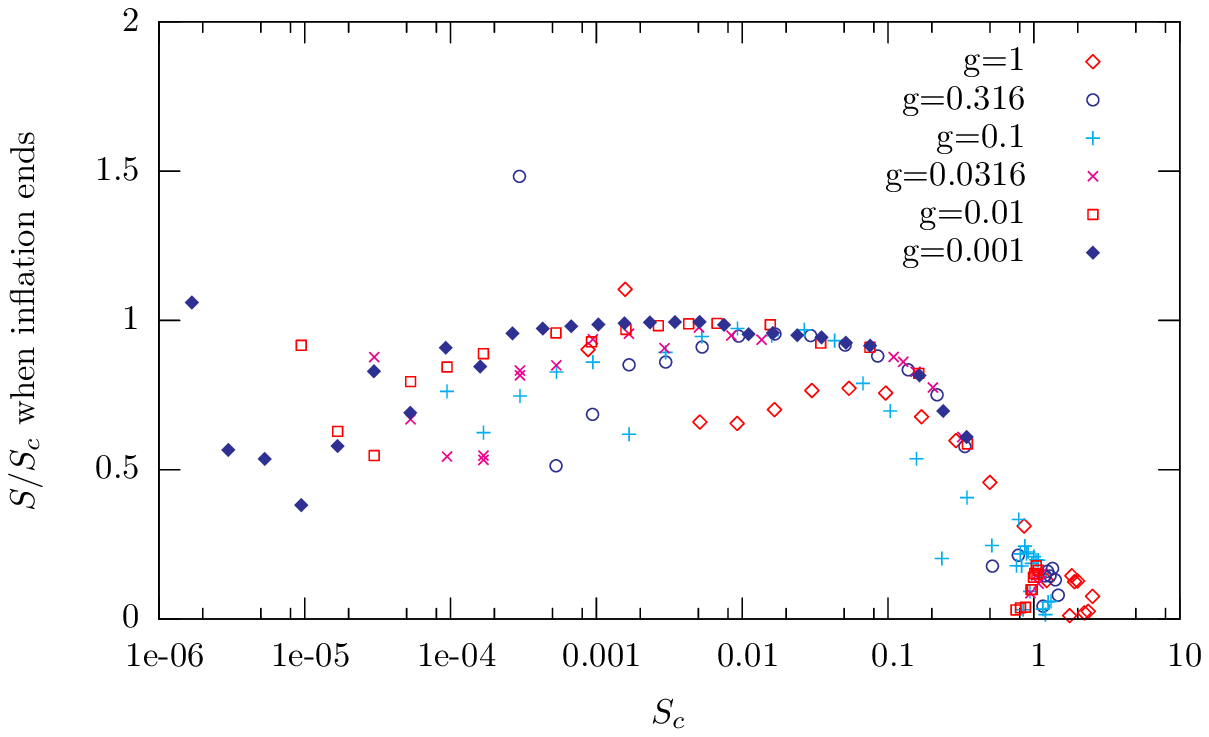}
\caption{\footnotesize{The $S/S_{c}$ field value at the end of inflation is plotted as a function of $S_{c}$. At around the predicted value of 0.28 $M_{Pl}$ this line goes steeply down, but also on the left of the picture there seems to be a possibility of a very flat potential up to around half-way down the S-axis. Most datapoints however lie very crowded on one curve.}}
\label{xideterminationplot_Sc-script}
\end{figure}

In section (\ref{seciassb}) we argued that, even classically, inflation can be extended on a large portion of the trajectory during spontaneous symmetry breaking, if $S_{c}$ becomes bigger than about 0.28 $M_{Pl}$, see equation (\ref{eqendinf}). Figure (\ref{xideterminationplot_Sc-script}) shows that this is indeed the case. Right around $S_{c}\approx 0.28 M_{Pl}$, the portion of the off-valley trajectory on which inflation still takes place, becomes substantial. This does not depend much on any of the other parameters.

\subsection{Spectral index}
\label{secspec}

The spectral index denotes how the amplitude of the tiny fluctuations depend on the scale. A spectral index of $n=1$ would correspond to a scale invariant perturbation. WMAP measurements reveal that this spectral index is very close to one. The 1-year results gave a value of $n=0.99 \pm 0.04$. The accuracy improved with the 3-year results (which include CMB polarisation measurements), giving a value of $n=0.961 \pm 0.017$ \cite{WMA}. So the spectral index is very close to one and probably a bit smaller ($n=1$ is at a $2\sigma$ deviation for the three-year results). Bayesian statistics give a somewhat larger interval as compared to the error estimate above \cite{bay}. 

During slow-roll there  is a simple equation for the spectral index \cite{lyt}:
\begin{equation}
n=1-6\epsilon+2\eta \; .
\end{equation}
In our computer-program this is evaluated between 58 and 60 efolds before the end of inflation. 

\begin{figure}
\center
\includegraphics[width=1.0\textwidth]{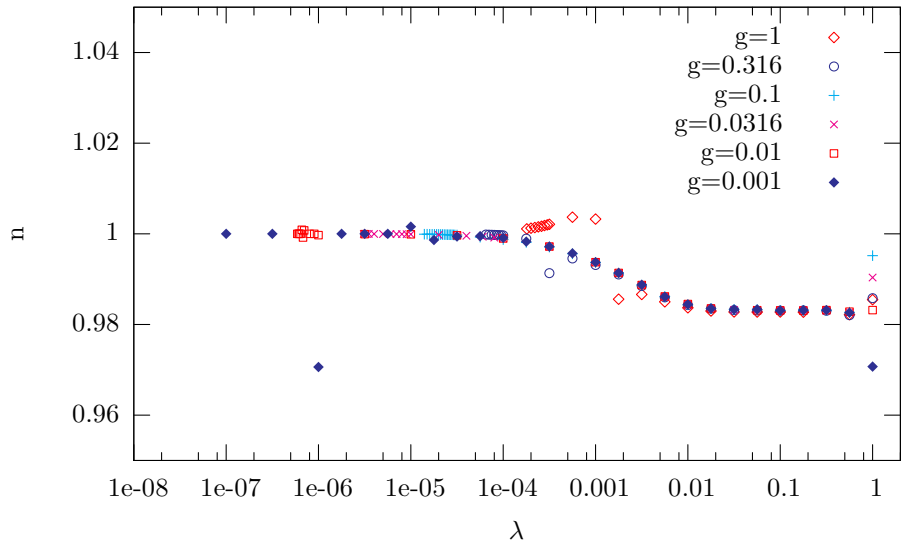}
\caption{\footnotesize{This figure depicts the values of the spectral index for all datapoints. They all lie on the same curve, with a dependence only on $\lambda$, except for a small portion of the $(g=1)$ dataset. So most points really exist of 6 datapoints right on top of each other. In Ref. \cite{rem} there was a large dependence on the gauge parameter $g$, which now has totally disappeared.}}
\label{xideterminationplot_n-script}
\end{figure}

The results emerging from our simulations are very simple. Again there is almost no dependence on the gauge-parameter $g$. The spectral index equals 1.0 for all $\lambda$ smaller than $10^{-4}$ and then goes to a value of $n=0.983$ and remains constant for $\lambda>0.01$, see figure (\ref{xideterminationplot_n-script}), compare with Ref. \cite{rem}, figure 2. The slight deviation for $g=1$, from all other $g$-values, is exactly in the same $\lambda$-region that was also slightly off in figure (\ref{xideterminationplotall_fit-script}).

The nice thing is that the model gives very precise predictions, so it can be falsified. In \cite{rem} for example the spectral index is equal to 1.0 on most of the domain and then suddenly goes down very fast, meaning that you can get any small value of $n$, but you have to fine-tune the parameters, which is not desirable. Our value of $n=0.983$ is at a $1.3\sigma$ deviation from WMAP measurements and $n=1.0$ is at $2.2\sigma$ deviation. So the values of $\lambda$ that are consistent with WMAP, are approximately $\lambda > 0.001$, although smaller values are not excluded.

The region $\lambda > 10^{-3}$, which gives the best spectral index according to the 3-year WMAP data, does not overlap with the region consistent with the 10\% cosmic-string contribution bound, which is approximately $\lambda<0.5 \cdot 10^{-4}$ or $\lambda<10^{-3}$ if cosmic strings can be effectively diluted, see section (\ref{diluting}). Of course, this can still be due to the statistical error in the WMAP-measurement (Bayesian statistics for WMAP do not exclude a spectral index equal to one). It might also be possible to get a spectral index smaller than one, by adding fluctuations to the path. Now we took the steepest path down, which is a sensible thing to do because of the large friction term in the equations of motion, but since the path on the inflationary valley is not necessarily long, it could be that small deviations from the steepest path have not damped out and these could give a spectral index other than one (since the spectral index has a large dependence on the second derivative of the path). Another possible, but maybe far-fetched solution, will be presented in section (\ref{pushing}).

\subsection{Diluting cosmic strings}
\label{diluting}

In section (\ref{numberofefolds}) we have seen that a large amount of off-valley inflation is possible for any value of the gauge coupling $g$. Cosmic strings will form at the end of inflation, because of the spontaneous breaking of a U(1) symmetry \cite{vil}. In this section we will see how off-valley inflation affects the cosmic string density.

\begin{figure}
\center
\includegraphics[width=1.0\textwidth]{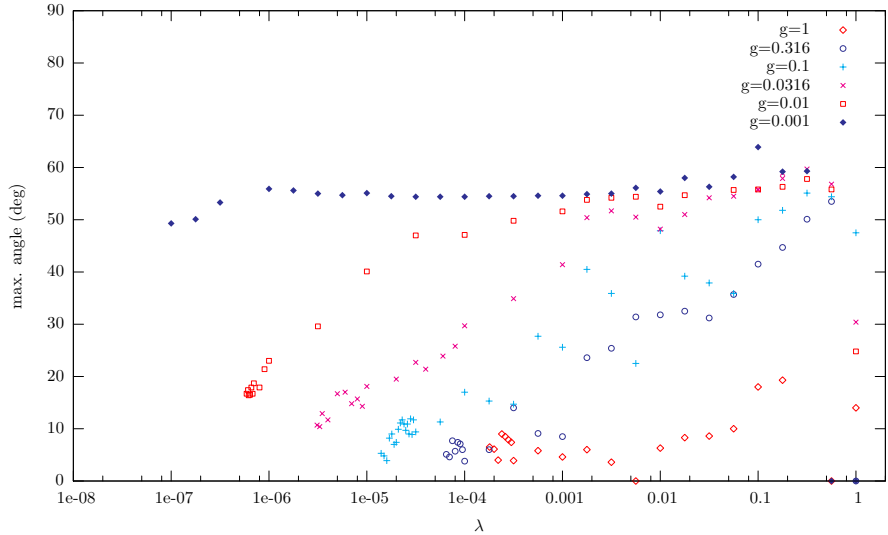}
\caption{\footnotesize{This figure denotes the maximum angle with respect to the $S$-axis of the steepest descent path, for all datapoints. Generally the fields leave the $S$-axis quite abruptly and after that the angle gradually becomes smaller.}}
\label{xideterminationplot_angle-script}
\end{figure}

We assume that the cosmic strings are formed at the moment that the $\Phi$ field leaves the inflationary valley. This happens if the process is relatively fast, in which case there will be a string production of approximately one length of string per Hubble volume. Figure (\ref{xideterminationplot_angle-script}) shows at which angle the $\Phi$-field will move from the $S$-axis. This angle always goes up very fast at the moment the fields leave the valley and then gradually goes down again. Since the leaving-angle is generally quite big, it is reasonable to assume that the strings are produced at the moment of going off-valley, but it is possible that the quantum fluctuations are still big enough to pull the field over the top of the Mexican hat for some time \cite{vil}, which would mean that the strings are effectively produced at some later instant, when the quantum-fluctuations become small as compared to the height of the Mexican hat. It would be interesting to do a precise analysis or simulation of this process.

Normally one assumes that one length of cosmic string is produced per Hubble volume at $S_{c}$. If there is inflation after the formation of strings the energy density stored in these cosmic strings will go down with the second power of the amount of inflation\footnote{Not the third power, because a cosmic string is a one dimensional object with a constant tension, which gets stretched by the expanding universe.}. So, for example 30 off-valley efolds, would reduce the energy density of the strings with a factor $e^{60} \approx 10^{26}$.

This certainly seems enough to get the string contribution below the 10\% bound, but this is not quite true. What matters is the string density at the moment of reionization. In between inflation and reionization the strings will interact and the string density will decrease. If the energy lost in this interaction goes into gravitational waves, then it does not contribute significantly to the WMAP anisotropy. As shown in Ref. \cite{qse} after a relevant cosmic string scale enters the Hubble radius again, it will go into a scaling regime, which makes the density of cosmic strings on a scale smaller than the Hubble radius approximately constant with respect to the Hubble radius. This means that the inter string distance grows with the Hubble radius. This is why the 10\% string contribution bound only depends on $\xi$ \cite{pog}: The density at reionization is always the same, but $\xi$ determines the tension of the string and therefore it determines the total energy-density stored in cosmic strings at reionization. The formula for the tension of a D-term string is very simple because it is a "critical coupling" or BPS string:
\begin{equation}
\mu=2\pi \xi \; .
\label{eqstrten}
\end{equation}

Now if the number of off-valley efolds exceeds 60, there will be no string contribution for sure, because there simply are no strings. Up to about 5 efolds less, the string contribution will still be very small, because the inter-string distance will not have entered the Hubble horizon at decoupling. For even smaller off-valley efolds the string contribution will go to the value as calculated before, due to the scaling. How fast the string contribution will go to this value depends on how fast the scaling will take place.

\subsection{Pushing the top of the Mexican hat down}
\label{pushing}

As shown in section (\ref{results}), for a restricted set of parameters it is possible to lower the top of the Mexican hat of the fields in the true vacuum state. This could lead to strings with less tension and since the effect seems to be the largest for the parameters with a nice spectral index of 0.983, it is worth looking at.

\begin{figure}
\center
\subfigure[all $g$ values]{ \includegraphics[width=0.8\textwidth]{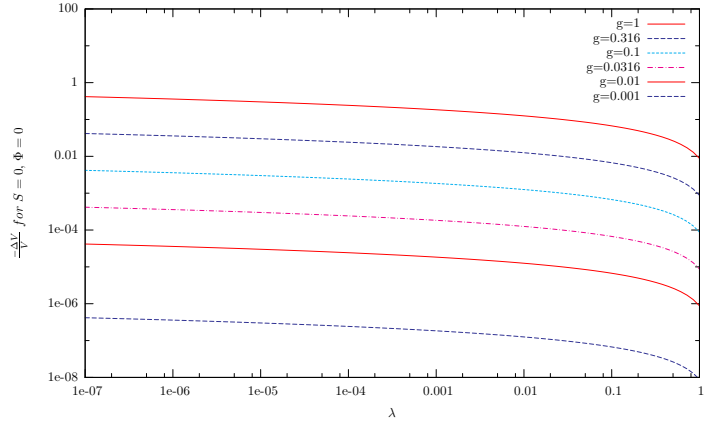}} \quad
\subfigure[$g$ close to one]{\includegraphics[width=0.8\textwidth]{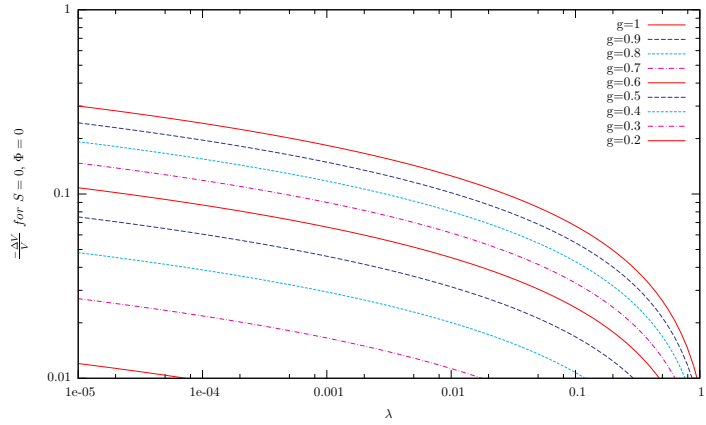}} \quad
\caption{\footnotesize{Relative size of the negative 1-loop corrections in the origin $(S=0, \Phi_{-}=0)$ for all datapoints (a), zoomed in on the relevant region (b). For almost all choices of $g$ we get a negligible 1-loop contribution. Only values of $g$ close to one can give a significant contribution.}}
\label{xideterminationplot_DeltaV-script}
\end{figure}

Since in the origin $(S=0, \Phi=0)$ many of the particles are massless, we can actually do the analysis analytically. Evaluating all the masses, by looking carefully at tables (\ref{tabelgau}), (\ref{tabelfer}) and (\ref{scamass}), we see that only four scalar masses contribute: $M^{2}(\phi_{+(1,2)})=g^{2}\xi$ and $M^{2}(\phi_{-(1,2)})=-g^{2}\xi$. The rest of the fields are massless. Inserting this in the Coleman-Weinberg formula gives:
\begin{equation}
\Delta V = \frac{1}{64\pi^{2}} \left(
2(g^{2}\xi)^{2} ln \left| \frac{g^{2}\xi}{\Lambda_{m}^{2}}  \right|+
2(-g^{2}\xi)^{2} ln \left| \frac{-g^{2}\xi}{\Lambda_{m}^{2}}  \right|
\right)
=\frac{g^{4}\xi^{2}}{16\pi^{2}} ln \left( \frac{g^{2}\xi}{\Lambda_{m}^{2}}  \right) \; .
\end{equation}
We should compare this with the tree-level potential in the origin, see equation (\ref{eqVVac}).
This gives a relative 1-loop correction in the origin of:
\begin{equation}
\frac{\Delta V}{V}=\frac{g^{2}}{8\pi^{2}} ln \left( \frac{g^{2}\xi}{\Lambda_{m}^{2}}  \right) \; .
\label{eqdvoverv}
\end{equation}
For our datapoints we used $\Lambda_{m}=S_{c}$. If we insert this into equation (\ref{eqdvoverv}) we get:
\begin{equation}
\frac{\Delta V}{V}=\frac{g^{2}}{8\pi^{2}} ln \left( \frac{\lambda^{2}}{2}  \right) \; .
\end{equation}
This will give a negative 1-loop contribution for $\lambda<\sqrt{2} M_{Pl}$, so all datapoints will have a 1-loop correction in the origin which counteracts the tree-level potential. Figure (\ref{xideterminationplot_DeltaV-script}) shows the relative size of these 1-loop corrections.

As you can see the 1-loop-contribution in the origin is very small indeed for almost all parameter combinations, but it can go up to about a maximum of 20\% for values of the gauge parameter $g$ very close to one. Of course we don't know what the two-loop contribution is going to do, so we cannot make a definite judgement to what extent this effect is important. The question remains open, whether the effect of this can be big enough to lower the string contribution below the 10\% bound for the datapoints at large $\lambda$. In any case, the best thing to do in order to get an expression for the tension of the strings, including this 1-loop correction, is to make a numerical model, find the string-solution and integrate the energy-density to get a corrected tension. Of course it will come out smaller than in equation (\ref{eqstrten}), but probably not small enough to get below the 10\% string contribution bound.

\section{Conclusion}
\label{conclusion}

In this paper we have looked in great depth into the D-term hybrid inflation model. The key ingredient we have added is a description of the fields after reaching the critical inflaton value and during spontaneous symmetry breaking, when the fields leave the inflationary valley. To analyse this we need a formula for the 1-loop quantum corrections to the potential energy density, outside the inflationary valley. In calculating this, problems arise related to the non-equilibrium Higgs mechanism. We presented a good way out of these problems, but there still is room for a thorough theoretical analysis of this phenomenon.

Using our method, the parameter bounds of the D-term hybrid inflation model, compatible with WMAP measurements, improve a lot. A nice and simple relation between $\lambda$ and $\xi$ and independent of $g$ is presented, that will give the correct CMB density perturbation. All parameter combinations with $\lambda<4.5 \cdot 10^{-5}$ give a cosmic string contribution less than 10\% and also for values of $\lambda$ approximately smaller than 0.001 there is the possibility of diluting the strings after formation, which could also give a string contribution less than 10\%. Moreover, where beforehand only $g<0.03$ was compatible, now all values of $g$ are possible. Results for the spectral index have no dependence on $g$. It lies between $n=0.983$ and $n=1.0$. The lower values are attained for approximately $\lambda$ bigger than 0.001. So the regions best compatible with the spectral index and with the cosmic string bound are exclusive.

For D-term inflation to be compatible with WMAP, either $n=1.0$ must prove to be the right value (it is currently at a $2\sigma$ deviation), or neglected dynamical effects must give a contribution to the spectral index, making it smaller than one for $\lambda<0.001$, or 1-loop effects which lower the top of the Mexican hat potential in the true vacuum must lower the string tension below the 10\% bound.

In the future, work should be done related to the precise length scale of string formation in this model, including quantum fluctuations. The scaling behaviour can be examined, in cases where there are a lot of off-valley e-foldings. And also the influence of insertion of the true equations of motion, instead of the steepest descent path, on the spectral index might be interesting. Of course one could try to calculate the tension of the resulting strings, which have a lowered top of the Mexican hat potential. We saw that we could not neglect 1-loop corrections, so maybe even two- or higher-loop corrections may prove to be significant. But the first thing to do is to extend the used method to F-term hybrid inflation.

So there are still a lot of improvements that can be made, but the model as a whole looks quite promising, which is something that could not be said before.

\appendix
\section{Contribution from massless particles}
\label{secmassless}

For some field-values, there will be massless particles contributing to the 1-loop-correction in (\ref{eqcw}). For example on the inflationary valley
the fields $B_{\mu}$ and $(S_{1} s_{2} - S_{2} s_{1})/|S|$ will be massless. For these massless loops there is an indeterminate expression in
(\ref{eqcw}), but taking a simple limit reveals that these massless particles really don't contribute to the loop-correction:
\begin{equation}
\lim_{M \to 0}\left(M^{2} ln(M/\Lambda_{m}^{2})\right)=0 \; ,
\end{equation}
where $M$ is the mass-squared.

\section{Contribution from imaginary masses}
\label{concave}

We have identified all field-dependent masses of the physical particles in the region ($\Phi_{+}=0$, $\Phi_{-}$ and $S$ arbitrary).
In plugging them into equation (\ref{eqcw}) there is still a problem. As can be seen in table (\ref{tabelsca}) the
$(mass)^{2}$-values are not always positive for all field values. Negative $(mass)^{2}$-values plugged into the natural logarithm will give
an imaginary contribution to the effective potential. This problem is resolved in Ref. \cite{wew}:

Quantum-mechanically the effective potential can be interpreted as the energy-density of the quantum-state that minimises
this energy-density, subject to the condition that the field expectation values (in this case $<\Phi_{\pm}>, <S>$) are as given. So for every point in field-space you can define
an effective potential value. For classical potentials that are convex everywhere, these quantum-states will be homogeneous, meaning:
concentrated around a point. But for partially concave classical potentials (which can give a spontaneous symmetry breaking)
these quantum-states will be mixed states (because,
if $V''$ is negative, the mixed state will certainly have a lower energy than a homogeneous state).

For our purposes we don't
want to know the energy density of these mixed states, but rather we want to know the energy density of a homogeneous state
centred around the same concave point in field-space.
Weinberg and Wu show in Ref. \cite{wew} that the encountered imaginary 1-loop-corrections to the effective potential correspond to these concave points (for mixed states).
The nice thing is that the real part of the 1-loop correction can be interpreted as the 1-loop correction to the energy density of the
homogeneous state\footnote{The imaginary part of the 1-loop corrections can be interpreted as a decay rate.}. For our calculations we should therefore only take the real part of the 1-loop effective
potential.

\section{Contribution from imaginary mass-squared values}
\label{imaginary}

As can be seen from tables (\ref{tabelsca}) and (\ref{tabelfer}) there is a possibility that some bosonic and fermionic $(mass)^{2}$ values are imaginary. This happens when the value under the square-root sign becomes negative. Therefore we'd better hope that this never happens.

For the fermionic contribution it is easy to see that, indeed, $E^{2}-8 \lambda^{2} g^{2} \Phi^{4}$ is strictly positive for all field values, since it can be written as a sum of squares.
\begin{equation}
E^{2}-8 \lambda^{2} g^{2} \Phi^{4}
=\lambda^{4} S^{4}+2 \lambda^{2}\left(\lambda^{2}+2 g^{2}\right) S^{2}\Phi^{2} +\left(\lambda^{2}-2 g^{2}\right)^{2}\Phi^{4} \ge 0 \; ,
\end{equation}
where $S$ is shorthand notation for $S_{1}^{2}+S_{2}^{2}$.

For the scalar contribution this is less trivial, because it is not clear how to write $B^{2}+4 C$ as a sum of squares.
\begin{displaymath}
B^{2}+4 C= \left(\frac{1}{2}\lambda^{2}S^{2}-g^{2}\xi\right)^{2}+\left(\frac{1}{2}(\lambda^{2}-3g^{2})\Phi^{2}+g^{2}\xi\right)^{2}
\end{displaymath}
\begin{equation}+\left(\frac{7}{2}\lambda^{4}+\frac{3}{2}\lambda^{2}g^{2}\right) S^{2}\Phi^{2}-g^{4}\xi^{2}
\end{equation}
However it is still possible that this function is strictly positive, because the three squares can not all be zero at the same time (fixing $S$  and $\Phi$ can make two squares vanish, but then the third will not vanish). Calculating the minimum of this function in the $(S,\Phi)$-plane for a given $\lambda,g$ and $\xi$, this indeed turns out to be strictly positive\footnote{The minimum is attained for:
$\Phi^{2}=\frac{1}{3}\frac{g^{2}}{\lambda^{2}+g^{2}}\xi
$ and $S^{2}=\left(\frac{g^{2}}{\lambda^{2}}-\frac{1}{3}\right)\frac{g^{2}}{\lambda^{2}+g^{2}}\xi $.}.
\begin{equation}
\left( B^{2}+4C \right)_{min}=\frac{4}{3}g^{4}\xi^{2}\left(\frac{\lambda^{2}}{\lambda^{2}+g^{2}}\right) \ge 0
\end{equation}
We conclude that there will never be imaginary mass-squared values in our computation, so we need not worry about this.

\section{Higgs mechanism for polar and Cartesian fields}
\label{apD}

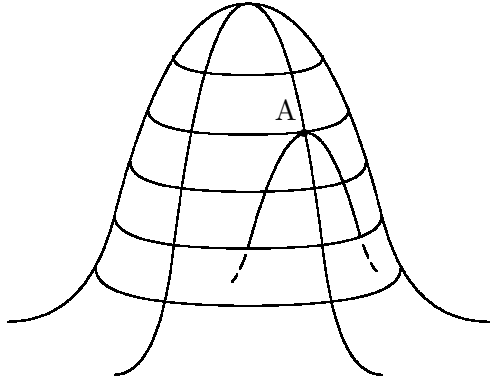
\begin{figure}[htbp]
\begin{center}
\begin{picture}(200,200)

\qbezier(100,180)(130,180)(150,100)
\qbezier(150,100)(160,60)(190,60)
\qbezier(100,180)(70,180)(50,100)
\qbezier(50,100)(40,60)(10,60)

\qbezier(128,160)(128,153)(100,153)
\qbezier(72,160)(72,153)(100,153)

\qbezier(137.5,140)(137.5,130.5)(100,130.5)
\qbezier(62.5,140)(62.5,130.5)(100,130.5)

\qbezier(144,120)(144,109)(100,109)
\qbezier(56,120)(56,109)(100,109)

\qbezier(150,100)(150,87.5)(100,87.5)
\qbezier(50,100)(50,87.5)(100,87.5)

\qbezier(157,80)(157,66)(100,66)
\qbezier(43,80)(43,66)(100,66)


\qbezier(100,180)(116.6,180)(128,87)
\qbezier(128,87)(133.5,40)(150,40)

\qbezier(100,180)(83.4,180)(72,87)
\qbezier(72,87)(66.5,40)(50,40)

\put(110,140){\makebox(40,0)[l]{A}}

\put(121,131){\circle*{3}}
\qbezier(121,131)(111,129)(100,88)
\qbezier(121,131)(131,133)(142,92)

\qbezier(99,85)(98,82)(97,80)
\qbezier(143,89)(144,86)(145,84)

\qbezier(96,78)(95,76)(94,75)
\qbezier(146,82)(147,80)(148,79)

\end{picture}
\caption{\footnotesize{This picture shows a U(1) invariant potential. At the point $A$ the
second derivative following the $\theta$-circle-trajectory will be zero, exactly because of the U(1) invariance of the potential, but
the genuine second derivative given by the intersection with a vertical plane will not be zero. Therefore the particle corresponding to the excitation of the field in this field-direction will
not be massless. }}
\label{doorsneden}
\end{center}
\end{figure}

Let us take a closer look at the definition of field-dependent particle masses in the case of a U(1) symmetry.
Take the simplest possible global U(1) invariant Lagrangian density:
\begin{equation}
\mathscr{L} = (\partial_{\mu}\phi)^{\dagger}\partial^{\mu}\phi-m^{2}|\phi|^{2} \; .
\end{equation}
We will look at it from two different coordinate-systems at exactly the same point in field space $\phi=1/\sqrt{2} \Phi$.
The Cartesian coordinates
\begin{equation}
\phi=\frac{1}{\sqrt{2}}(\Phi+\phi_{1}+i \phi_{2})
\end{equation}
give the Lagrangian density
\begin{equation}
\mathscr{L} = \frac{1}{2}\partial_{\mu}\phi_{1}\partial^{\mu}\phi_{1}+\frac{1}{2}\partial_{\mu}\phi_{2}\partial^{\mu}\phi_{2}
-\frac{1}{2}m^{2}(\phi_{1}^{2}+\phi_{2}^{2}+2\phi_{1}\Phi+\Phi^{2}) \; .
\end{equation}
Both scalar fields have normal kinetic terms and both have a classical mass $m$, because the tree-level propagator (being the inverse of the quadratic part of the Lagrangian) in momentum space has a pole at $k^{2}=m^{2}$ . If we look at the same Lagrangian in polar coordinates
\begin{equation}
\phi=\frac{1}{\sqrt{2}}(\Phi+\rho) e^{i \theta}
\end{equation}
we get
\begin{equation}
\mathscr{L} = \frac{1}{2}\partial_{\mu}\rho\partial^{\mu}\rho+\frac{1}{2}(\Phi+\rho)\partial_{\mu}\theta\partial^{\mu}\theta
-\frac{1}{2}m^{2}(\rho^{2}+2\rho\Phi+\Phi^{2}) \; .
\end{equation}
The $\rho$-field enters in exactly the same way as the $\phi_{1}$-field before, but we can't read of the $\theta$-mass directly, because of the non-standard kinetic term, so this looks like a bad choice of coordinates. There is no term proportional to $\theta^{2}$, because the second derivative of the potential along the circle trajectory corresponding to the $\theta$-field is zero. The real second derivative in the straight $\theta$-direction corresponds to the $\phi_{2}$-mass and is certainly not zero, see figure (\ref{doorsneden}). The trace of the $(mass)^{2}$-matrix is just equal to the Laplacian\footnote{The Laplacian is the sum of all second derivatives in perpendicular directions.}, which is of course coordinate independent, as long as the coordinate trajectories are straight lines. This all seems obvious, and people
will not interpret the $\theta$-field as a massless field in this case, because of the non-standard kinetic term.

However if we make the global U(1) symmetry local, the gaugefield will enter. It's coupling with the $\theta$-field will cancel all $\theta$-dependence and the masses can be easily evaluated again. Indeed using $B_{\mu}=A_{\mu}+\frac{1}{2}\partial_{\mu}\theta$ the Lagrangian density becomes:
\begin{equation}
\mathscr{L} = \frac{1}{2}\partial_{\mu}\rho\partial^{\mu}\rho+\frac{1}{2}g^{2}(\Phi+\rho)^{2} B_{\mu} B^{\mu} -\frac{1}{2}m^{2}(\Phi+\rho)^{2} -\frac{1}{4}F_{\mu\nu}F^{\mu\nu} \; .
\label{eq42}
\end{equation}
The $\theta$-mass has disappeared. 

If instead we stay in the Cartesian basis, in which the masses were easy to read of before making the global U(1) invariance local, then the second order term for the `Goldstone'\footnote
{Maybe it is not right to call this a Goldstone boson, because this terminology is normally used in the global minimum (of, for example, the Mexican hat potential). In
this minimum the Goldstone boson will always be massless, even for the global case.} boson does not disappear. We can either take exactly the same definition of $B_{\mu}$ as before, then the
Lagrangian density becomes:
\begin{displaymath}
\mathscr{L} = \frac{1}{2}\frac{(\partial_{\mu}\phi_{1}(\Phi+\phi_{1})+\partial_{\mu}\phi_{2} \phi_{1})^{2}}{(\Phi+\phi_{1})^{2}+\phi_{2}^{2}}
\end{displaymath}
\begin{equation}
+\frac{1}{2}g^{2}((\Phi+\phi_{1})^{2}+\phi_{2}^{2})B_{\mu}B^{\mu}-\frac{1}{2}m^{2}((\Phi+\phi_{1})^{2}+\phi_{2}^{2})-\frac{1}{4}F_{\mu\nu}F^{\mu\nu} \; .
\end{equation}
Or we can use a locally equivalent definition of the Higgs-mixing $A_{\mu}=C_{\mu}-\frac{1}{g}\frac{1}{\Phi}\partial_{\mu}\phi_{2}$ yielding:
\begin{equation}
\mathscr{L} = \frac{1}{2}\partial_{\mu}\phi_{1}\partial^{\mu}\phi_{1}+\frac{1}{2}g^{2}\Phi^{2}C_{\mu}C^{\mu}-\frac{1}{2}m^{2}((\Phi+\phi_{1})^{2}+\phi_{2}^{2})-\frac{1}{4}F_{\mu\nu}F^{\mu\nu}+... \; ,
\end{equation}
where the dots stand for third- and higher order terms in the fields.
In both cases the second order term for the Goldstone boson does not disappear in applying the Higgs mechanism. However we cannot associate a pole in the propagator with this term, because there is no kinetic term for the $\phi_{2}$-field. Where in the global case one could either define the classical mass to be used in the Coleman Weinberg equation (\ref{eqf}) by the pole in the propagator or by the second derivative in the diagonal field direction. For the local case, including the Higgs mixing, both definitions no longer agree. In fact the second derivative in the `Goldstone' field direction is the same as in the global case, but the `Goldstone' mass coming from poles in the propagator arising from equation (\ref{eq42}) has vanished.

\end{document}